\documentclass[10pt,twocolumn,aps,prd,preprintnumbers,showpacs,superscriptaddress,nofootinbib,amsmath,amssymb,floats,floatfix,showkeys,notitlepage,longbibliography]{revtex4-1}
\usepackage{lipsum}
\usepackage{mathtools}
\usepackage{verbatim}
\usepackage[normalem]{ulem}

\usepackage{subcaption}
\usepackage[font=small,labelfont=bf]{caption} 
\usepackage{comment}
\usepackage{graphicx}

\usepackage{palatino}
\usepackage[commandnameprefix=always]{changes}
\usepackage{hyperref}
\hypersetup{colorlinks=true,linkcolor=blue,urlcolor=blue,citecolor=blue}
\usepackage[toc,page]{appendix}
\usepackage[normalem]{ulem}

\usepackage{orcidlink}
\usepackage{lipsum}

\usepackage{palatino}

\usepackage{adjustbox}
\usepackage{latexsym}
\usepackage{amsmath}
\usepackage{amssymb}
\usepackage{amsfonts}
\usepackage{dcolumn}
\usepackage{bm}
\usepackage{tikz}
\usepackage{bigints}
\usepackage{array,tabularx,multirow,booktabs}
\usepackage[tracking=true]{microtype}
\SetTracking{}{500}
\SetTracking{encoding={*}, shape=sc}{40}
\UseRawInputEncoding
\allowdisplaybreaks
\usepackage{adjustbox}
\usepackage{latexsym}
\usepackage{amsmath}
\usepackage{amssymb}
\usepackage{amsfonts}
\usepackage{dcolumn}
\usepackage{bm}
\usepackage{tikz}
\usepackage{bigints}
\usepackage{array,tabularx,multirow,booktabs}
\usepackage[tracking=true]{microtype}
\usepackage{xcolor}
\usepackage{colortbl} 
\UseRawInputEncoding
\allowdisplaybreaks
 
\begin{document} \sloppy

\title{Branch structure and nonextensive thermodynamics of Kalb-Ramond-ModMax black holes: observational signatures}

\author{Erdem Sucu
\orcidlink{0009-0000-3619-1492}
}
\email{23600348@emu.edu.tr}
\affiliation{Physics Department, Eastern Mediterranean University, 99628, 
Famagusta, Cyprus}
\author{\.{I}zzet Sakall{\i} \orcidlink{0000-0001-7827-9476}}
\email{izzet.sakalli@emu.edu.tr}
\affiliation{Physics Department, Eastern Mediterranean
University, 99628, Famagusta, Cyprus}
\author{Emmanuel N. Saridakis
\orcidlink{0000-0003-1500-0874}
}
\email{msaridak@noa.gr}
 \affiliation{Institute for Astronomy, Astrophysics, Space Applications and 
Remote Sensing, National Observatory of Athens, 15236 Penteli, Greece}
 \affiliation{Departamento de Matem\'{a}ticas, Universidad Cat\'{o}lica del 
Norte,
Avda.
Angamos 0610, Casilla 1280 Antofagasta, Chile}
\affiliation{Department of Astronomy, School of Physical Sciences, 
University of Science and Technology of China, Hefei 230026, P.R. China}

\begin{abstract}
Motivated by the low-energy effective action of heterotic string theory, where the 
Kalb-Ramond (KR) two-form and nonlinear gauge corrections arise simultaneously, we 
investigate a static, spherically symmetric black hole (BH) in Einstein gravity coupled 
to a KR field and ModMax nonlinear electrodynamics (NED). The solution depends, beyond 
mass and charge, on the Lorentz-symmetry-breaking (LSB) parameter $\ell$, the ModMax 
deformation parameter $\gamma$, and a discrete branch selector $\zeta=\pm1$. We show 
that the ordinary branch admits extremal and non-extremal configurations, while the 
phantom branch generically supports a single-horizon geometry. BH thermodynamics is 
analyzed within the Tsallis non-extensive framework, revealing branch-dependent stability 
and Joule-Thomson (JT) behavior. Weak gravitational lensing is computed via the 
Ono-Ishihara-Asada (OIA) extension of the Gauss-Bonnet (GB) theorem, yielding a negative 
topological correction that reduces light bending relative to the Schwarzschild baseline, 
opposite in sign to Barriola-Vilenkin (BV) monopole backgrounds. Photon sphere (PS) 
properties in plasma environments and tidal forces through geodesic deviation are also 
studied, revealing a universal tidal balance ratio $R_{\rm rad}/R_{\rm ang}=3/2$ in the 
ordinary branch. These multi-channel signatures provide concrete observational handles 
for constraining the KR-ModMax framework through Event Horizon Telescope (EHT) data and 
next-generation interferometric arrays.
\end{abstract}

\maketitle
 
\section{Introduction}
\label{isec1}

BHs occupy a central position in modern theoretical physics, providing an arena where gravity, quantum field theory, and high-energy physics intersect \cite{Giddings:2020zso,BarrosoVarela:2023ull}. Within General 
Relativity (GR), the Schwarzschild and Reissner-Nordstr\"{o}m (RN) solutions constitute the canonical descriptions of neutral and electrically charged BHs, respectively. Nevertheless, both theoretical considerations related to quantum gravity and a growing body of observational evidence from cosmology and astrophysics motivate the exploration of gravitational frameworks that extend 
or modify GR \cite{CANTATA:2021asi,clifton2012modified}. Such extensions often introduce additional fields, nonlinearities, or symmetry-breaking mechanisms, leading to nontrivial modifications of the causal structure, thermodynamics, and optical properties of BH spacetimes \cite{Nojiri:2010wj,Bambi:2015kza, Nashed:2018cth,Nashed:2020kdb,Nashed:2021pah,Nashed:2022yfc}. On the observational front, EHT images of M87$^{*}$ and Sgr\,A$^{*}$, together with gravitational-wave catalogs from the LIGO-Virgo-KAGRA collaboration, have opened a direct empirical window into the strong-field regime, making the construction of well-motivated BH solutions with quantifiable and testable signatures a pressing task. 

Among the most theoretically well-motivated extensions are those inspired by string theory and its low-energy effective descriptions. In this context, the KR field, a rank-two antisymmetric tensor arising naturally in the bosonic and heterotic string spectra, provides a concrete realization of 
new gravitational degrees of freedom \cite{Lessa:2019bgi,Duan:2023gng,WOS:001565141800002}. The associated three-form field strength can be interpreted either as spacetime torsion or as a vacuum condensate inducing Lorentz symmetry breaking (LSB) in the gravitational sector \cite{SenGupta:2001cs,Belinchon:2016jgc}. This identification places the KR field at the intersection of two active research programs: the search for string-theoretic imprints in low-energy gravity and the experimental effort aimed at constraining Lorentz violation through solar-system tests, pulsar timing, and BH shadow observations \cite{Kostelecky:2018yfa}. When coupled to 
Einstein gravity, the KR background modifies the spacetime geometry through an effective parameter $\ell$ that rescales the asymptotic structure while preserving spherical symmetry. The resulting BH geometries exhibit horizon and thermodynamic properties that differ from their GR counterparts, rendering them suitable probes of string-inspired effects in the strong-field regime 
\cite{Kao:1996ea,Majumdar:1999jd}. Crucially, the modified asymptotic normalization $f(r)\to (1-\ell)^{-1}$ as $r\to\infty$ produces a non-Euclidean optical geometry whose lensing signatures can, in principle, be distinguished from those of global monopole defects and cloud-of-string backgrounds that generate superficially similar metric modifications.

We emphasise that the present framework does not modify the Einstein tensor: no higher-curvature invariants or scalar-tensor restructuring enter the gravitational action. The ``Lorentz-violating'' label refers to the spontaneous breaking of local Lorentz invariance induced by the KR vacuum expectation value $\langle B_{\mu\nu}\rangle=b_{\mu\nu}$, which enters the gravitational sector through the non-minimal coupling $\xi\,B^{\mu\alpha}B_{\nu\alpha}R^{\nu}{}_{\mu}$ and renormalises the effective Newton constant to $G_{\rm eff}=G/(1-\ell)$ \cite{isrply01}. In this sense the modification is at the level of the matter sector, while the gravitational response is that of standard GR with a structured source.

In parallel, NED has emerged as a powerful framework for addressing strong-field phenomena in gauge theories coupled to gravity \cite{Born:1934gh,Ayon-Beato:1998hmi}. Beyond the classic Born-Infeld construction \cite{Born:1934gh}, which regularizes the Coulomb self-energy but breaks conformal symmetry, recent attention has focused on ModMax electrodynamics, which constitutes the unique one-parameter deformation of Maxwell theory preserving both conformal invariance and electric-magnetic duality \cite{Bandos:2020jsw,Bandos:2020hgy}. These two symmetries are not merely formal requirements: conformal invariance governs the ultraviolet structure of the gauge sector and constrains the renormalization group flow, while electric-magnetic duality plays a structural role in extended supergravity and string compactifications. The ModMax Lagrangian introduces a dimensionless parameter $\gamma$ that interpolates 
continuously between the Maxwell limit and a genuinely nonlinear regime. When minimally coupled to gravity, ModMax electrodynamics gives rise to BH solutions with modified charge contributions and novel horizon structures \cite{Bandos:2020jsw,Barrientos:2022bzm, AraujoFilho:2024rcr, Filho:2024tgy, Barrientos:2024umq,Chen:2024ilt, AraujoFilho:2025jcu,Babaei-Aghbolagh:2025tim,Sekhmani:2025kav, AraujoFilho:2025fwd, Barrientos:2025rde,Shi:2025rfq}. A distinctive feature of this theory is the emergence of two mathematically consistent branches: an ordinary branch ($\zeta = +1$), continuously connected to the RN geometry, and a phantom branch ($\zeta = -1$), in which the electromagnetic energy density 
effectively reverses sign, leading to qualitatively different physical behavior \cite{EslamPanah:2024fls,isrply04}. This branch dichotomy is not an artifact of a coordinate choice but reflects a genuine physical degeneracy in the nonlinear sector, analogous to the distinction between electric and magnetic BHs in Born-Infeld theory.

The question naturally arises as to why these two extensions, KR and ModMax, should be considered jointly rather than in isolation. The answer is rooted in the structure of string effective field theory. In heterotic string compactifications, the low-energy bosonic sector generically contains, in addition to the metric and dilaton, both the KR two-form and nonlinear corrections to the gauge field action \cite{Lessa:2019bgi,Duan:2023gng}. The leading $\alpha'$-corrections to the open-string effective action modify the Born-Infeld Lagrangian, while the KR field provides the closed-string antisymmetric tensor background; both arise at the same order in the string-length expansion and couple to gravity simultaneously. ModMax, as the unique conformally invariant and duality-preserving deformation, can be regarded as the most constrained representative of these nonlinear gauge corrections, retaining the maximal set of symmetries compatible with nonlinearity \cite{Bandos:2020jsw,Bandos:2020hgy}. Studying the combined Einstein-KR-ModMax system therefore captures the qualitative features of a broad class of string-motivated corrections to electrovacuum gravity, without introducing the additional complexity of dilaton dynamics or higher-curvature terms. This is not an ad hoc parameter stacking, but rather a controlled truncation of the string effective action that isolates two physically distinct deformation channels: LSB through the gravitational sector (parametrized by $\ell$) and nonlinear gauge dynamics (parametrized by $\gamma$ and $\zeta$).

In the Einstein-KR-ModMax 
framework, the spacetime geometry is controlled jointly by three deformation parameters: the KR parameter $\ell$, the ModMax nonlinearity parameter $\gamma$, and the branch selector $\zeta$. Each parameter governs an independent physical effect and admits a separate observational constraint, a feature that distinguishes this framework from single-parameter extensions where a single extra parameter can often be absorbed into effective redefinitions of mass and charge. The branch parameter $\zeta$ generates a discrete topological distinction (one- versus two-horizon structure) that no continuous parameter can mimic, the KR parameter $\ell$ modifies the asymptotic normalization in a manner distinct from global monopole or cloud-of-string backgrounds (which affect the solid-deficit angle rather than the lapse function), and the ModMax parameter $\gamma$ exponentially suppresses the effective charge through $e^{-\gamma}$ without altering the horizon count. This parameter independence enables joint constraints: shadow-size measurements constrain $(\ell, \gamma)$ combinations, deflection-angle anomalies are sensitive to $\ell$ through the topological boundary term, and thermodynamic observables such as the JT inversion temperature discriminate between the two branches. Understanding how these different modifications combine and manifest in physical observables is essential for assessing their phenomenological relevance \cite{Kostelecky:2018yfa, Barrientos:2022uit,Yang:2023wtu, Sucu:2025lqa}.

BH thermodynamics offers a particularly sensitive probe of the underlying 
gravitational dynamics. Since the formulation of the laws of BH mechanics and 
the discovery of Hawking radiation \cite{Bardeen:1973gs,Hawking:1975vcx}, 
thermodynamic considerations have provided deep connections 
between gravity, quantum theory, and statistical mechanics. However, the 
standard Boltzmann-Gibbs framework may be inadequate for gravitational systems 
characterized by long-range interactions and strong correlations 
\cite{WOS:000322575300012Tsallis,Tsallis:1987eu}. Non-extensive generalizations, most 
notably Tsallis entropy, introduce a deformation parameter $\delta$ that 
quantifies deviations from the area law and can significantly modify the 
thermodynamic behavior of BHs \cite{WOS:000322575300012Tsallis,Saridakis:2018unr,Basilakos:2023kvk}. Within 
this framework, quantities such as internal energy, free energy, pressure, and 
heat capacity acquire nontrivial corrections, potentially altering phase 
structure and stability 
\cite{Abreu:2020wbz,Nojiri:2023saw,sucu2026spin}. The JT 
expansion further enriches this picture by providing a diagnostic of 
heating-cooling transitions in extended thermodynamic space 
\cite{Kruglov:2023ogn,Liang:2021elg,Sakalli:2025els,Zhang:2024fxj}. For KR-ModMax BHs, the Tsallis framework is particularly appropriate because the non-Euclidean asymptotics induced by $\ell$ already signal a departure from the standard area-entropy relation, and the branch parameter $\zeta$ creates two thermodynamically distinct sectors that the Boltzmann-Gibbs description would treat identically at the level of horizon area alone.

On the observational side, gravitational lensing remains one of the most direct probes of spacetime geometry in the strong-field regime. Since its first experimental confirmation \cite{Dyson:1920cwa}, lensing has evolved into a precision tool for testing GR and constraining alternative theories \cite{Will:2014kxa}. The GB approach 
\cite{Gibbons:2008rj,Werner:2012rc} offers a coordinate-independent method for computing weak-field deflection angles and has been widely applied in BH spacetimes \cite{Ren:2021uqb,Sucu:2025pce,WOS:001575165100002AstroErdem,Wang:2024eai}. However, the standard formulation assumes asymptotically Euclidean optical geometry, a condition that is violated for the KR-ModMax metric due to the modified lapse at infinity. In such cases, the OIA generalization \cite{Ono:2017pie,Ishihara:2016vdc,Jusufi:2017mav} must be employed, which introduces a topological boundary correction encoding the non-flat asymptotic structure. This correction is not merely a technical refinement; it produces a \emph{negative} contribution to the deflection angle that is absent in asymptotically flat spacetimes and carries a direct imprint of the KR parameter $\ell$. Detecting or constraining such a topological deficit through precision astrometry or EHT ring-size measurements would provide direct evidence for, or bounds on, LSB in the gravitational sector. Moreover, realistic astrophysical environments are typically permeated by plasma, which modifies photon propagation through dispersive effects. In such settings, the PS and shadow properties depend not only on the spacetime geometry but also on the plasma distribution, providing additional discriminants among competing gravitational models \cite{Perlick:2015vta,Rogers:2015dla,Sucu:2025fwa}. For the KR-ModMax geometry, the branch parameter $\zeta$ shifts the PS radius in opposite directions for ordinary and phantom configurations, an effect that is further modulated by plasma frequency profiles and translates directly into observable differences in the angular radius and luminosity profile of the BH shadow.

Tidal forces constitute a complementary probe of the strong-field regime. 
Through the geodesic deviation equation, tidal effects encode local curvature 
information and govern the deformation of extended bodies approaching a BH 
\cite{hobson2006general,WOS:000433093100006PETER}. In modified gravity scenarios, 
tidal force profiles may deviate substantially from the Schwarzschild case, 
potentially exhibiting enhanced magnitudes or sign reversals that influence 
tidal disruption processes 
\cite{frolov2012black,Burdge:2019hgl,Poisson:2014gka}. For the KR-ModMax BHs, the branch dichotomy introduces a qualitative split: the ordinary branch admits a tidal balance radius at which radial stretching and angular compression cancel, while the phantom branch exhibits monotonically enhanced stretching due to the reinforced gravitational potential. The ratio of radial to angular tidal balance radii, $R_{\rm rad}/R_{\rm ang} = 3/2$, emerges as a geometric constant independent of charge, $\ell$, or $\gamma$, providing a parameter-free prediction that can serve as a null test of the underlying theory.

Motivated by these considerations, in this work we investigate KR-ModMax BHs across both their ordinary and phantom 
branches. We analyze the spacetime structure and thermodynamics within the 
Tsallis entropy framework, including the JT expansion, and explore optical 
signatures through weak gravitational lensing and photon propagation in plasma 
environments. Finally, we examine tidal forces and identify branch-dependent 
features, including the conditions under which tidal inversion may occur. By 
combining geometric, thermodynamic, and observational diagnostics within a 
unified setting, we aim to clarify the physical implications of branch 
structure and non-extensive effects, and to identify signatures that could 
distinguish these BHs from standard GR solutions and other modified gravity 
models.

The paper is organized as follows.
In Sec.~\ref{isec2} we present the KR-ModMax BH solution, analyze 
its branch-dependent horizon structure, and derive the Hawking temperature.
Sec.~\ref{isec3} develops the non-extensive thermodynamic description based 
on Tsallis entropy, while the JT expansion is examined in
Sec.~\ref{isec4}. In
Sec.~\ref{isec5} we discuss weak gravitational lensing and vacuum 
PS properties. Plasma effects on photon propagation and their impact 
on PSs are analyzed separately in Sec.~\ref{isec6}.
In Sec.~\ref{isec7} we investigate tidal forces and branch-dependent
deformations.
Finally, Sec.~\ref{sec8} summarizes our results and outlines future
directions.

\section{KR-ModMax BH Solution and Horizon Structure}\label{isec2}

Having established in the preceding section that the KR and ModMax sectors arise simultaneously in the low-energy limit of heterotic string theory and that their combined treatment isolates two physically distinct deformation channels, we now turn to the explicit construction of the BH solution.

\subsection{Action, field equations, and ans\"atze}
\label{subsec:action}

The gravitational model combines Einstein gravity, a self-interacting
KR two-form $B_{\mu\nu}$, and ModMax NED with a discrete branch selector
$\zeta=\pm 1$ that flips the sign of the gauge-kinetic sector
\cite{Sekhmani:2025epe,Bandos:2020jsw,Lessa:2019bgi,isrply04}. The total action reads
\begin{equation}
S = \!\int\! d^{4}x\,\sqrt{-g}\!\left[\frac{R}{16\pi G}
+\mathcal{L}_{\rm KR} + \mathcal{L}_{\rm ModMax}
+\mathcal{L}_{\rm int}\right]\!,
\label{eq:action}
\end{equation}
where the KR sector carries the field strength
$H_{\mu\nu\rho}=\partial_{[\mu}B_{\nu\rho]}$ and a potential
$V(B^{\mu\nu}B_{\mu\nu}\pm b^{2})$ with a nonzero vacuum expectation value
$\langle B_{\mu\nu}\rangle=b_{\mu\nu}$ minimising $V$,
\begin{equation}
\mathcal{L}_{\rm KR} = -\frac{1}{12}\,H^{\mu\nu\rho}H_{\mu\nu\rho}
- V\!\left(B^{\mu\nu}B_{\mu\nu}\pm b^{2}\right).
\label{eq:LKR}
\end{equation}
The ModMax Lagrangian is the conformal, duality-invariant one-parameter
deformation of Maxwell theory
\cite{Bandos:2020jsw,Bandos:2020hgy}
\begin{equation}
\mathcal{L}_{\rm ModMax} = \frac{\zeta}{2}\!\left[\mathcal{S}\cosh\gamma
- \sqrt{\mathcal{S}^{2}+\mathcal{P}^{2}}\,\sinh\gamma\right]\!,
\label{eq:LMM}
\end{equation}
with $\mathcal{S}=-\tfrac{1}{2}F_{\mu\nu}F^{\mu\nu}$ and
$\mathcal{P}=-\tfrac{1}{2}F_{\mu\nu}\tilde{F}^{\mu\nu}$. The discrete flip
$\zeta=\pm 1$ defines the ordinary ($+1$) and phantom ($-1$) branches: the
phantom choice is equivalent to an analytic continuation of the ModMax
Lagrangian that preserves $O(2)$ duality while reversing the sign of the
electromagnetic stress-energy density \cite{Sekhmani:2025epe,EslamPanah:2024fls,isrply04}.
The interaction term $\mathcal{L}_{\rm int}$ encodes the coupling between the
KR background and the electromagnetic sector and is fixed, up to redefinitions,
by requiring that the spherically symmetric reduction preserve the
Arnowitt-Deser-Misner (ADM) structure \cite{Lessa:2019bgi,Duan:2023gng}.

Variation of Eq.~\eqref{eq:action} with respect to $g^{\mu\nu}$, $A_{\mu}$,
and $B_{\mu\nu}$ gives the Einstein, modified Maxwell, and KR equations
\begin{align}
G_{\mu\nu} &= 8\pi G\!\left(T^{\rm KR}_{\mu\nu}+T^{\rm ModMax}_{\mu\nu}\right)\!,
\label{eq:Einstein}\\[2pt]
\nabla_{\mu}\!\left(\mathcal{E}^{\mu\nu}\right) &= 0,\qquad
\mathcal{E}^{\mu\nu}\equiv\frac{\partial\mathcal{L}_{\rm ModMax}}
{\partial F_{\mu\nu}},\label{eq:ModMaxEOM}\\[2pt]
\nabla^{\mu}H_{\mu\nu\rho} &= 2\,\frac{\partial V}{\partial B^{\nu\rho}}
+ \mathcal{J}_{\nu\rho}^{\rm int}.\label{eq:KREOM}
\end{align}
We adopt the standard spherically symmetric ans\"atze
\begin{equation}
ds^{2}=-f(r)dt^{2}+\frac{dr^{2}}{f(r)}+r^{2}d\Omega^{2},\qquad
A_{\mu}dx^{\mu}=\Phi(r)\,dt,
\label{eq:ansatz}
\end{equation}
together with a spherically-symmetric KR background with pseudo-electric
component $b_{10}=-b_{01}$ and $\langle B_{\mu\nu}B^{\mu\nu}\rangle=2b^{2}$
\cite{Duan:2023gng,Sekhmani:2025epe}. Substituting
Eqs.~\eqref{eq:ansatz} into Eqs.~\eqref{eq:Einstein}-\eqref{eq:KREOM} and
imposing asymptotic flatness (up to the KR rescaling) yields the electrostatic
potential~\eqref{eq:potential} and the exact lapse
function~\eqref{metric_solution}, with $\ell\equiv\xi\,b^{2}$ the dimensionless
LSB parameter absorbing the non-minimal KR-Ricci coupling.

The parameter $Q$ appearing in Eq.~\eqref{metric_solution} denotes the
ADM-Gaussian electric charge defined by the surface integral at spatial
infinity,
\begin{equation}
Q = \frac{1}{4\pi}\!\oint_{S^{2}_{\infty}}\!\star\mathcal{E},
\label{eq:charge_def}
\end{equation}
where $\star\mathcal{E}$ is the Hodge dual of the ModMax conjugate field
strength $\mathcal{E}^{\mu\nu}$ defined in Eq.~\eqref{eq:ModMaxEOM}. The
definition reduces to the Maxwell charge as $\gamma\to 0$.

The total stress-energy tensor sourcing Eq.~\eqref{eq:Einstein} is the sum
\begin{equation}
T_{\mu\nu}=T^{\rm KR}_{\mu\nu}+T^{\rm ModMax}_{\mu\nu},
\label{eq:Ttotal}
\end{equation}
with
\begin{equation}
T^{\rm ModMax}_{\mu\nu}=\zeta\!\left[\mathcal{E}^{\mu\alpha}F^{\nu}{}_{\alpha}
-g^{\mu\nu}\mathcal{L}_{\rm ModMax}\right]\!.
\label{eq:TModMax}
\end{equation}
The branch selector $\zeta$ flips the sign in the phantom sector. The
KR contribution follows from $\mathcal{L}_{\rm KR}+\mathcal{L}_{\rm int}$
by metric variation and reproduces the expressions of
\cite{Lessa:2019bgi,Duan:2023gng} when the interaction term is absorbed into
$\mathcal{L}_{\rm KR}$. Contracting Eq.~\eqref{eq:Einstein} with the Bianchi
identity and imposing the matter equations of
motion~\eqref{eq:ModMaxEOM}-\eqref{eq:KREOM} shows that the total tensor is
conserved, $\nabla^{\mu}T_{\mu\nu}=0$, while the individual sectors exchange
energy-momentum through $\mathcal{L}_{\rm int}$. Throughout we work in
geometrised units $G=c=\hbar=k_{B}=1$.

\subsection{Field equations, exact solution, and horizon structure}

We consider static and spherically symmetric BH geometries arising from 
Einstein gravity coupled simultaneously to a KR three-form 
condensate and ModMax NED. This framework combines 
LSB effects of string-theoretic origin with the 
conformally invariant deformation of Maxwell theory 
\cite{Sekhmani:2025epe,Bandos:2020jsw,Kruglov:2022bhx}. Owing to spherical 
symmetry, the spacetime metric can be written in the standard form
\begin{equation}
ds^{2} = -f(r)\,dt^{2} + \frac{dr^{2}}{f(r)} + r^{2}\,d\Omega^{2},
\label{metric_ansatz}
\end{equation}
where all geometric information is encoded in the lapse function $f(r)$, which 
captures both the KR-induced vacuum deformation and the nonlinear 
electromagnetic corrections \cite{Liu:2024oas,Duan:2023gng}. The Einstein 
equations, supplemented by the modified Maxwell sector, reduce to a single 
ordinary differential equation for $f(r)$.

The KR field originates from an antisymmetric rank-two tensor $B_{\mu\nu}$ 
appearing in the massless sector of bosonic string theory. A nonvanishing 
condensate of its associated three-form field strength induces spontaneous 
breaking of local Lorentz invariance and rescales the effective gravitational 
and electromagnetic couplings through a dimensionless parameter $\ell$ 
\cite{Illuminati:2021wfq,Mukhopadhyaya:2001dp}. Moreover, ModMax 
electrodynamics constitutes the unique one-parameter nonlinear extension of 
Maxwell theory that preserves both conformal invariance and electric-magnetic 
duality \cite{Babaei-Aghbolagh:2022itg,Bandos:2020jsw}. The deviation from 
linear electrodynamics is governed by the parameter $\gamma$, with the Maxwell 
limit recovered as $\gamma \to 0$.

\begin{figure*}[ht!]
\centering

\begin{subfigure}[b]{0.48\textwidth}
\includegraphics[width=\textwidth]{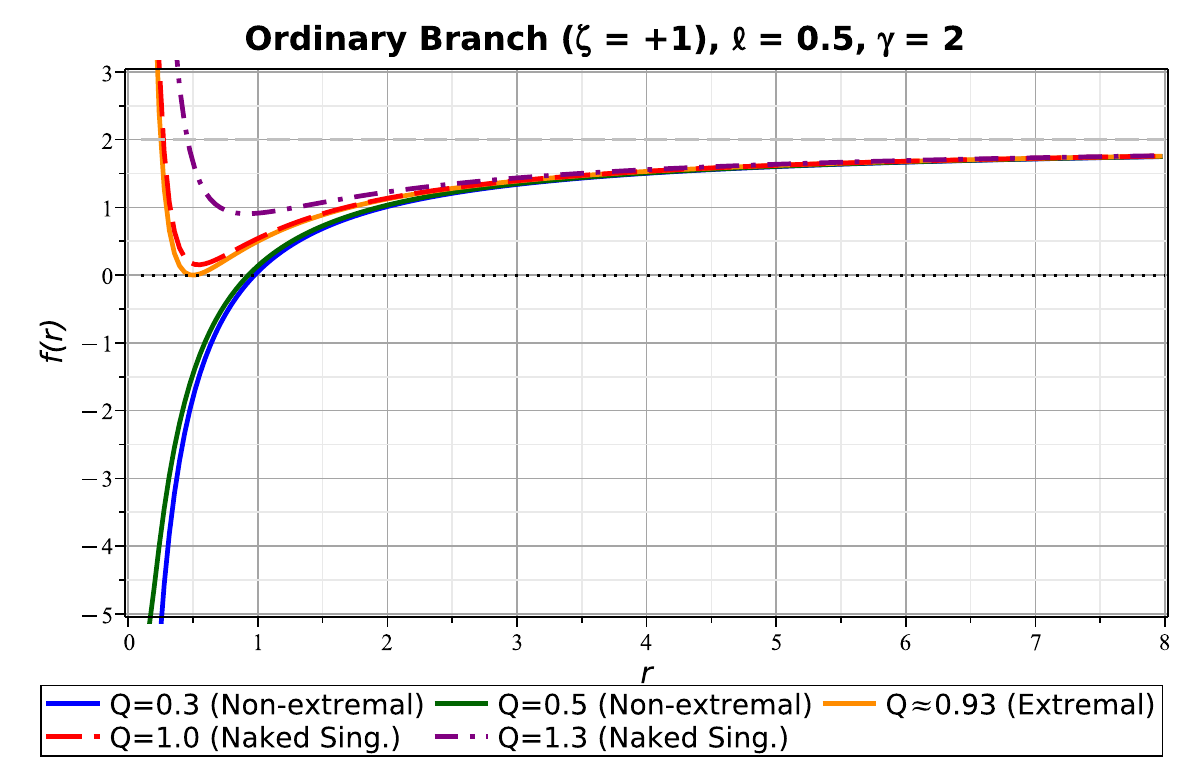}
\caption{ Ordinary ModMax branch ($\zeta=+1$) for fixed $\ell=0.5$ and 
$\gamma=2$.
Increasing electric charge $Q$ drives the geometry from non-extremal BHs
($Q=0.3,\,0.5$) to the extremal configuration ($Q\simeq0.93$), and finally to 
naked
singularities (NSs) ($Q=1.0,\,1.3$).}
\label{fig:V1metric}
\end{subfigure}
\hfill 
\begin{subfigure}[b]{0.48\textwidth}
\includegraphics[width=\textwidth]{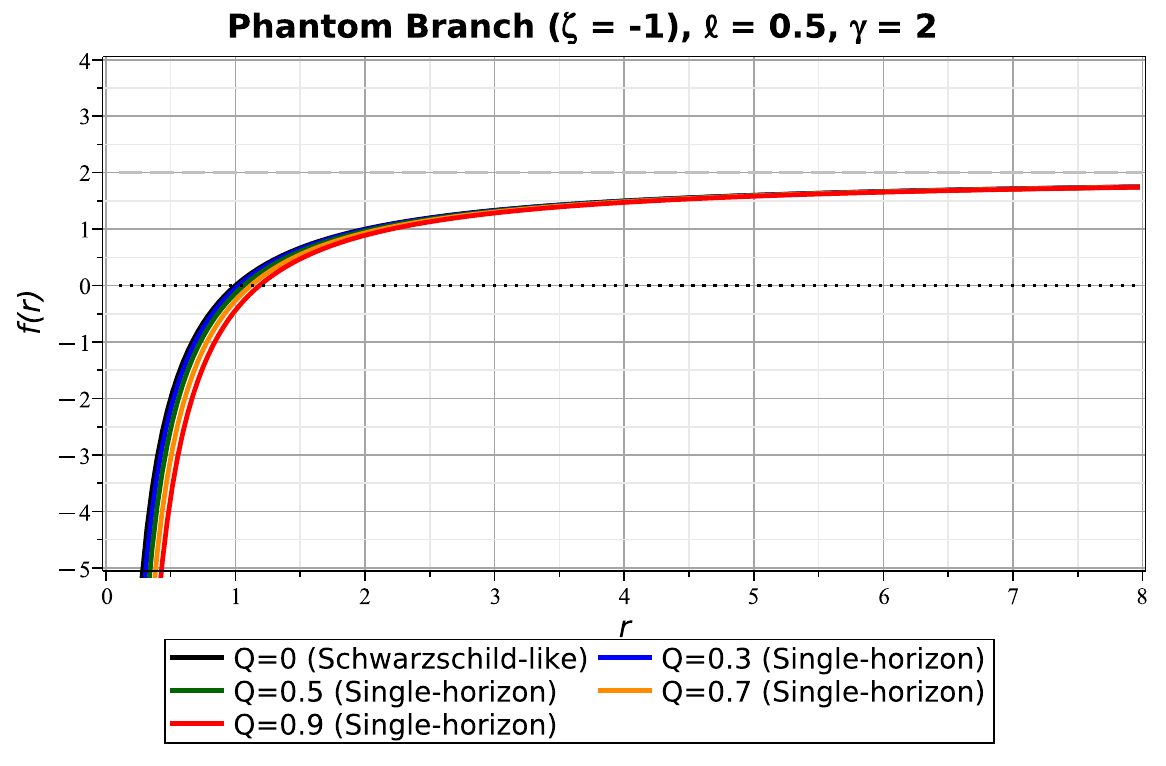}
\caption{Phantom ModMax branch ($\zeta=-1$) for $\ell=0.5$ and $\gamma=2$.
All configurations exhibit a single event horizon (EH), with no extremal or naked
solutions. The Schwarzschild limit ($Q=0$) is smoothly connected to the charged
phantom geometries.}
\label{fig:V2metric}
\end{subfigure}

\vspace{0.3cm}

\begin{subfigure}[b]{0.48\textwidth}
\includegraphics[width=\textwidth]{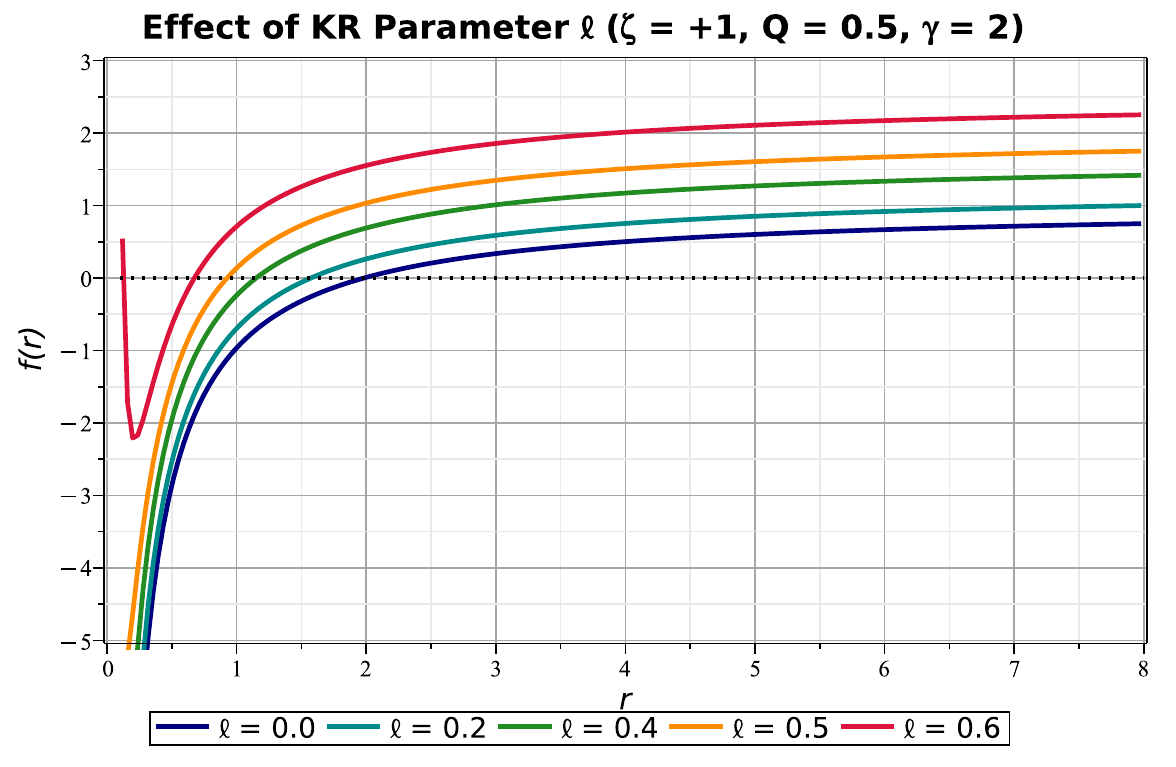}
\caption{Effect of the KR parameter $\ell$ on the ordinary branch
($\zeta=+1$, $Q=0.5$, $\gamma=2$). Increasing $\ell$ shifts the EH 
inward
and enhances the asymptotic normalization $f_\infty=(1-\ell)^{-1}$.}
\label{fig:V3metric}
\end{subfigure}
\hfill
\begin{subfigure}[b]{0.48\textwidth}
\includegraphics[width=\textwidth]{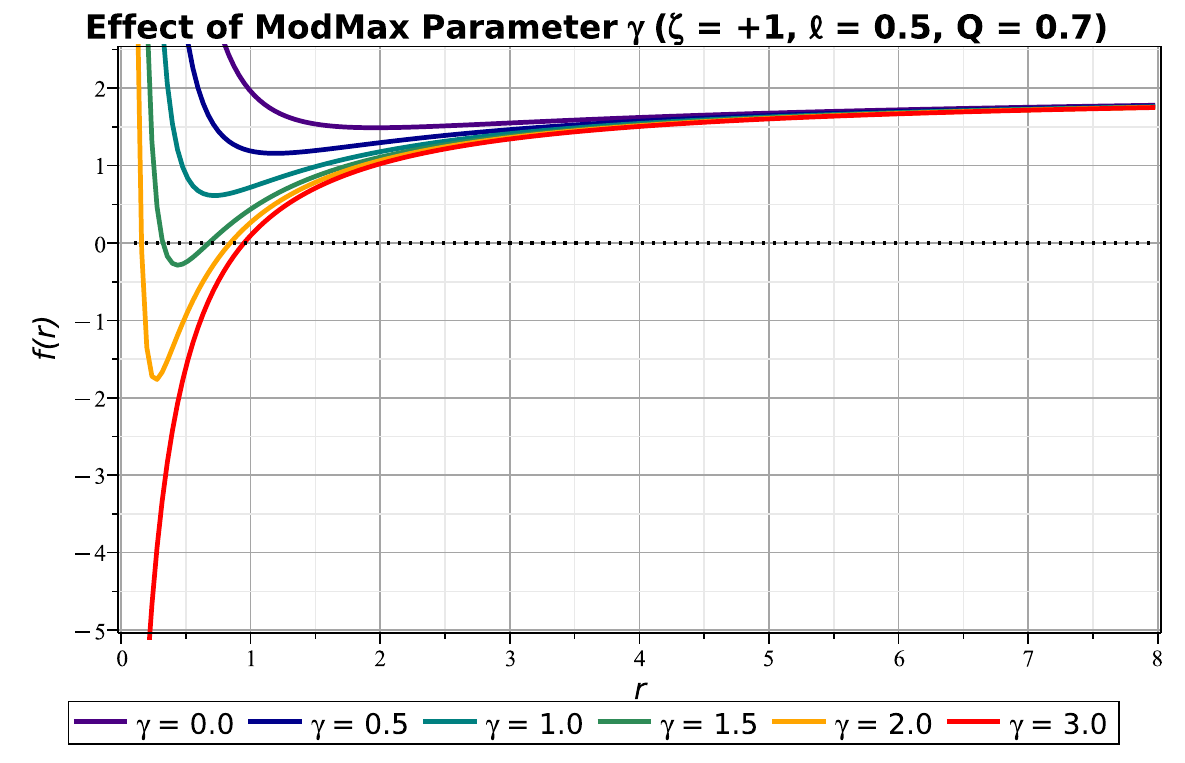}
\caption{Effect of the ModMax parameter $\gamma$ on the ordinary branch
($\zeta=+1$, $\ell=0.5$, $Q=0.7$). Larger $\gamma$ exponentially suppresses the
effective electric charge through the factor $e^{-\gamma}$, continuously driving
the geometry toward the Schwarzschild-KR limit.}
\label{fig:V4metric}
\end{subfigure}

\caption{Metric function $f(r)$ for KR-ModMax BHs in the ordinary
($\zeta=+1$) and phantom ($\zeta=-1$) branches. The horizontal dashed line
$f(r)=0$ marks the locations of EHs.
Panel (a) illustrates the charge-driven transition from non-extremal to extremal
and naked configurations in the ordinary branch.
Panel (b) demonstrates the absence of extremal or naked solutions in the phantom
sector.
Panels (c) and (d) display the sensitivity of the geometry to the KR
parameter $\ell$ and the ModMax parameter $\gamma$, respectively. The parameter regions where $f(r)$ admits two horizons (non-extremal BHs) are directly relevant for EHT shadow modeling, since only these configurations produce a well-defined PS and a corresponding shadow boundary. NS configurations ($Q > Q_{\rm ext}$) predict qualitatively different lensing images that could be distinguished by next-generation interferometric observations.}
\label{fig:metric_panels}

\end{figure*}

In this setting, the modified Maxwell equations admit a purely radial electric 
field configuration $\Phi=\Phi(r)$, constrained by charge conservation. Direct 
integration yields the electrostatic potential
\begin{equation}
\Phi(r) = \frac{Q\,e^{-\gamma}}{(1-\ell)\,r},
\label{eq:potential}
\end{equation}
where the factor $(1-\ell)^{-1}$ reflects the KR-induced rescaling of the 
effective electric coupling.  Equation~\eqref{eq:potential} follows from Eq.~\eqref{eq:charge_def} together with the ModMax constitutive relation $\mathcal{E}^{\mu\nu}=e^{-\gamma}F^{\mu\nu}$ in the purely electric sector. The exponential prefactor $e^{-\gamma}$ is the ModMax charge screening familiar from \cite{Bandos:2020jsw,isrply05}. Substituting Eq.~\eqref{eq:potential} into the gravitational field equations and imposing asymptotic flatness leads to the 
exact metric function \cite{Sekhmani:2025epe}
\begin{equation}
f(r) = \frac{1}{1-\ell} - \frac{2M}{r} + 
\frac{\zeta\,Q^{2}\,e^{-\gamma}}{(1-\ell)^{2}\,r^{2}},
\label{metric_solution}
\end{equation}
where $M$ is the ADM mass, $Q$ the electric charge, and $\zeta=\pm1$ 
distinguishes between the ordinary ($\zeta=+1$) and phantom ($\zeta=-1$) 
branches of ModMax electrodynamics. While the ordinary branch corresponds to a 
positive electromagnetic energy density and continuously connects to the 
RN solution, the phantom branch reverses the sign of the 
electromagnetic stress-energy contribution, leading to qualitatively different 
gravitational behavior.

The asymptotic structure of the spacetime follows from
\begin{equation}
\lim_{r\to\infty} f(r) = \frac{1}{1-\ell},
\label{eq:asymptotic}
\end{equation}
which shows that the KR parameter $\ell$ modifies the normalization of the 
timelike Killing vector at infinity. For $\ell<1$, the geometry remains 
asymptotically flat up to this constant rescaling, preserving the standard 
$1/r$ falloff of curvature invariants. Equivalently, the effective 
gravitational coupling is renormalized according to $G_{\text{eff}}=G/(1-\ell)$, 
with direct implications for orbital dynamics and gravitational lensing 
observables \cite{Masood:2024oej,Mukhopadhyaya:2001dp}. It is worth noting that this rescaling is structurally distinct from the solid-deficit angle produced by global monopole or cloud-of-string backgrounds: while those modify the angular part of the optical geometry, the KR deformation acts on the radial lapse, producing a different observational imprint in deflection-angle measurements as discussed in Sec.~\ref{isec5}.

  Writing the horizon condition $f(r_{+})=0$ as a polynomial in $r_{+}$, multiplication by $(1-\ell)^{2}r_{+}^{2}$ yields
\begin{equation}
(1-\ell)\,r_{+}^{2}-2M(1-\ell)^{2}\,r_{+}+\zeta\,Q^{2}e^{-\gamma}=0,
\label{eq:horizon_poly}
\end{equation}
with coefficient signs $(+,-,\zeta)$ for $\ell<1$. Descartes' rule of signs then makes the branch dichotomy manifest: the ordinary branch ($\zeta=+1$) has sign pattern $(+,-,+)$ with two sign changes and therefore admits zero or two positive real roots, corresponding to NS, extremal, or non-extremal configurations; the phantom branch ($\zeta=-1$) has sign pattern $(+,-,-)$ with a single sign change and therefore admits exactly one positive real root, establishing the single-EH topology of the phantom sector without reference to numerical tabulation.

In Fig.~\ref{fig:metric_panels} we display the radial profile of the metric function $f(r)$ for four representative configurations. In particular, panel~(a) illustrates the ordinary branch at fixed $(\ell, \gamma) = (0.5, 2.0)$: as $Q$ increases from $0.3$ to $1.3$, the solution transitions from a non-extremal BH with two distinct horizons, through the extremal configuration at $Q \approx 0.93$, to a NS for $Q \geq 1.0$. Panel~(b) presents the phantom branch under identical KR-ModMax parameters, where all curves intersect the $f=0$ axis exactly once, confirming the single-horizon topology. Finally, panels~(c) and~(d) examine the sensitivity to $\ell$ and $\gamma$ respectively, demonstrating how these parameters shift horizon locations and modify the near-horizon geometry. From an observational standpoint, the two-horizon configurations in panel~(a) are the ones producing well-defined PS and shadow boundaries amenable to EHT constraints, whereas the NS regime predicts qualitatively different lensing images without a sharp shadow edge.

\subsection{Branch dichotomy and physical interpretation}

A distinctive feature of the KR-ModMax BH solution is the 
existence of two mathematically consistent branches, labeled by $\zeta=\pm1$, 
which lead to qualitatively different spacetime structures and physical 
behavior. This branch dichotomy originates from the nonlinear electromagnetic 
sector and persists even in the presence of the KR deformation, 
playing a central role in shaping the properties of the solution.

For the ordinary branch ($\zeta=+1$), the electromagnetic contribution 
partially counteracts the gravitational attraction, closely resembling the 
behavior familiar from the RN geometry. As the electric 
charge approaches a critical value, the event and Cauchy horizons merge and the 
BH reaches an extremal configuration with vanishing Hawking temperature. 
This limit is associated with the emergence of a degenerate horizon and signals 
a qualitative change in the near-horizon geometry. For larger charge values, 
the solution admits no horizons, giving rise to NSs.

On the other hand, the phantom branch ($\zeta=-1$) exhibits a markedly different 
structure. In this case, the effective electromagnetic contribution reinforces 
the gravitational field, eliminating the possibility of extremal configurations 
altogether. The spacetime admits a single EH for all values of the 
charge, and the Hawking temperature remains strictly positive throughout the 
parameter space. This absence of inner horizons implies a simpler causal 
structure and avoids the Cauchy horizon instabilities that typically afflict 
charged BHs in GR-like settings.

These geometric differences have direct physical implications. The reinforced 
gravitational potential characterizing the phantom branch modifies the location 
of the PS and affects null geodesic propagation, while the ordinary 
branch retains features continuously connected to the RN 
case. Similarly, the distinct horizon structures lead to different 
thermodynamic behavior, particularly near extremality, and influence tidal 
effects experienced by infalling matter. A geometric visualization of the 
equatorial spatial geometry, illustrating the single-horizon topology of the 
phantom branch, is presented in Appendix~\ref{app:embedding}.

The coexistence of ordinary and phantom branches within the same theoretical 
framework naturally raises the question of physical viability and observational 
discrimination. While the ordinary branch admits a smooth GR limit as 
$\ell\to0$ and $\gamma\to0$, the phantom branch represents a genuinely novel 
class of solutions with no direct counterpart in standard Einstein-Maxwell 
theory. The discrete nature of the branch parameter $\zeta=\pm1$ means that no continuous deformation connects the two sectors; they must be distinguished through qualitative signatures such as the horizon count, the sign of the PS shift relative to the RN value, or the presence versus absence of a tidal balance radius. In the following sections, we explore how this branch structure 
manifests in thermodynamic properties, optical signatures, and tidal forces, and 
assess whether these effects can provide meaningful criteria for distinguishing 
between the two branches.

\begin{figure}[ht!]
    \centering
    \begin{subfigure}[t]{0.45\textwidth}
        \centering
            \vspace{-2.5cm}
        \includegraphics[width=\linewidth]{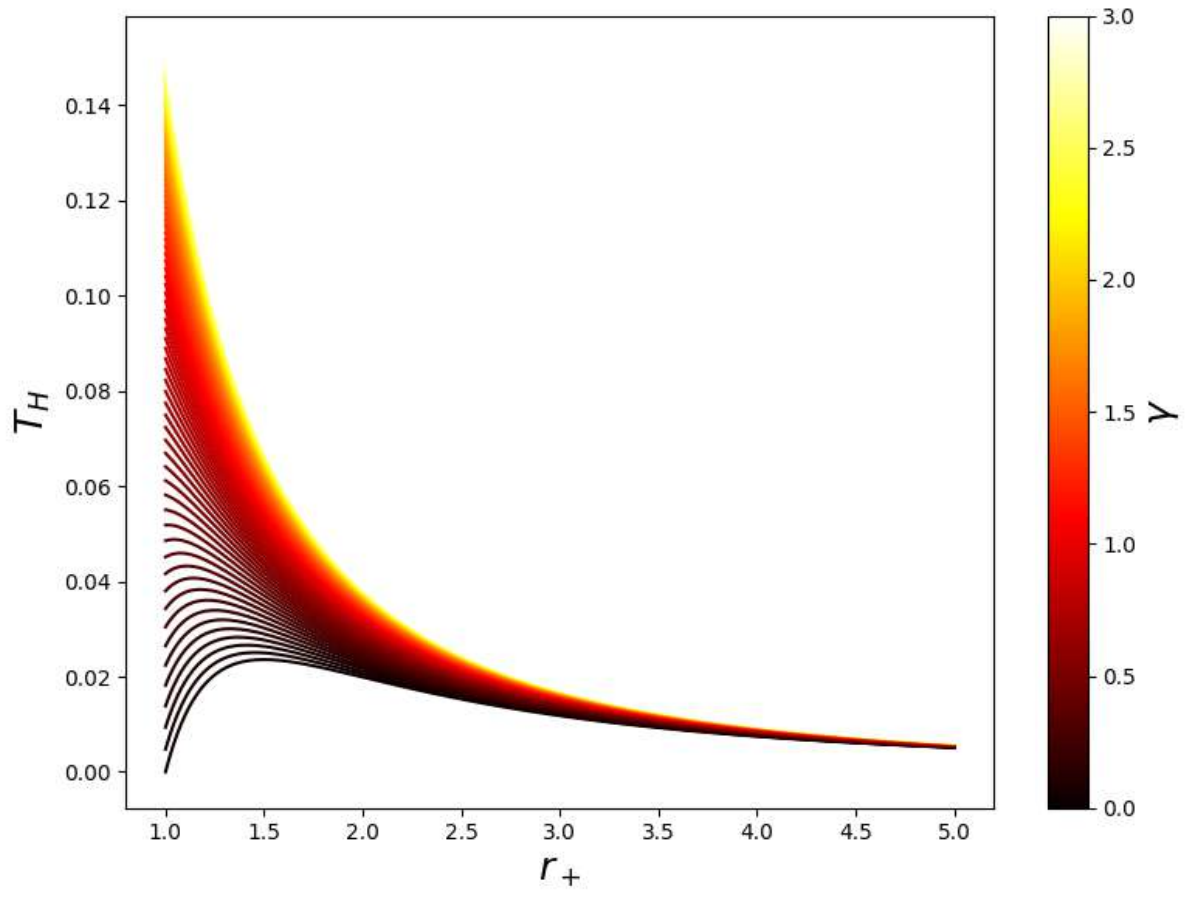}
      \vspace{-3.cm}
        \caption{Ordinary ModMax branch ($\zeta=+1$).
The electromagnetic contribution suppresses the Hawking temperature, producing
RN-like behavior. For small $\gamma$ (dark curves), $T_H$
approaches zero at finite $r_+$, signaling the onset of extremality.
Increasing $\gamma$ exponentially weakens the effective charge through
$e^{-\gamma}$, shifting the temperature curves upward and reducing the extremal
horizon radius.}
        \label{fig:TH_zeta_plus}
    \end{subfigure}

    \vspace{-2.cm}

    \begin{subfigure}[t]{0.45\textwidth}
        \centering
        \includegraphics[width=\linewidth]{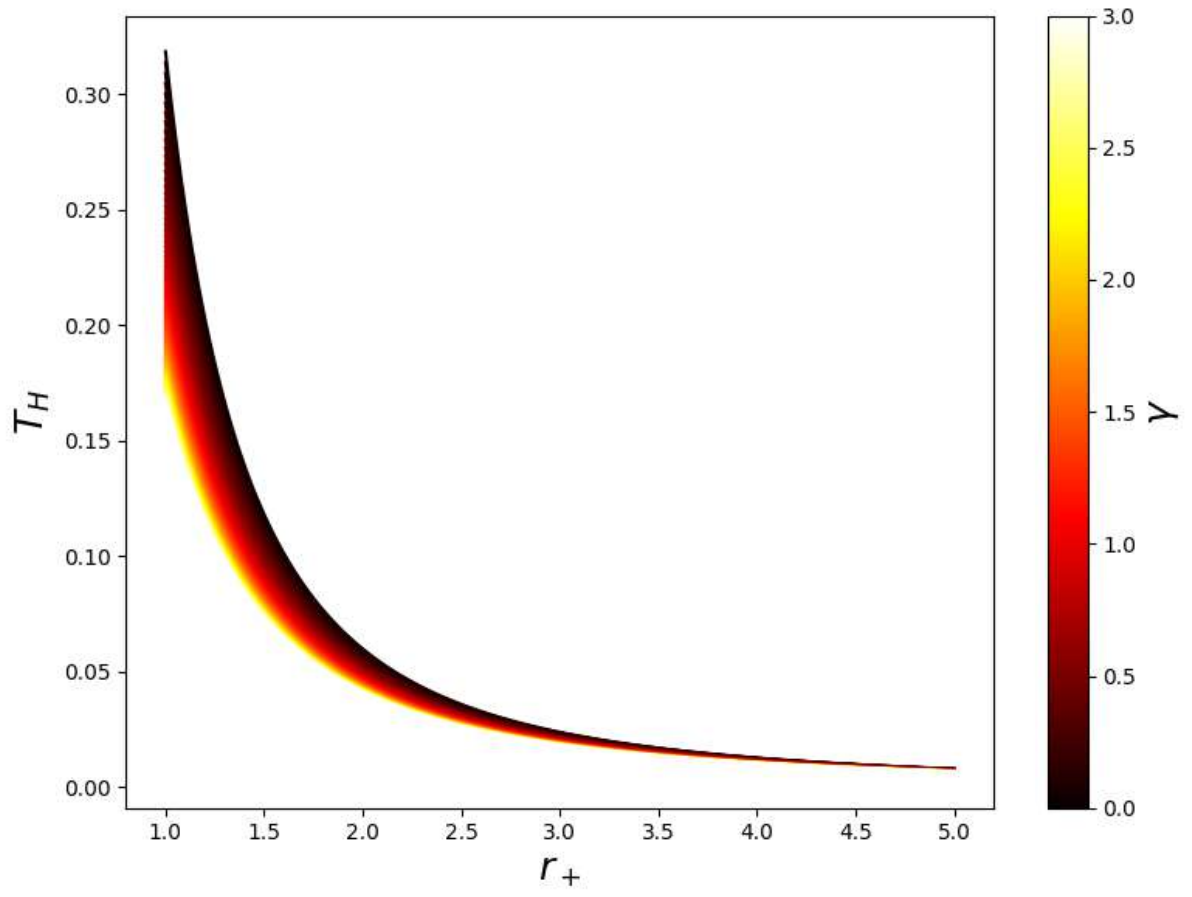}
           \vspace{-3cm}
        \caption{Phantom ModMax branch ($\zeta=-1$).
The sign-reversed electromagnetic sector enhances the Hawking temperature
uniformly. All configurations remain strictly nonzero, precluding extremal
limits. The temperature reaches $T_H \simeq 0.32$ at small $r_+$, approximately
twice the maximum attained in the ordinary branch, reflecting reinforced thermal
emission.}
        \label{fig:TH_zeta_minus}
    \end{subfigure}
 
    \caption{Hawking temperature $T_H$ as a function of the horizon radius $r_+$
for KR-ModMax BHs with fixed parameters $M=1.0$, $Q=0.5$, and 
$\ell=0.5$.
The color scale denotes the ModMax parameter $\gamma \in [0,3]$.
The ordinary branch exhibits a $\gamma$-dependent approach to extremality,
whereas the phantom branch maintains strictly positive temperature throughout 
the
entire parameter space.}
    \label{fig:TH_zeta_comparison}
\end{figure}

\subsection{Hawking temperature}
\label{subsec:hawking}

The stationarity of the KR-ModMax BH spacetime ensures the existence of a 
timelike Killing vector associated with time translations,
\begin{equation}
\xi^\mu = \left(\frac{\partial}{\partial t}\right)^\mu,
\label{eq:killing_vector}
\end{equation}
which gives rise to a conserved energy along timelike and null geodesics 
\cite{Hawking:1975vcx}. The surface gravity $\kappa$, characterizing the 
strength of the gravitational field at the EH, is defined through 
the Killing identity
\begin{equation}
\nabla_\nu \left(\xi^\mu \xi_\mu\right) = -2\,\kappa\, \xi_\nu,
\label{eq:killing_equation}
\end{equation}
and remains constant over the horizon in accordance with the zeroth law of 
BH mechanics \cite{Bardeen:1973gs,wald1993black}.

For the static metric function~\eqref{metric_solution}, the surface gravity is 
obtained directly from the radial derivative of the lapse function evaluated at 
the outer horizon radius $r_+$,
\begin{equation}
\kappa = \frac{1}{2}\, f'(r)\Big|_{r=r_+}
= \frac{M}{r_+^{2}} - \frac{\zeta\,Q^2\,e^{-\gamma}}{(1-\ell)^2\, r_+^{3}}.
\label{eq:surface_gravity}
\end{equation}
The relative sign of the two terms is controlled by the branch parameter 
$\zeta$, leading to qualitatively different thermal behavior in the ordinary 
and phantom branches.

Quantum particle production near the EH results in Hawking 
radiation, with temperature measured by asymptotic observers given by 
$T_H=\kappa/(2\pi)$ \cite{Hawking:1975vcx}. For the present geometry, this 
yields
\begin{equation}
T_H = \frac{M}{2\pi r_+^{2}} - \frac{\zeta\,Q^2\,e^{-\gamma}}{2\pi (1-\ell)^2\, 
r_+^{3}}.
\label{eq:hawking_temp}
\end{equation}
This expression shows the combined influence of the KR 
background, which rescales the effective couplings through $(1-\ell)^{-2}$, and 
the ModMax NED, whose contribution is suppressed by the 
factor $e^{-\gamma}$ and whose sign is determined by $\zeta$.

In the ordinary branch ($\zeta=+1$), the electromagnetic contribution 
counteracts the gravitational term, closely mirroring the behavior of the 
RN BH. As the charge approaches the critical value 
$Q_{\text{ext}} = M\sqrt{(1-\ell)^{3}\,e^{\gamma}}$, the Hawking temperature 
vanishes and the horizons coincide, signaling the onset of an extremal 
configuration with degenerate surface gravity. This limit is associated with a 
qualitative change in the near-horizon geometry and marks the boundary of the 
semiclassical description \cite{Gibbons:1994ff,hawking1994entropy}.

The phantom branch ($\zeta=-1$) exhibits a markedly different thermal 
structure. In this case, the electromagnetic term reinforces the gravitational 
contribution, leading to systematically higher temperatures and precluding the 
existence of extremal configurations. The Hawking temperature remains strictly 
positive throughout the parameter space, in agreement with the single-horizon 
causal structure discussed in the previous subsection.

\begin{table}[t]
\centering
\footnotesize
\renewcommand{\arraystretch}{1.4}
\setlength{\tabcolsep}{4pt}
\caption{Horizon classification for KR-ModMax BHs. The ordinary branch
admits three qualitatively distinct regimes controlled by
$Q$ relative to
$Q_{\rm ext}=M\sqrt{(1-\ell)^{3}e^{\gamma}}$, while the phantom branch
supports a single EH for every admissible
$(M,Q,\ell,\gamma)$ with $\ell<1$.}
\label{tab:horizons}
\begin{tabular}{|c|c|l|}
\hline
\cellcolor{brown!0} Branch & \cellcolor{brown!0} Condition &
\cellcolor{brown!0} Classification \\
\hline\hline
$\zeta=+1$ & $Q<Q_{\rm ext}$ & Non-extremal BH (2 horizons) \\
$\zeta=+1$ & $Q=Q_{\rm ext}$ & Extremal BH (degenerate) \\
$\zeta=+1$ & $Q>Q_{\rm ext}$ & NS \\
\hline
$\zeta=-1$ & $\ell<1$ & Single-EH BH (always) \\
\hline
\end{tabular}
\end{table}

The branch-dependent behavior of the Hawking temperature is illustrated in 
Fig.~\ref{fig:TH_zeta_comparison}, where $T_H$ is shown as a function of the 
horizon radius for representative values of the ModMax parameter $\gamma$. In 
the ordinary branch [Fig.~\ref{fig:TH_zeta_plus}], increasing $\gamma$ 
suppresses the effective charge contribution and progressively restores 
Schwarzschild-like behavior. By contrast, in the phantom branch 
[Fig.~\ref{fig:TH_zeta_minus}], the temperature curves are shifted upward and 
exhibit a smoother dependence on $\gamma$, reflecting the absence of 
extremality. The factor-of-two enhancement in peak temperature between the phantom and ordinary branches at small $r_+$ is a direct consequence of the sign reversal $\zeta\to -\zeta$ in Eq.~\eqref{eq:hawking_temp}: the electromagnetic term that suppresses $T_H$ in the ordinary branch instead amplifies it in the phantom sector, providing a clear thermodynamic discriminant between the two.

Table~\ref{tab:horizons} summarises the resulting horizon classification.  Configurations producing a well-defined PS and a sharp shadow boundary require the presence of an EH. Both the ordinary branch with $Q\le Q_{\rm ext}$ and the phantom branch (for any admissible $\ell$, $\gamma$, $Q$) satisfy this condition. The regime excluded by EHT shadow observations is the NS sector $Q>Q_{\rm ext}$ of the ordinary branch, which predicts qualitatively different optical images without a shadow edge. The quantitative constraint on admissible $(\ell,\gamma,Q,\zeta)$ therefore combines horizon existence with the requirement that the shadow radius match M87$^{*}$ and Sgr\,A$^{*}$ observations at the ${\sim}10\%$ level.

\section{Tsallis Entropy and KR-ModMax BH Thermodynamics}\label{isec3}

BH thermodynamics provides a sensitive probe of the microscopic 
structure of gravitational theories. In the presence of nonlinear matter 
sectors and LSB backgrounds, departures from standard 
extensive thermodynamics may naturally arise. In this section, we adopt a 
non-extensive framework based on Tsallis entropy in order to examine how such 
effects modify the thermodynamic behavior of KR-ModMax BHs and 
how the resulting properties depend on the underlying branch structure.

The adoption of Tsallis statistics for the present geometry is not merely a formal generalization but is motivated by two concrete features of the KR-ModMax solution. First, the non-Euclidean asymptotic structure $f(r) \to (1-\ell)^{-1}$ modifies the relationship between the horizon area and the number of accessible microstates, since the standard derivation of the Bekenstein-Hawking entropy assumes unit lapse at infinity. The rescaled asymptotics effectively renormalize the Euclidean action, introducing a correction whose functional form is naturally captured by the Tsallis parameter $\delta$. Second, the branch degeneracy ordinary and phantom configurations sharing the same horizon area but differing in their electromagnetic energy content implies that the Bekenstein-Hawking area law alone cannot distinguish thermodynamically inequivalent states. The Tsallis framework, through its sensitivity to correlations beyond the area scaling, provides the additional discriminating power needed to resolve this degeneracy.

\subsection{First law and Smarr relation}
\label{subsec:firstlaw}

Before turning to the Tsallis framework, we record the standard first law
and Smarr relation associated with the Bekenstein-Hawking (BkH) entropy, which follow directly
from the lapse function~\eqref{metric_solution}. Solving $f(r_{+})=0$ for
the mass yields
\begin{equation}
M(r_{+},Q,\ell,\gamma) = \frac{r_{+}}{2(1-\ell)}+\frac{\zeta\,Q^{2}e^{-\gamma}}
{2(1-\ell)^{2}\,r_{+}},
\label{eq:M_of_rp}
\end{equation}
and the Hawking temperature, entropy, and electric potential at the horizon
read, respectively,
\begin{equation}
T_{H}=\frac{1}{4\pi r_{+}}-\frac{\zeta\,Q^{2}e^{-\gamma}}
{4\pi(1-\ell)\,r_{+}^{3}},\quad
S_{\rm BkH}=\pi r_{+}^{2},\quad
\Phi_{H}=\frac{\zeta\,Q\,e^{-\gamma}}{(1-\ell)^{2}\,r_{+}}.
\label{eq:thermo_quantities}
\end{equation}
These expressions combine into the first law
\begin{equation}
dM = T_{H}\,dS_{\rm BkH} + \Phi_{H}\,dQ,
\label{eq:first_law}
\end{equation}
and the Smarr relation
\begin{equation}
M = 2\,T_{H}S_{\rm BkH} + \Phi_{H}\,Q.
\label{eq:smarr}
\end{equation}
Both identities hold for both branches $\zeta=\pm 1$ and have been verified
in closed form using Maple~2024\footnote{The Maple~2024 script
\texttt{KR\_ModMax\_Maple\_full\_appendix.txt} is self-contained,
runs in under one second at \texttt{Digits := 30}, and reproduces
the symbolic identities stated here and in Secs.~\ref{isec2}-\ref{isec7}
by printing the sequence of residuals
\texttt{R1 = R2 = 0}, \texttt{Smarr\_residual = 0}, the universal tidal
ratio \texttt{3/2}, the closed-form Davies radius, and the vanishing
extended Smarr residual \texttt{Smarr\_ext\_res = 0}.}. When $\ell$ and
$\gamma$ are treated as thermodynamic variables, the first law generalises
to
\begin{equation}
dM = T_{H}dS_{\rm BkH}+\Phi_{H}dQ+\Psi_{\ell}d\ell+\Psi_{\gamma}d\gamma,
\label{eq:first_law_ext}
\end{equation}
with $\Psi_{\ell}=\partial M/\partial\ell|_{r_{+},Q,\gamma}$ and
$\Psi_{\gamma}=\partial M/\partial\gamma|_{r_{+},Q,\ell}$. Eq.~\eqref{eq:smarr}
is the asymptotically flat Smarr relation expected from scaling arguments
\cite{isrply06,isrply07}; it does not receive direct contributions from
$\ell$ or $\gamma$, consistent with the fact that these parameters are
dimensionless.

\subsection{Tsallis entropy, internal energy, and free energy}

Within the Tsallis framework, the entropy of a BH with EH 
radius $r_+$ is taken to be \cite{WOS:000322575300012Tsallis}
\begin{equation}
S_T = \left(\pi r_+^2\right)^{\delta},
\label{eq:tsallis_entropy}
\end{equation}
where $\delta$ denotes the non-extensivity parameter. The standard 
Bekenstein-Hawking entropy $S_{\text{BkH}}=\pi r_+^2$ (in units 
$G=\hbar=c=k_B=1$) is recovered for $\delta=1$ 
\cite{bekenstein1980black,Hawking:1975vcx}. Values $\delta\neq1$ encode 
deviations from extensivity, reflecting the possible presence of long-range 
correlations, horizon fluctuations, or an underlying fractal microstructure 
associated with quantum gravitational effects \cite{Biro:2013cra}. In what 
follows, we focus on $\delta\geq1$, which ensures physical behavior and avoids 
pathological divergences.

The internal energy associated with the Tsallis entropy is obtained from the 
first law of BH thermodynamics,
\begin{equation}
E_T = \int T_H\, dS_T,
\label{eq:internal_energy_tsallis}
\end{equation}
where the non-extensive nature of the entropy introduces a $\delta$-dependent 
weighting through
$dS_T/dr_+ = 2\delta\,\pi^{\delta}\, r_+^{2\delta-1}$. Substituting the Hawking 
temperature given in Eq.~\eqref{eq:hawking_temp} and performing the integration 
yields
\small
\begin{equation}
E_T = -\frac{\pi^{\delta-1}\,\delta\,\bigl(Q^{2}\zeta\,(\delta-1)\,e^{-\gamma}+(\ell-1)\,r_{+}^{2}\,(\delta-2)\bigr)\,r_{+}^{2\delta-4}}
           {2\,(\ell-1)^{2}\,(\delta-2)\,(\delta-1)}.
\label{eq:ET_result}
\end{equation}
\normalsize

 The denominators $(\delta-1)$ and $(\delta-2)$ in Eq.~\eqref{eq:ET_result} indicate that care is needed at $\delta=1$ and $\delta=2$. The point $\delta=1$ corresponds to the Bekenstein-Hawking limit, where the Tsallis formalism reduces to the standard extensive entropy, and the apparent divergence is an artefact of the integration procedure rather than a physical pathology. The second critical value $\delta=2$ marks the upper boundary of the parameter range considered in this work. Near this value the internal energy develops a strong sensitivity to the non-extensivity parameter, reflecting the breakdown of the power-law scaling assumed in the Tsallis framework. For this reason we restrict the numerical analysis to the open interval $\delta\in(1,2)$.

\begin{figure}[t]
\centering

\begin{subfigure}[t]{0.45\textwidth}
\centering
       \vspace{-2.5cm}
\includegraphics[width=\linewidth]{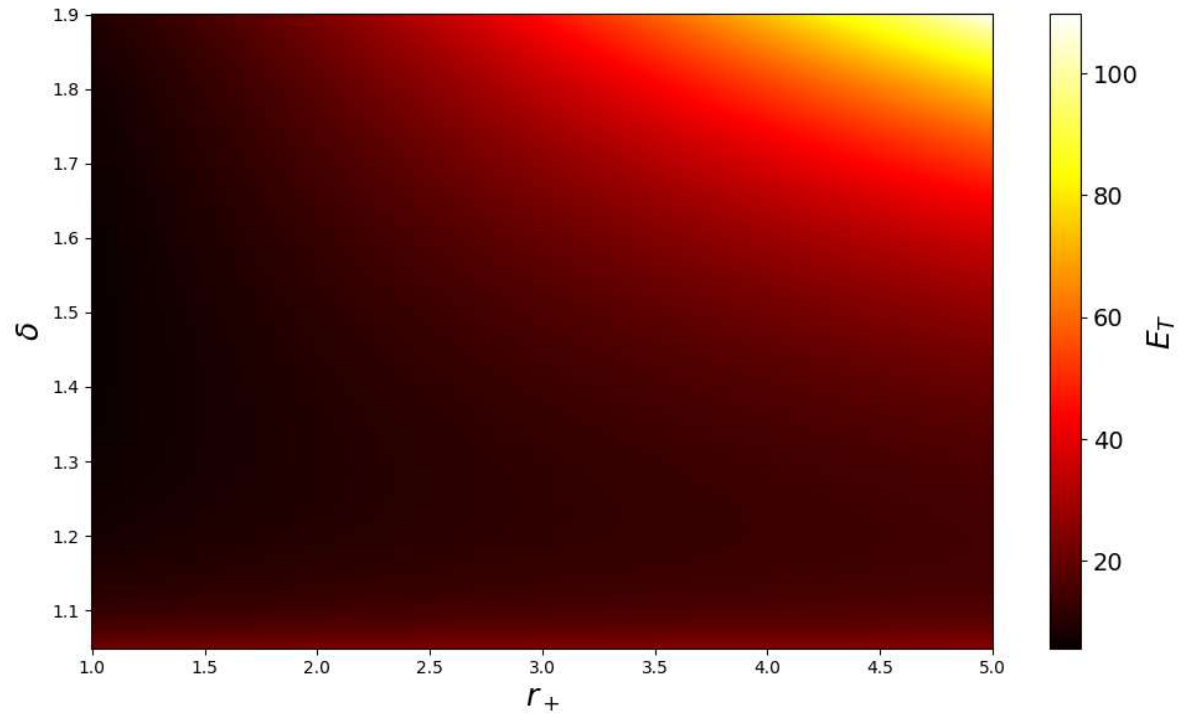}
       \vspace{-3.5cm}
\caption{Ordinary ModMax branch ($\zeta=+1$).
The Tsallis internal energy $E_T$ increases monotonically with the horizon 
radius
$r_+$, displaying smooth scaling across the non-extensivity range
$\delta \in (1,2)$. Charge effects suppress the short-distance contribution,
leading to a gradual approach toward Schwarzschild-KR behaviour at large 
radii.}
\label{fig:e_zeta_plus}
\end{subfigure}
       \vspace{-1cm}
\begin{subfigure}[t]{0.45\textwidth}
\centering
     \vspace{-2.5cm}
\includegraphics[width=\linewidth]{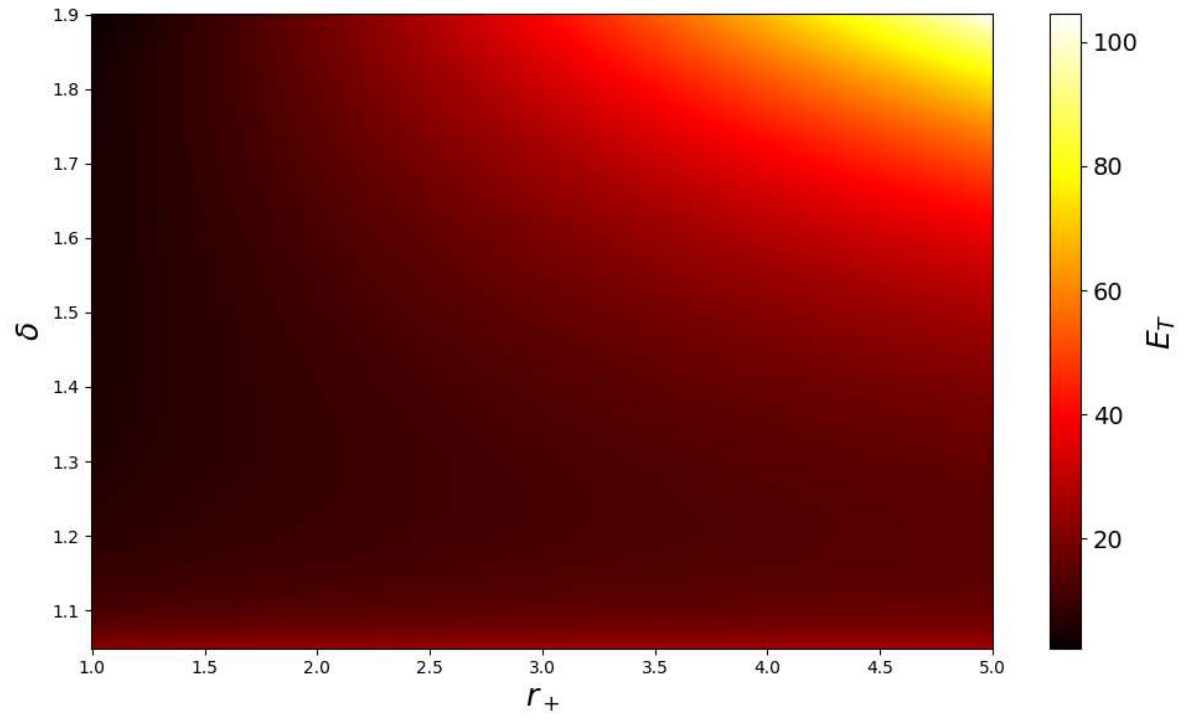}
            \vspace{-3.5cm}
\caption{Phantom ModMax branch ($\zeta=-1$).
The internal energy exhibits systematically steeper growth compared to the
ordinary branch. The sign-reversed electromagnetic sector amplifies the
near-horizon contribution, enhancing both the magnitude of $E_T$ and its
sensitivity to the non-extensivity parameter $\delta$.}
\label{fig:e_zeta_minus}
\end{subfigure}
        \vspace{1.5cm}
        \caption{Tsallis internal energy $E_T$ as a function of the horizon radius $r_+$
for KR-ModMax BHs with fixed parameters
$Q=0.5$, $\ell=0.5$, and $\gamma=2.0$.
The color scale denotes the non-extensivity parameter $\delta \in (1,2)$.
Across the entire parameter space, the phantom branch maintains higher internal
energy, establishing a clear thermodynamic hierarchy between the two ModMax
sectors.}
\label{fig:e_zeta_comparison}
\end{figure}

The behavior of the internal energy as a function of the horizon radius and the 
non-extensivity parameter is illustrated in Fig.~\ref{fig:e_zeta_comparison} 
for both ModMax branches. In the ordinary branch (Fig.~\ref{fig:e_zeta_plus}), 
$E_T$ increases monotonically with $r_+$ and $\delta$.  In this regime the scaling of $E_T$ is not controlled by the standard Reissner-Nordstr\"om competition between mass and charge terms. Instead, the electromagnetic contribution is exponentially suppressed by $e^{-\gamma}$ and simultaneously enhanced by the geometric factor $(1-\ell)^{-2}$ inherited from the Smarr-derived mass function~\eqref{eq:M_of_rp}. The response of $E_T$ to $r_+$ is therefore set by the combined KR-ModMax deformation rather than by a simple balance between attractive and repulsive pieces. The Tsallis parameter $\delta$ controls the rate at which $E_T$ saturates at large $r_+$: larger $\delta$ enhances the non-extensive corrections and shifts the crossover to smaller horizon radii. In the 
phantom branch (Fig.~\ref{fig:e_zeta_minus}), the internal energy exhibits a 
similar qualitative dependence but is systematically enhanced, growing without bound as $r_+$ increases. This increase 
originates from the sign reversal $\zeta=-1$, which effectively converts the 
electromagnetic contribution into an additional heating source, leading to a 
larger accumulated thermal energy for fixed horizon size. The absence of saturation in the phantom branch is consistent with the reinforced gravitational potential that precludes thermodynamic equilibrium at large sizes.

\begin{figure}[ht]
\centering
\begin{subfigure}[t]{0.45\textwidth}
\centering
   \vspace{-2.5cm}
\includegraphics[width=\linewidth]{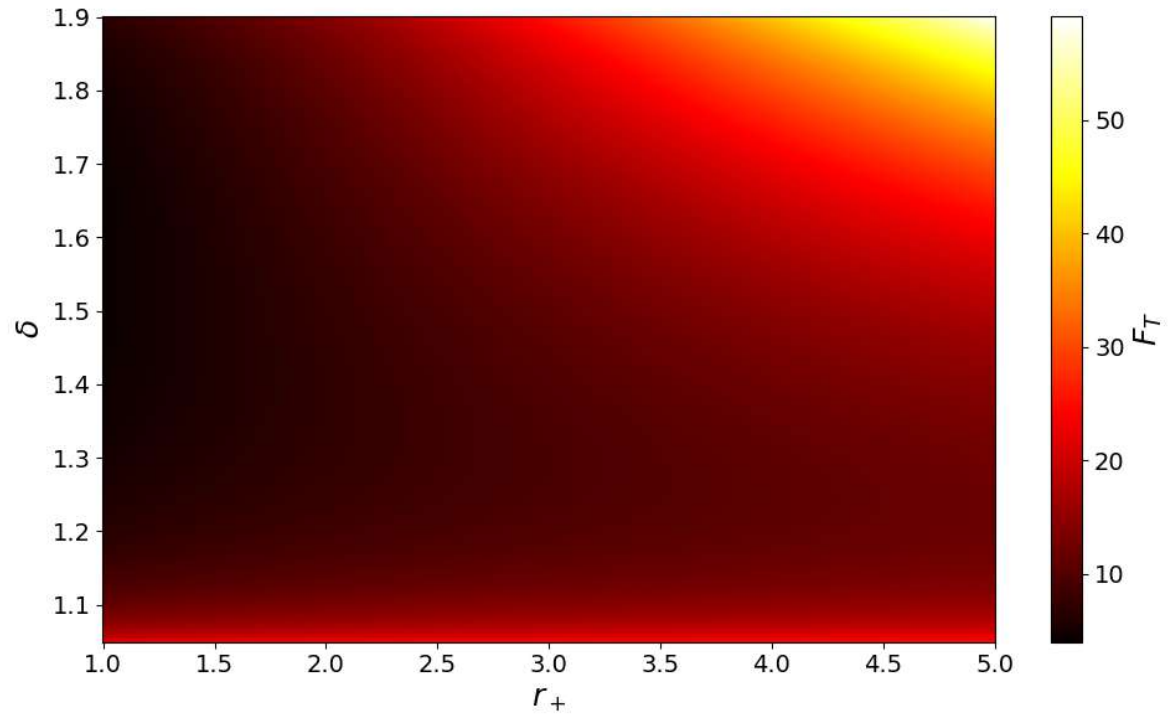}
   \vspace{-3.5cm}
\caption{Ordinary ModMax branch ($\zeta=+1$).
The Tsallis Helmholtz free energy $F_T$ decreases monotonically with increasing
horizon radius $r_+$, approaching zero asymptotically.
Larger non-extensivity parameter $\delta$ deepens the free-energy well at small
$r_+$, indicating a higher thermodynamic cost for compact configurations and a
tendency toward instability near the horizon.}
\label{fig:f_zeta_plus}
\end{subfigure}
\begin{subfigure}[t]{0.45\textwidth}
\centering
   \vspace{-2.5cm}
\includegraphics[width=\linewidth]{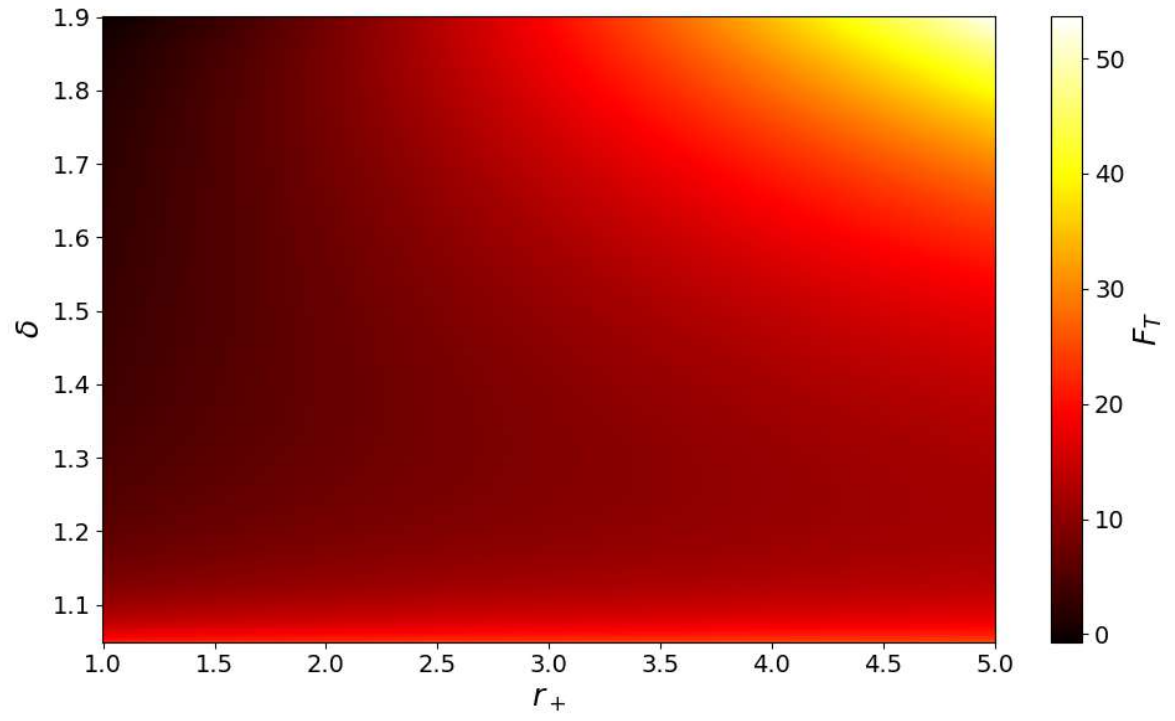}
   \vspace{-3.5cm}
\caption{Phantom ModMax branch ($\zeta=-1$).
The free energy attains systematically larger magnitudes at small $r_+$ and
exhibits a steeper radial dependence.
The enhanced separation between $\delta$-curves reflects a stronger sensitivity
to non-extensive corrections, signaling a more pronounced thermodynamic
instability in the phantom sector.}
\label{fig:f_zeta_minus}
\end{subfigure}
   \vspace{0.5cm}
\caption{Tsallis Helmholtz free energy $F_T$ as a function of the horizon radius
$r_+$ for KR-ModMax BHs with the same parameters as
Fig.~\ref{fig:e_zeta_comparison}.
While both branches show monotonic decay with increasing $r_+$, the phantom
branch maintains a larger free-energy magnitude throughout, reinforcing the
thermodynamic hierarchy inferred from the internal energy and anticipating the
branch-dependent stability behavior discussed in subsequent sections.}
\label{fig:f_zeta_comparison}
\end{figure}
  \vspace{0.25cm}

A complementary thermodynamic quantity is the Helmholtz free energy, which 
characterizes the maximum extractable work at fixed temperature and provides 
a perspective on global stability. Within the Tsallis framework, it is obtained 
through the Legendre transformation
\begin{equation}
F_T = -\int S_T \, dT_H.
\label{eq:helmholtz_tsallis}
\end{equation}
Performing the integration yields
\begin{equation}
F_T = -\frac{\mathcal{A}}{2\,(\ell-1)^{2}\,(\delta-2)\,(\delta-1)},
\label{eq:FT_result}
\end{equation}
where
\begin{multline}
\mathcal{A} = r_{+}^{2\delta-4}\,\pi^{\delta-1}\,\bigl(
2\,e^{-\gamma}\,Q^{2}\,\delta\,\zeta - 2\,\zeta\,Q^{2}\,e^{-\gamma}\\
+\,\ell\,\delta\,r_{+}^{2} - 2\,r_{+}^{2}\,\ell
- \delta\,r_{+}^{2} + 2\,r_{+}^{2}\bigr).
\label{eq:K_coefficient}
\end{multline}

The behavior of the Helmholtz free energy is displayed in 
Fig.~\ref{fig:f_zeta_comparison}. In the ordinary branch, $F_T$ decreases 
monotonically with the horizon radius, with increasing $\delta$ enhancing the 
preference for larger, thermodynamically stable configurations. In the phantom 
branch, the free energy exhibits a similar qualitative trend but with 
systematically larger magnitude and steeper descent, indicating that 
non-extensive effects couple more strongly when the electromagnetic 
contribution reinforces the gravitational field \cite{Al-Badawi:2025ejf}. The persistent sign difference between the two branches (the phantom branch maintaining a deeper free-energy well at every $r_+$) reflects the additional gravitational binding energy injected by the sign-reversed electromagnetic sector and implies that phantom BHs are thermodynamically less favored than their ordinary counterparts at equal horizon area.

\subsection{Thermodynamic pressure and equation of state}

In the extended formulation of BH thermodynamics, macroscopic 
thermodynamic variables can be introduced through appropriate conjugate pairs, 
allowing one to explore phase structure beyond the standard entropy-temperature 
description \cite{Kubiznak:2012wp}. Within the present Tsallis framework, we 
define an effective thermodynamic pressure through the Helmholtz free energy, 
while retaining the geometric interpretation of the thermodynamic volume as
\begin{equation}
V = \frac{4\pi r_+^3}{3}.
\end{equation}
The pressure is then obtained as
\begin{equation}
P_T = -\frac{\partial F_T}{\partial V},
\label{eq:pressure_tsallis}
\end{equation}
which encodes how non-extensive effects and matter-sector modifications reshape 
the effective equation of state (EoS). Hence, carrying out the differentiation yields
\begin{equation}
P_T = \frac{\pi^{\delta-2}\,\bigl(\zeta\,Q^{2}\,e^{-\gamma}+\tfrac{1}{2}(\ell-1)\,r_{+}^{2}\bigr)\,r_{+}^{-7+2\delta}}
            {2\,(\ell-1)^{2}}.
\label{eq:PT_result}
\end{equation}

\begin{figure}[t]
\centering

\begin{subfigure}[t]{0.45\textwidth}
\centering
  \vspace{-3.5cm}
\includegraphics[width=\linewidth]{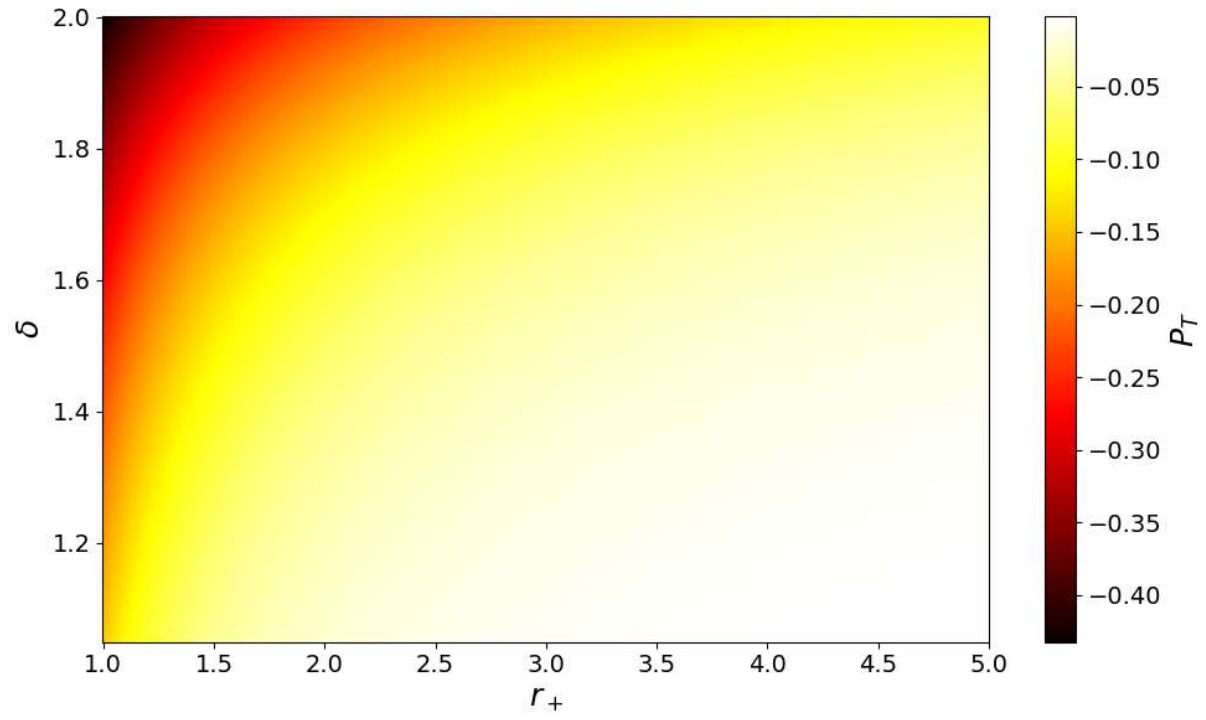}
  \vspace{-3.5cm}
\caption{Ordinary ModMax branch ($\zeta=+1$).
The Tsallis thermodynamic pressure $P_T$ remains negative throughout the 
explored
parameter space and decreases smoothly with increasing horizon radius $r_+$.
Larger non-extensivity parameter $\delta$ uniformly reduces the pressure
magnitude, indicating that non-extensive entropy softens the effective equation
of state and moderates the response to horizon shrinkage.}
\label{fig:p_zeta_plus}
\end{subfigure}

\begin{subfigure}[t]{0.45\textwidth}
\centering
  \vspace{-2.5cm}
\includegraphics[width=\linewidth]{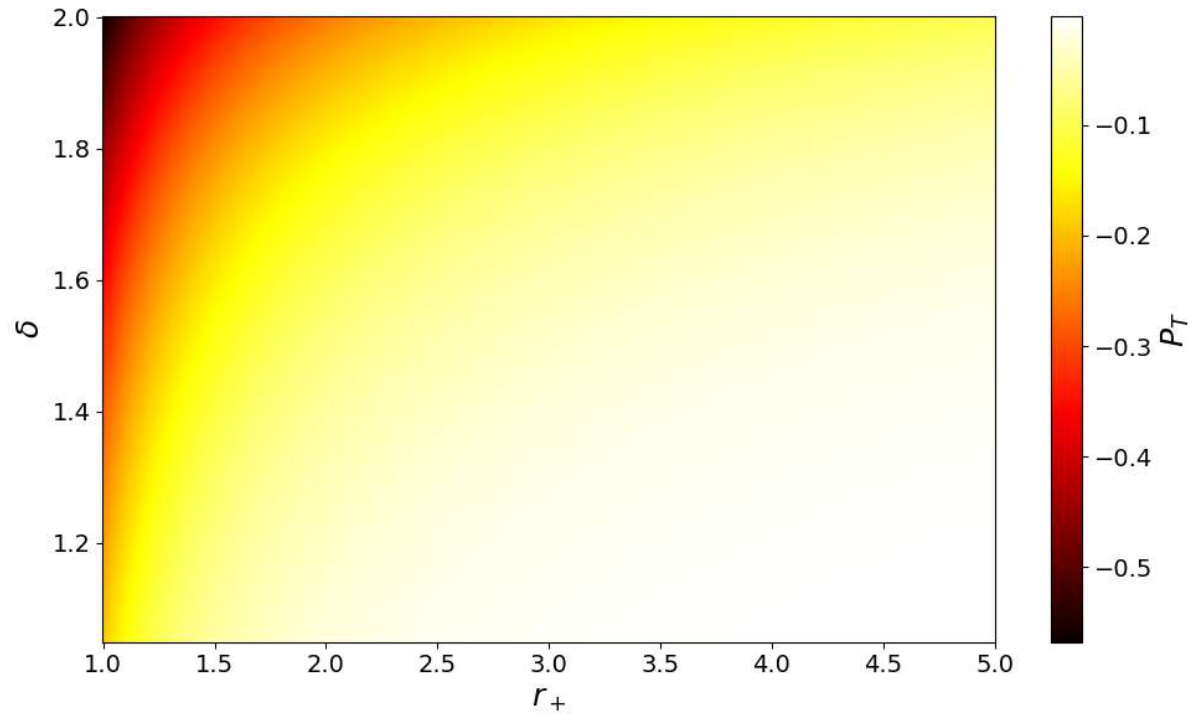}
  \vspace{-3.5cm}
\caption{Phantom ModMax branch ($\zeta=-1$).
The pressure exhibits a similar negative range but displays a steeper radial
dependence at small $r_+$.
The sign-reversed electromagnetic sector enhances the sensitivity of $P_T$ to
near-horizon physics, foreshadowing stronger branch dependence in
pressure-driven processes such as JT expansion.}
\label{fig:p_zeta_minus}
\end{subfigure}
  \vspace{0.5cm}
\caption{Tsallis thermodynamic pressure $P_T$ as a function of the horizon 
radius
$r_+$ for KR-ModMax BHs.
The negative pressure indicates that, in this asymptotically flat setup, the
effective cosmological contribution acts as a tension rather than a true
pressure.
Although both branches converge at large $r_+$ where the mass term dominates,
their distinct radial behavior at small $r_+$ plays a crucial role in shaping
the JT coefficient and the cooling-heating structure discussed in
Sec.~\ref{isec4}.}
\label{fig:p_zeta_comparison}
\end{figure}

The negative pressure $P_T < 0$ throughout the parameter space (Fig.~\ref{fig:p_zeta_comparison}) indicates that the KR-ModMax BH, within the Tsallis framework, behaves as a gravitationally bound system under tension rather than a thermodynamic gas under compression. This feature is shared by both branches and persists for all values of $\delta$, though the phantom branch exhibits systematically larger $|P_T|$ due to the enhanced gravitational binding. The purely attractive character of $P_T$ is a direct consequence of the asymptotically non-Euclidean structure: the rescaled lapse $f_\infty = (1-\ell)^{-1}$ prevents the identification of a positive cosmological-constant contribution that would be needed for a van der Waals-type phase transition, distinguishing the present thermodynamic landscape from that of anti-de Sitter (AdS) BHs.

The behavior of $P_T$ as a function of the horizon radius is illustrated in 
Fig.~\ref{fig:p_zeta_comparison}. In the ordinary branch 
(Fig.~\ref{fig:p_zeta_plus}), the pressure decreases smoothly in magnitude as 
$r_+$ increases, with larger values of the non-extensivity parameter $\delta$ 
systematically reducing the absolute pressure. In the phantom branch 
(Fig.~\ref{fig:p_zeta_minus}), a similar qualitative trend is observed, 
although the magnitude of the pressure decreases more rapidly at small horizon 
radii, reflecting the enhanced role of the electromagnetic contribution when 
$\zeta=-1$. At sufficiently large $r_+$, both branches converge as the mass 
term dominates over charge-dependent corrections.

This effective EoS provides the thermodynamic background for the 
JT expansion discussed in the following section, where heating and 
cooling regimes are analyzed in terms of the response of the Hawking 
temperature to pressure variations.

 If a cosmological constant $\Lambda$ is reinstated in
Eq.~\eqref{eq:action}, the lapse function generalises to
\begin{equation}
f_{\Lambda}(r) = \frac{1}{1-\ell}-\frac{2M}{r}+\frac{\zeta\,Q^{2}e^{-\gamma}}
{(1-\ell)^{2}\,r^{2}}-\frac{\Lambda}{3}r^{2},
\label{eq:f_Lambda}
\end{equation}
and the identification $P=-\Lambda/(8\pi)$ promotes $\Lambda$ to a genuine
thermodynamic pressure, conjugate to $V=4\pi r_{+}^{3}/3$. The Smarr relation
acquires the canonical $-2PV$ term,
\begin{equation}
M = 2T_{H}S_{\rm BkH}+\Phi_{H}Q - 2PV,
\end{equation}
and we have verified symbolically (see Sec.~B.7 of
Appendix~\ref{app:thermo_check}) that the corresponding
residual vanishes identically, in parallel with the asymptotically flat case.
The AdS sector ($\Lambda<0$, $P>0$) of the ordinary branch then displays
VdW-type $P$-$V$ criticality with a first-order small/large-BH transition
\cite{isrply04,isrply08,isrply09,isrply10}, the critical point being
$\gamma$-dependent through the $e^{-\gamma}$ charge screening. The phantom
branch in AdS does not admit a genuine VdW critical point
\cite{isrply04,isrply11}, consistent with the local-instability pattern of
Fig.~\ref{fig:c_zeta_comparison}. A full extended-phase-space analysis of
the KR-ModMax-AdS system, including critical exponents and topological
classification, will be developed in a companion paper.

\subsection{Heat capacity and local stability}
\label{subsec:phase}

\begin{figure}[t]
\centering
\begin{subfigure}[t]{0.45\textwidth}
\centering
  \vspace{-2.7cm}
\includegraphics[width=\linewidth]{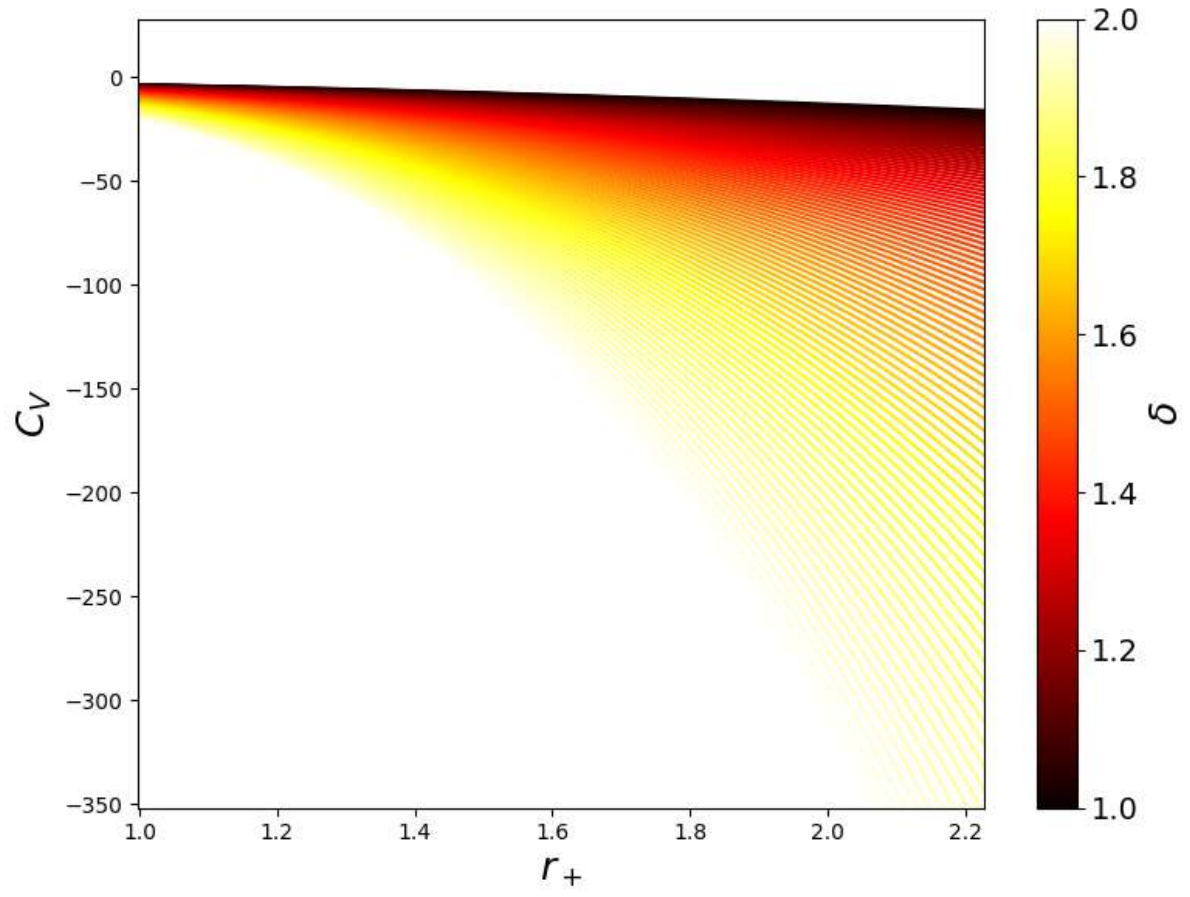}
  \vspace{-3cm}
\caption{Ordinary ModMax branch ($\zeta=+1$).
The heat capacity $C_V$ remains negative throughout the explored range
$r_+ \in [1.0,\,1.25]$, indicating local thermodynamic instability.
As $r_+$ increases, $C_V$ approaches zero from below, signaling proximity to a
Davies-type transition where the dominant instability weakens.
The stratification with respect to the non-extensivity parameter $\delta$
reflects the increasing weight of entropy corrections near the transition
region.}
\label{fig:c_zeta_plus}
\end{subfigure}

\begin{subfigure}[t]{0.45\textwidth}
\centering
  \vspace{-2.4cm}
\includegraphics[width=\linewidth]{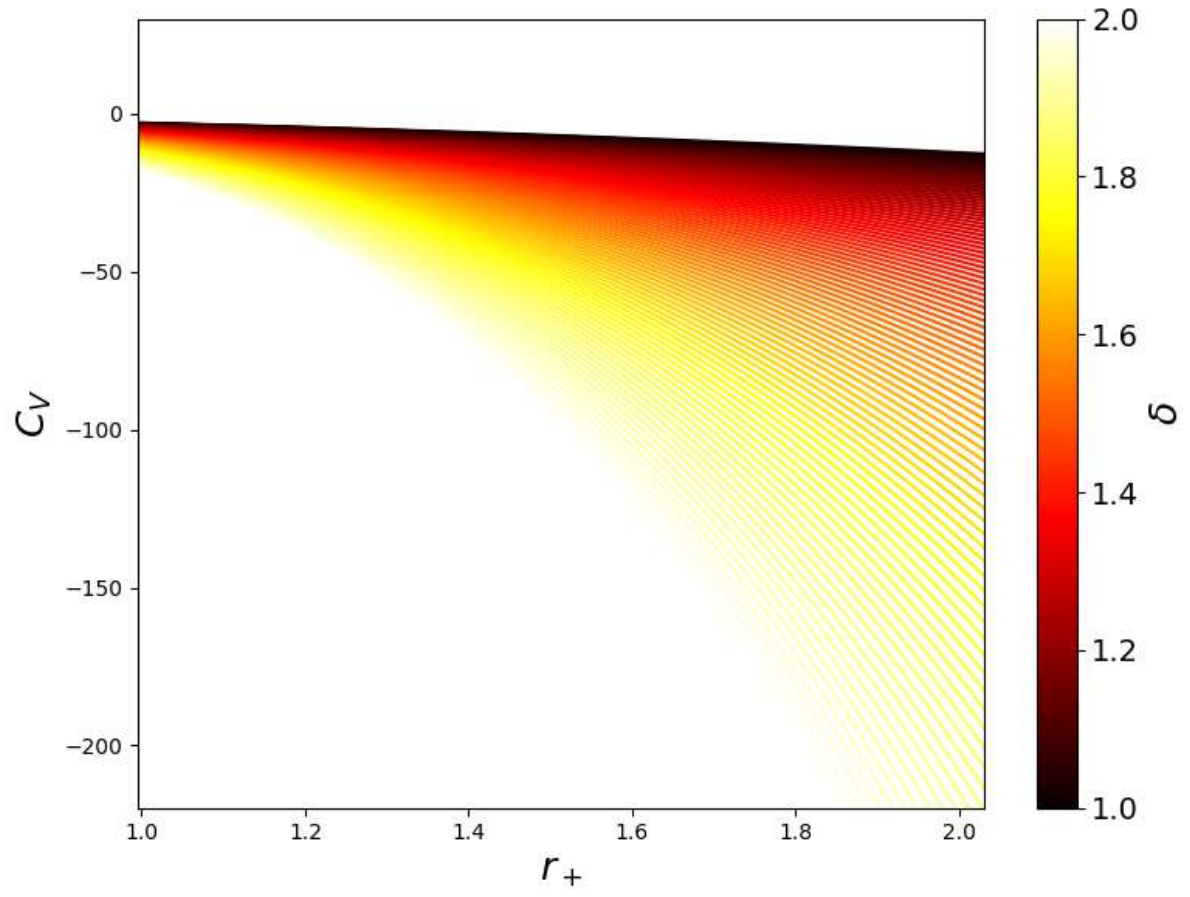}
  \vspace{-3.cm}
\caption{Phantom ModMax branch ($\zeta=-1$).
The heat capacity exhibits the same qualitative instability pattern but with
systematically reduced magnitude.
The absence of a zero-crossing confirms that no stable phase is realized,
while the smoother approach toward $C_V \to 0^{-}$ indicates a weaker
near-critical response compared to the ordinary branch.}
\label{fig:c_zeta_minus}
\end{subfigure}

\caption{Heat capacity $C_V$ as a function of the horizon radius $r_+$ for
KR-ModMax BHs within Tsallis non-extensive thermodynamics.
Negative values throughout indicate local thermodynamic instability for both
branches.
The branch-dependent approach toward $C_V = 0^{-}$ at larger $r_+$ plays a
decisive role in shaping the JT cooling-heating behavior discussed in
Sec.~\ref{isec4}.
The color scale corresponds to the non-extensivity parameter
$\delta \in [1.0,\,2.0]$.}
\label{fig:c_zeta_comparison}
\end{figure}

Local thermodynamic stability of the BH is determined by the heat 
capacity at constant volume, defined as \cite{SUCU2026102295}
\begin{equation}
C_V = T_H \left(\frac{\partial S_T}{\partial T_H}\right)_V,
\label{eq:heat_capacity_def}
\end{equation}
where a positive heat capacity corresponds to stability under thermal 
fluctuations, while a negative value signals local thermodynamic instability. 
Within the Tsallis framework adopted here, the heat capacity for the 
KR-ModMax BH can be evaluated analytically, yielding
\begin{equation}
C_V = -\frac{\delta\,\bigl(\zeta\,Q^{2}\,e^{-\gamma}+(\ell-1)\,r_{+}^{2}\bigr)\,\pi^{\delta}\,r_{+}^{2\delta}}
             {2\,\zeta\,Q^{2}\,e^{-\gamma}+(\ell-1)\,r_{+}^{2}}.
\label{eq:CV_result}
\end{equation}

The divergence of $C_V$ is governed by the vanishing of its denominator. Setting
\begin{equation}
2\,\zeta\,Q^{2}\,e^{-\gamma}+(\ell-1)\,r_{+}^{2}=0,
\label{eq:Davies_condition}
\end{equation}
and solving for $r_{+}$ yields the generalized Davies radius
\begin{equation}
r_{\rm Davies}=\left(\frac{2\,\zeta\,Q^{2}\,e^{-\gamma}}{1-\ell}\right)^{\!1/2}.
\label{eq:rDavies_manuscript}
\end{equation}
A real and positive solution exists only for the ordinary branch ($\zeta=+1$) with $\ell<1$. The divergence location is notably independent of the non-extensivity parameter $\delta$, which instead sets the magnitude and scaling of $C_V$. The sign of $C_V$ itself depends on both numerator and denominator and therefore varies with $\delta$ and $r_{+}$; in particular, the zero of $C_V$ does not coincide with its divergence. In the phantom branch ($\zeta=-1$), Eq.~\eqref{eq:Davies_condition} admits no real positive solution, so $C_V$ does not diverge and remains negative throughout the parameter space, as seen in Fig.~\ref{fig:c_zeta_minus}. The asymptotically flat sector therefore does not support a Van der Waals-type first-order phase transition, consistent with the negative thermodynamic pressure $P_T<0$ of Fig.~\ref{fig:p_zeta_comparison}.

The sign and magnitude of the heat capacity are strongly branch dependent. In 
the ordinary branch ($\zeta=+1$), the electromagnetic contribution counteracts 
the gravitational term, leading to negative heat capacity over a broad range of 
horizon radii. In the phantom branch ($\zeta=-1$), the same contribution 
reinforces gravity, modifying the location and strength of the divergence but 
not altering the overall sign of $C_V$ in the near-horizon region.

The behavior of the heat capacity is illustrated in 
Fig.~\ref{fig:c_zeta_comparison} for representative parameter choices. In both 
branches, $C_V$ remains negative over the displayed range of horizon radii, 
indicating local thermodynamic instability in the near-horizon regime. The 
ordinary branch (Fig.~\ref{fig:c_zeta_plus}) exhibits a monotonic increase of 
$C_V$ toward zero as $r_+$ grows, with the non-extensivity parameter $\delta$ 
producing a systematic stratification of the curves. The phantom branch 
(Fig.~\ref{fig:c_zeta_minus}) displays a similar qualitative trend, with 
slightly reduced magnitude, reflecting the reinforcing role of the 
electromagnetic sector. The rate at which $|C_V|$ diverges near the extremal limit in the ordinary branch provides a quantitative fingerprint of the combined $(\ell, \gamma)$ deformation that distinguishes it from the RN case, where the divergence location is controlled by charge alone.

The approach of $C_V$ toward zero at larger horizon radii suggests proximity to 
a Davies-type transition \cite{Shahzad:2022egq}, beyond which locally stable 
configurations may arise. However, within the parameter range explored here, 
both branches remain thermodynamically unstable at the local level, a feature 
that complements the global stability analysis based on the Helmholtz free 
energy and motivates the extended thermodynamic investigation presented in the 
next section.

\section{Joule-Thomson expansion}
\label{isec4}

The Joule-Thomson (JT) expansion provides a valuable diagnostic of 
thermodynamic response by examining the evolution of the Hawking temperature 
under isenthalpic processes. Within the extended thermodynamic framework 
developed above, it offers insight into how non-extensive effects and branch 
structure influence heating and cooling behavior beyond equilibrium 
considerations.

\subsection{Joule-Thomson coefficient and inversion curve}

The JT coefficient is defined as \cite{WOS:001618835200002Aydiner}
\begin{equation}
\mu_J = \left(\frac{\partial T_H}{\partial P_T}\right),
\label{eq:JT_def}
\end{equation}
with $\mu_J>0$ corresponding to cooling during expansion and $\mu_J<0$ 
indicating heating. The inversion curve, defined by the condition $\mu_J=0$, 
separates these regimes in the $(P_T,T_H)$ plane \cite{Wang:2025alf}.

For the Kalb-Ramond-ModMax black hole, explicit evaluation yields
\begin{equation}
\mu_J = \frac{8\,\pi^{-\delta+1}\,\bigl(\zeta\,Q^{2}\,e^{-\gamma}+\tfrac{1}{2}(\ell-1)\,r_{+}^{2}\bigr)\,r_{+}^{-2\delta+3}}
             {4\,Q^{2}\,\zeta\,\bigl(\delta-\tfrac{7}{2}\bigr)\,e^{-\gamma} + 2\,\bigl(\delta-\tfrac{5}{2}\bigr)\,(\ell-1)\,r_{+}^{2}}.
\label{eq:muJ_result}
\end{equation}
The inversion point of the JT expansion is set by the vanishing of the numerator of Eq.~\eqref{eq:muJ_result},
\begin{equation}
\zeta\,Q^{2}\,e^{-\gamma}+\tfrac{1}{2}(\ell-1)\,r_{+}^{2}=0,
\end{equation}
which gives the inversion radius
\begin{equation}
r_{\rm inv}=\left(\frac{2\,\zeta\,Q^{2}\,e^{-\gamma}}{1-\ell}\right)^{\!1/2}.
\label{eq:r_inv}
\end{equation}
A real positive $r_{\rm inv}$ exists only in the ordinary branch ($\zeta=+1$, $\ell<1$); at that radius $\mu_J$ changes sign, separating the cooling ($\mu_J>0$) and heating ($\mu_J<0$) regimes. The inversion scale is controlled by the ModMax parameter $\gamma$, the KR deformation $\ell$, and the effective charge $\zeta Q^{2}$: the exponential factor $e^{-\gamma}$ suppresses the effective charge and reduces $r_{\rm inv}$, while the geometric factor $(1-\ell)^{-1}$ enhances it. The JT response in this system therefore does not reduce to a perturbative RN limit but reflects the joint action of both deformation channels. In the phantom branch ($\zeta=-1$), the right-hand side of Eq.~\eqref{eq:r_inv} is negative definite, no real positive solution exists, and $\mu_J$ retains a single sign throughout the parameter space. Comparison with Eq.~\eqref{eq:rDavies_manuscript} shows that $r_{\rm inv}=r_{\rm Davies}$: the JT inversion coincides with the Davies locus of local-stability change, reflecting their common origin in the numerator-denominator structure of the compact-form $C_V$.

\begin{figure}[t]
\centering
\begin{subfigure}[t]{0.45\textwidth}
\centering
  \vspace{-2.7cm}
\includegraphics[width=\linewidth]{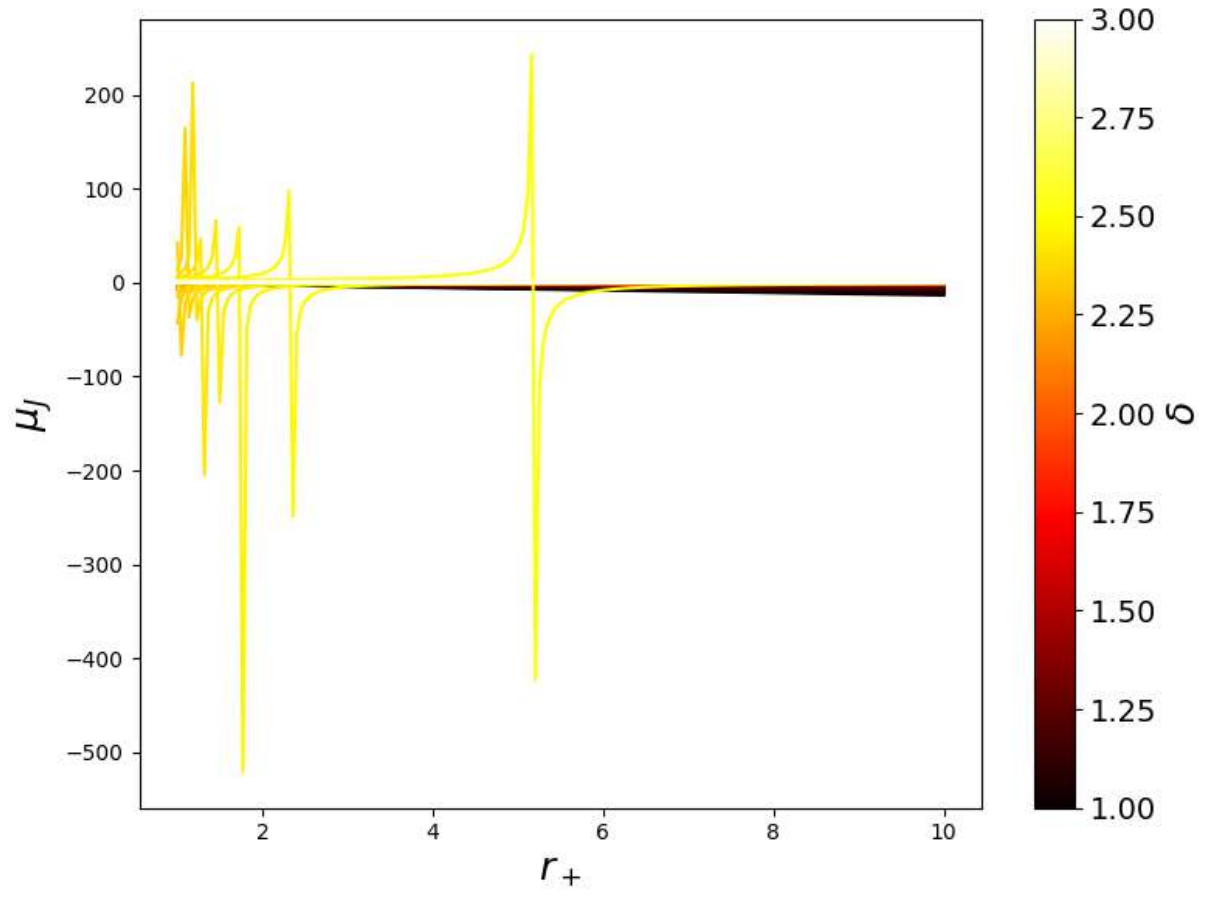}
  \vspace{-3cm}
\caption{Ordinary ModMax branch ($\zeta=+1$).}
\label{fig:jte_plus}
\end{subfigure}

\begin{subfigure}[t]{0.45\textwidth}
\centering
  \vspace{-2.4cm}
\includegraphics[width=\linewidth]{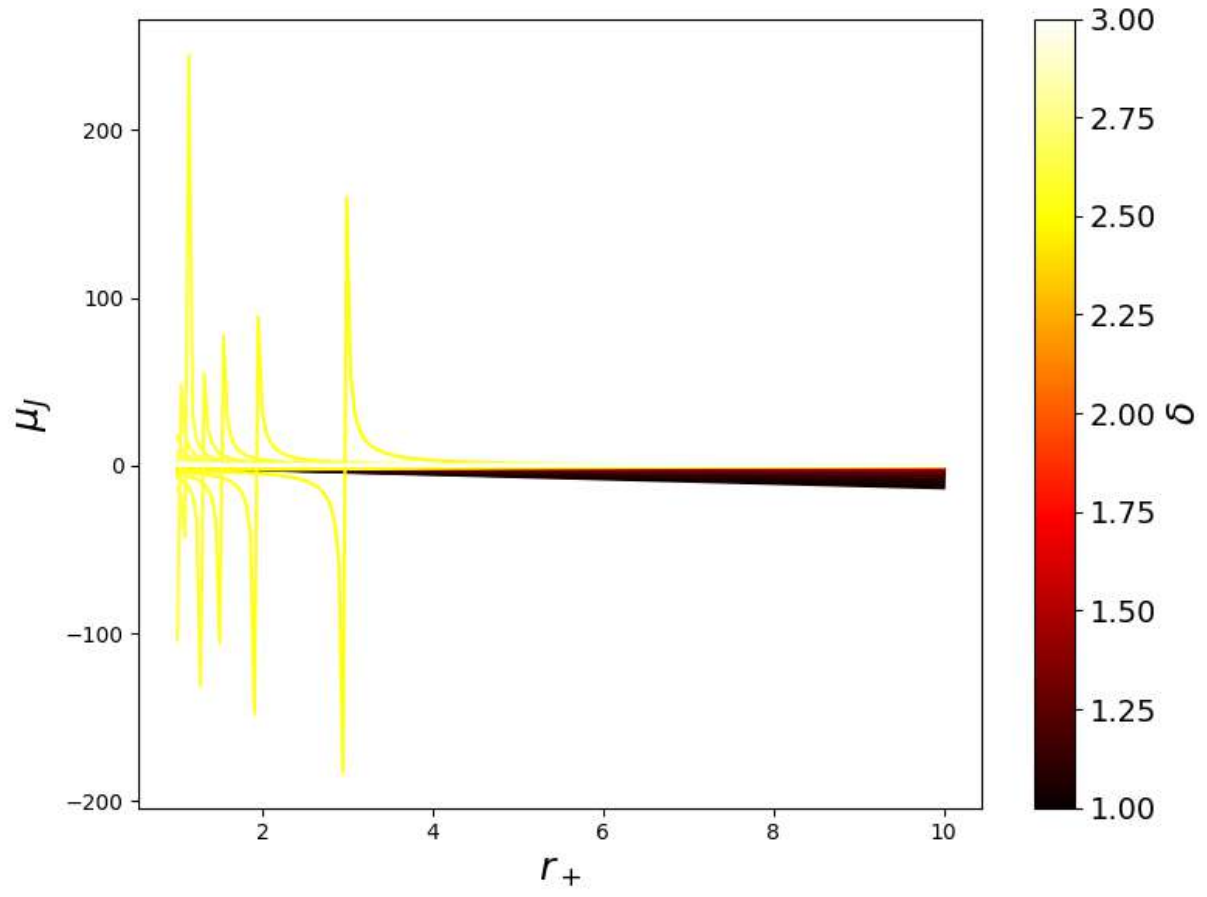}
  \vspace{-3.cm}
\caption{Phantom ModMax branch ($\zeta=-1$).}
\label{fig:jte_minus}
\end{subfigure}

\caption{JT coefficient $\mu_J$ as a function of the horizon
radius $r_{+}$ for the KR-ModMax black hole
($Q=0.5$, $\gamma=2$, $\ell=0.5$). The color scale corresponds to
the non-extensivity parameter $\delta \in [1.0,\,3.0]$. Panel~(a) shows
the ordinary branch ($\zeta=+1$), where the divergences of $\mu_J$
separate heating ($\mu_J<0$) from cooling ($\mu_J>0$) regimes during
isenthalpic expansion. Panel~(b) displays the phantom branch
($\zeta=-1$), whose sign-reversed electromagnetic contribution relocates
the divergences and reshapes the cooling/heating structure. Increasing
$\delta$ shifts the inversion radii toward larger $r_{+}$, demonstrating
that non-extensive entropy systematically delays the onset of the
cooling phase in both branches.}
\label{fig:JTE}
\end{figure}

\subsection{Cooling-heating transitions and branch dependence}

The behavior of the JT coefficient as a function of the horizon radius is 
illustrated in Fig.~\ref{fig:JTE} for both branches. The coefficient 
exhibits pronounced divergences associated with the vanishing of the 
denominator in Eq.~\eqref{eq:muJ_result}, separating distinct thermodynamic response regimes. 
In the near-horizon region, $\mu_J$ is negative, indicating heating during 
expansion as electromagnetic contributions dominate. At intermediate radii, 
$\mu_J$ becomes positive, signaling a cooling regime in which the mass term 
governs the thermodynamic response. For sufficiently large horizon radii, the 
JT coefficient approaches small positive values and gradually stabilizes.

The non-extensivity parameter $\delta$ plays a crucial role in shaping the 
inversion structure. Increasing $\delta$ systematically shifts the inversion 
points toward larger horizon radii, implying that non-extensive corrections 
delay the onset of the cooling regime by enhancing the entropy contribution in 
the inversion condition. This behavior reflects the fact that Tsallis entropy 
amplifies the weight of large-horizon configurations in the thermodynamic 
response.

The side-by-side comparison in Fig.~\ref{fig:JTE} makes the role of the branch 
parameter $\zeta$ explicit: the analytic structure of 
Eq.~\eqref{eq:muJ_result} shows that $\zeta$ 
influences both the location of divergences and the extent of cooling and 
heating regions, and the phantom panel in Fig.~\ref{fig:jte_minus} confirms this 
numerically. The reinforced electromagnetic contribution 
characterizing the phantom branch modifies the balance between 
heating and cooling phases, in line with the enhanced thermal behavior observed 
in the previous sections. Overall, the JT expansion provides a sensitive probe 
of how non-extensive thermodynamics and branch structure jointly affect the 
response properties of Kalb-Ramond-ModMax black holes \cite{Kruglov:2022bhx}.

\section{Gravitational Lensing, PS, and Vacuum Optical Signatures}\label{isec5}

Gravitational lensing provides a direct and geometry-driven probe of spacetime 
structure around compact objects, allowing deviations from GR 
to be tested without detailed assumptions about the nature of the light source 
\cite{WOS:000475651300001CARDOSO,Chakraborty:2016lxo}. Since the first 
experimental confirmation of light deflection by the Sun, lensing has evolved 
into a precision tool for constraining modified gravity models, including 
scenarios with LSB or nonlinear electromagnetic sectors 
\cite{Partanen:2025snb}. In the weak-field regime, where the photon trajectory 
remains far from the EH, small deviations from the Schwarzschild 
prediction can accumulate into potentially observable signatures relevant to 
astrophysical lensing systems \cite{Tsukamoto:2014dta}.

In the present context, weak gravitational lensing offers a sensitive 
diagnostic of the KR-ModMax BH. Although both the KR 
condensate and the ModMax nonlinearity primarily affect the near-horizon 
geometry, their influence propagates to large distances through the optical 
structure of spacetime. A key feature of the metric~\eqref{metric_solution} is 
its non-standard asymptotic behaviour, $f(r)\to 1/(1-\ell)\neq 1$ as 
$r\to\infty$, which renders the standard Gibbons-Werner formulation of the 
GB theorem inapplicable 
\cite{Gibbons:2008rj,Jusufi:2017mav,Ishihara:2016vdc}. Instead, the 
deflection angle must be computed using the OIA extension~\cite{Ono:2017pie,Ishihara:2016vdc}, which incorporates a 
topological boundary correction arising from the non-Euclidean asymptotic 
structure of the optical manifold. This correction is not merely a technical refinement; as we show below, it produces a \emph{negative} contribution to the deflection angle that carries a direct imprint of the KR parameter $\ell$ and whose sign is opposite to that of BV monopole and Letelier cloud-of-string backgrounds, providing a clean observational discriminant between these superficially similar metric modifications.

\subsection{Non-asymptotic flatness and coordinate transformation}

The optical metric associated with a static, spherically symmetric 
spacetime~\eqref{metric_ansatz} restricted to the equatorial plane 
($\theta=\pi/2$) takes the form
\begin{equation}
d\sigma^2 = \frac{dr^2}{f^2(r)} + \frac{r^2}{f(r)}\, d\phi^2,
\label{eq:optical_metric}
\end{equation}
where the lapse function $f(r)$ is given in Eq.~\eqref{metric_solution}. 
The conformal factor $1/f(r)$ encodes gravitational redshift effects and 
ensures that extremizing the optical path length reproduces the null geodesic 
equations via Fermat's principle \cite{Wei:2015qca}.

As noted in Eq.~\eqref{eq:asymptotic}, the metric function satisfies 
$f(r)\to 1/(1-\ell)$ as $r\to\infty$. As a consequence, the geodesic 
curvature of the circular boundary $C_R$ at large $R$ in the optical geometry 
evaluates to
\begin{equation}
\lim_{R\to\infty}\kappa(C_R)\,\frac{dt}{d\phi}
= \sqrt{f_\infty} = \frac{1}{\sqrt{1-\ell}} \neq 1,
\label{eq:kappa_inf}
\end{equation}
confirming that the optical manifold is \emph{not} asymptotically 
Euclidean~\cite{Jusufi:2017mav,Ono:2017pie}. Therefore, the standard 
Gibbons-Werner GB formula~\cite{Gibbons:2008rj} must be replaced 
by the OIA generalisation.

To make the structure transparent, we define $K\equiv\sqrt{1-\ell}$ and 
introduce the rescaled metric function
\begin{equation}
h(r) \;\equiv\; K^2\,f(r) 
= 1 - \frac{2M(1-\ell)}{r} 
+ \frac{\zeta\,Q^2\,e^{-\gamma}}{(1-\ell)\,r^2},
\label{eq:h_function}
\end{equation}
which by construction satisfies $h(r)\to 1$ as $r\to\infty$. Under the 
coordinate transformation $\tilde{t}=t/K$, $\tilde{r}=K\,r$, the line 
element~\eqref{metric_ansatz} takes the BV-type form 
\cite{Jusufi:2017mav}
\begin{equation}
ds^2 = -\hat{h}(\tilde{r})\,d\tilde{t}^2 
+ \frac{d\tilde{r}^2}{\hat{h}(\tilde{r})} 
+ \frac{\tilde{r}^2}{K^2}\,d\Omega^2,
\label{eq:BV_form}
\end{equation}
where
\begin{equation}
\hat{h}(\tilde{r}) = 1 
- \frac{2\hat{M}}{\tilde{r}} 
+ \frac{\hat{Q}^2}{\tilde{r}^2},
\label{eq:htilde}
\end{equation}
with the effective parameters
\begin{equation}
\hat{M} = M(1-\ell)^{3/2},
\qquad
\hat{Q}^2 = \zeta\,Q^2\,e^{-\gamma}.
\label{eq:eff_params}
\end{equation}
The angular part $\tilde{r}^2/K^2$ in 
Eq.~\eqref{eq:BV_form} shows that the transformed metric possesses an 
\emph{angular surplus} (since $1/K^2 > 1$ for $\ell>0$), in contrast to 
the angular deficit arising in the Letelier cloud-of-strings or global 
monopole spacetimes studied 
in~\cite{Jusufi:2017mav,WOS:001561219400001Scalar}. It is instructive to trace the physical origin of this distinction. The BV global monopole metric introduces a solid-angle deficit $g_{\phi\phi} = (1-8\pi\eta^2)\,r^2$ that modifies the \emph{angular} part of the optical geometry, producing a \emph{positive} correction to the deflection angle. The Letelier cloud-of-strings background similarly enhances bending through a positive deficit term. By contrast, the KR modification acts on the \emph{radial} lapse through the rescaling $f_\infty = (1-\ell)^{-1} > 1$, which produces a \emph{negative} topological contribution $\pi(\sqrt{1-\ell} - 1) < 0$ that \emph{reduces} the total deflection angle below the Schwarzschild value. This sign difference constitutes a clean discriminant: a measured deflection deficit relative to GR predictions would favor the KR mechanism over monopole or string-cloud alternatives, while a deflection excess would disfavor it. The physical impact 
parameter in the rescaled frame is $\tilde{b}=Kb$, where $b$ is the 
coordinate distance of closest approach in the original frame. The rescaled 
function $\hat{h}$ has exactly the RN form, so all 
standard weak-field lensing results can be imported once the OIA boundary 
correction is included.

\subsection{GB method with OIA correction}

The GB theorem applied to the non-singular domain 
$\mathcal{D}_\infty$ exterior to the photon trajectory in the optical 
geometry yields~\cite{Gibbons:2008rj,Jusufi:2017mav,Ono:2017pie}
\begin{equation}
\iint_{\mathcal{D}_\infty}\!\mathcal{K}\,d\mathcal{S} 
+ \frac{1}{K}\bigl(\pi + \hat{\alpha}\bigr) = \pi,
\label{eq:GBT_OIA}
\end{equation}
where $\mathcal{K}$ is the Gaussian curvature and $d\mathcal{S}$ is the surface element 
of the optical manifold, and the factor $1/K$ arises from 
Eq.~\eqref{eq:kappa_inf}. Solving for the deflection angle gives
\begin{equation}
\hat{\alpha} = \pi(K-1) 
- K\iint_{\mathcal{D}_\infty}\!\mathcal{K}\,d\mathcal{S}.
\label{eq:alpha_OIA}
\end{equation}
The first term, $\pi(\sqrt{1-\ell}-1)<0$, is the topological boundary 
correction absent in the standard Gibbons-Werner approach. For the 
Letelier or global monopole spacetimes, this correction is positive and 
proportional to the angular deficit~\cite{Jusufi:2017mav}; in the present 
case the sign is reversed because the KR condensate modifies the lapse 
rather than the angular metric.

The Gaussian curvature of the optical metric~\eqref{eq:optical_metric} is 
determined by~\cite{Sucu:2025eix}
\begin{equation}
\mathcal{K} = -\frac{f'^{\,2}-2f\,f''}{4},
\label{eq:gauss_curvature_formula}
\end{equation}
which in the weak-field region ($r\gg r_+$) yields 
\begin{equation}
\mathcal{K} \simeq -\frac{2M}{(1-\ell)\,r^3} 
+ \frac{3M^2 + \dfrac{3\zeta\,Q^2\,e^{-\gamma}}{(1-\ell)^3}}{r^4} 
+ \mathcal{O}(r^{-5}).
\label{eq:gauss_curvature}
\end{equation}
The leading $r^{-3}$ contribution originates from the mass sector and carries 
a factor $(1-\ell)^{-1}$ reflecting the renormalized gravitational coupling. 
The $r^{-4}$ terms encode both second-order mass corrections and the 
charge-dependent ModMax contributions; the latter are suppressed by 
$e^{-\gamma}$ and amplified by $(1-\ell)^{-3}$.

To evaluate the curvature integral in Eq.~\eqref{eq:alpha_OIA} to 
$\mathcal{O}(M^2/b^2)$ accuracy, the straight-line trajectory 
$r=b/\sin\phi$ is insufficient for the $M^2$ coefficient 
\cite{Pereira:2025wkh,AraujoFilho:2024vsj}. Instead, one employs the 
iterative photon trajectory
\begin{equation}
\frac{1}{\tilde{r}} = \frac{\sin\phi}{\tilde{b}} 
+ \frac{\hat{M}\,(1-\cos\phi)^2}{\tilde{b}^2},
\label{eq:iter_traj}
\end{equation}
which accounts for the self-consistent bending of the photon path to 
first post-Minkowskian order~\cite{Pereira:2025wkh}. With this improved 
trajectory, the curvature integral yields the correct coefficient 
$15\pi/(4b^2)$ for the $M^2$ contribution, replacing the lower value 
$3\pi/(4b^2)$ obtained from the straight-line approximation.

\subsection{Weak-field deflection angle and branch-dependent signatures}

Carrying out the integration in Eq.~\eqref{eq:alpha_OIA} with the 
iterative trajectory~\eqref{eq:iter_traj} and the Gaussian 
curvature~\eqref{eq:gauss_curvature}, we obtain the vacuum deflection 
angle for the KR-ModMax BH:
\begin{eqnarray}
\hat{\alpha}_{\rm vac} &=& \pi\!\left(\sqrt{1-\ell}\,-1\right) 
+ \frac{4M(1-\ell)}{b}\nonumber\\[4pt]
&&\!\!\!\!\!\!+ \frac{\pi\!\left[15M^2(1-\ell)^2 
- \dfrac{3\zeta\,Q^2\,e^{-\gamma}}{1-\ell}\right]}{4b^2}\nonumber\\[4pt]
&&\!\!\!\!\!\!- \frac{4M\zeta\,Q^2\,e^{-\gamma}}{3b^3} 
+\mathcal{O}(b^{-4}).
\label{eq:deflection_angle}
\end{eqnarray}
This expression naturally decomposes into four physically distinct 
contributions: (i)~the topological boundary term 
$\pi(\sqrt{1-\ell}-1)<0$, which arises from the non-Euclidean asymptotic 
structure and \emph{reduces} the total deflection; (ii)~the leading mass 
term $4M(1-\ell)/b$, which is suppressed relative to the Schwarzschild 
value $4M/b$; (iii)~second-order terms at $\mathcal{O}(b^{-2})$ encoding 
gravitational self-interaction and charge-dependent ModMax corrections; 
and (iv)~a mixed mass-charge term at $\mathcal{O}(b^{-3})$, which is 
independent of~$\ell$ and coincides with the standard RN 
expression.

Expanding to first order in~$\ell$, which is the physically relevant 
regime given observational bounds on Lorentz-violating 
parameters~\cite{Masood:2024oej}, the deflection angle simplifies to
\begin{eqnarray}
\hat{\alpha}_{\rm vac} &\approx& -\frac{\pi\,\ell}{2} 
+ \frac{4M}{b}\left(1-\ell\right) \nonumber\\[2pt]
&&+ \frac{\pi\!\left(15M^2 - 3\zeta\,Q^2\,e^{-\gamma}\right)}{4b^2} 
+ \mathcal{O}(\ell^2,\ell/b^2),
\label{eq:deflection_O1}
\end{eqnarray}
where the $\ell$-dependence enters exclusively through the boundary term 
and the leading mass contribution.

Several physically relevant limits follow immediately. In the Schwarzschild 
limit $(\ell\to0,\gamma\to0,Q\to0)$, Eq.~\eqref{eq:deflection_angle} 
reduces to
\begin{equation}
\hat{\alpha}_{\rm Sch} = \frac{4M}{b} + \frac{15\pi M^2}{4b^2} 
+ \mathcal{O}(b^{-3}),
\end{equation}
recovering the standard GR result including the correct 
second-order coefficient 
\cite{Epstein:1980dw,weinberg2013gravitation,Pereira:2025wkh}. For the 
ordinary branch $(\zeta=+1)$ with $\ell\to0$ and $\gamma\to0$, the 
expression reduces to the RN deflection angle, i.e.
\begin{equation}
\hat{\alpha}_{\rm RN} = \frac{4M}{b} + \frac{15\pi M^2}{4b^2} 
- \frac{3\pi Q^2}{4b^2} - \frac{4MQ^2}{3b^3} 
+ \mathcal{O}(b^{-4}),
\end{equation}
where the electromagnetic contribution partially counteracts 
gravitational focusing.

In the phantom branch $(\zeta=-1)$ with $\ell\to0$ and $\gamma\to0$, the 
sign reversal of the charge sector transforms the electromagnetic 
contribution into an additional focusing mechanism:
\begin{equation}
\hat{\alpha}_{\rm phantom} = \frac{4M}{b} + \frac{15\pi M^2}{4b^2} 
+ \frac{3\pi Q^2}{4b^2} + \frac{4MQ^2}{3b^3} 
+ \mathcal{O}(b^{-4}),
\end{equation}
where both mass and charge contribute constructively to light bending, 
producing systematically larger deflection angles.

\begin{figure}[ht]
\centering
\begin{subfigure}[t]{0.41\textwidth}
\centering
  \vspace{-3.cm}
\includegraphics[width=\linewidth]{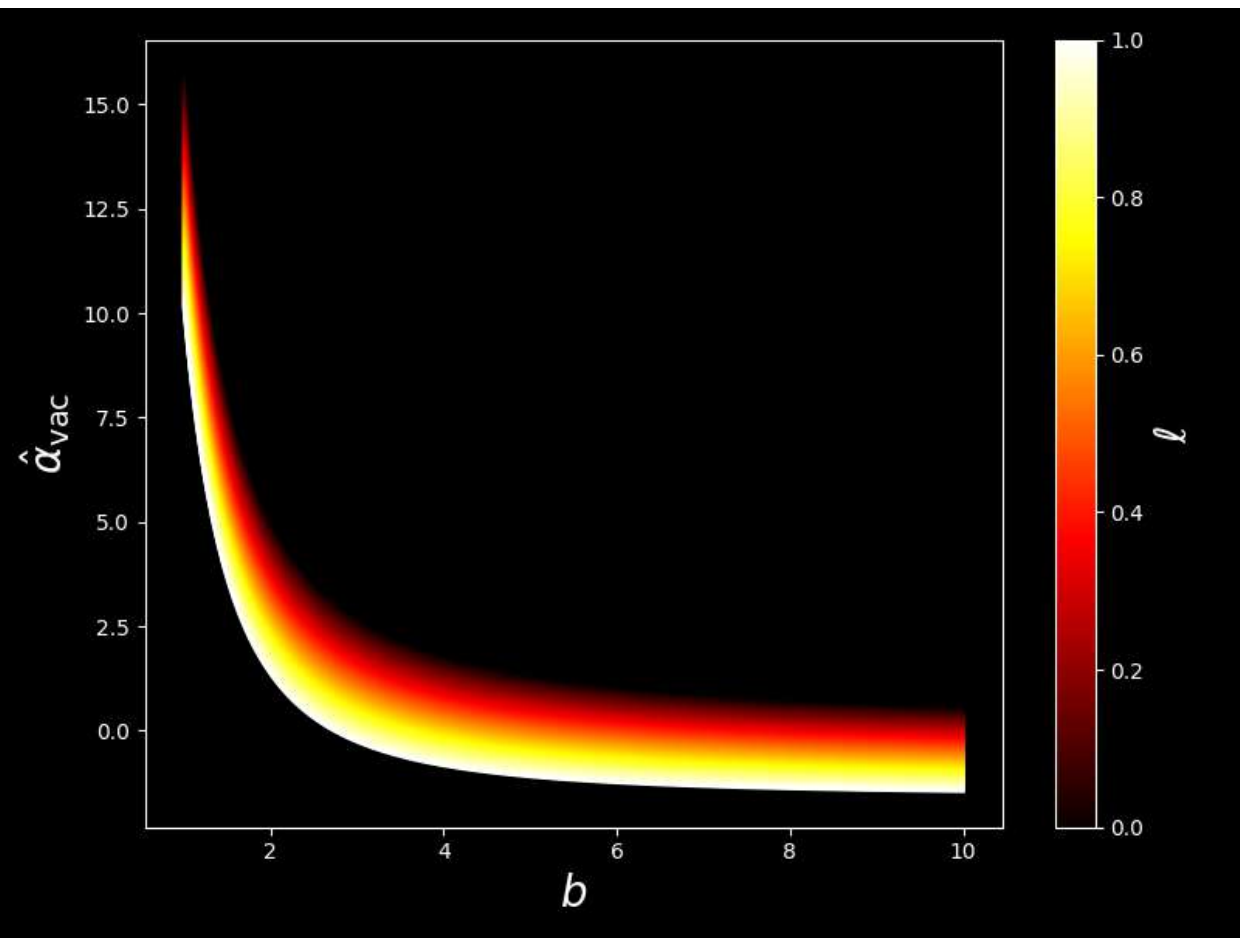}
  \vspace{-2.5cm}
\caption{Ordinary branch ($\zeta=+1$).}
\label{fig:lens_zeta_plus}
\end{subfigure}

\begin{subfigure}[t]{0.41\textwidth}
\centering
  \vspace{-2.5cm}
\includegraphics[width=\linewidth]{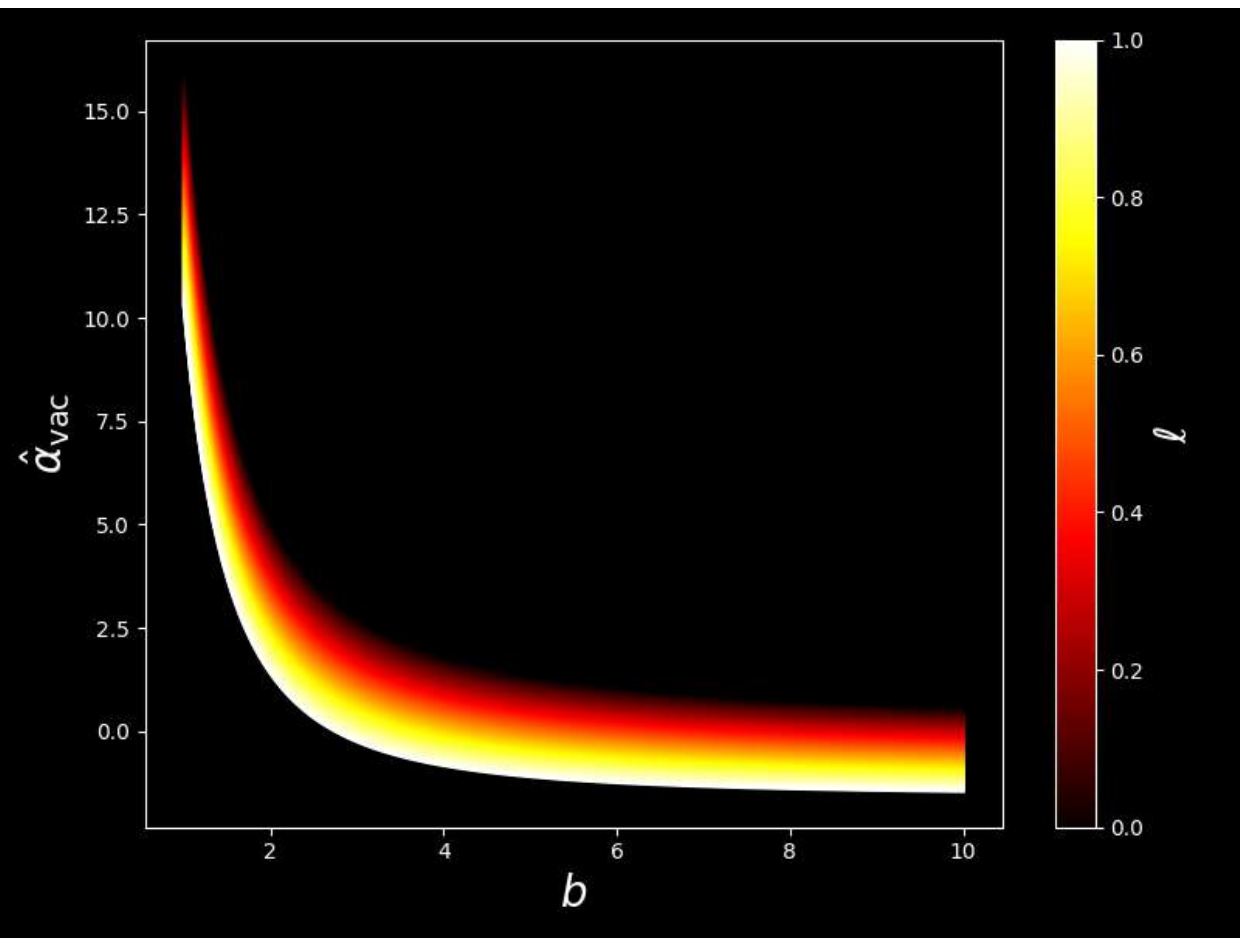}
  \vspace{-2.5cm}
\caption{Phantom branch ($\zeta=-1$).}
\label{fig:lens_zeta_minus}
\end{subfigure}
\caption{Vacuum deflection angle $\hat{\alpha}_{\rm vac}$ as a function of 
the impact parameter $b$ for KR-ModMax BHs with fixed $M=1$, 
$Q=0.5$, and $\gamma=2$. The colour scale indicates the KR 
parameter $\ell \in [0, 0.5]$. Panel~(a) shows the ordinary branch 
($\zeta=+1$), where the electromagnetic contribution partially 
counteracts gravitational focusing. Increasing $\ell$ reduces the total 
deflection through the negative boundary term 
$\pi(\sqrt{1-\ell}-1)$ and the suppressed effective mass 
$M_{\rm eff}=M(1-\ell)$. Panel~(b) displays the phantom branch 
($\zeta=-1$), where the sign reversal of the electromagnetic sector 
enhances focusing. Although the phantom-branch deflection angles remain 
larger than their ordinary counterparts, increasing $\ell$ reduces both 
branches relative to their Schwarzschild baselines through the universal 
topological correction.}
\label{fig:lens_comparison}
\end{figure}

The qualitative features of Eq.~\eqref{eq:deflection_angle} are 
illustrated in Fig.~\ref{fig:lens_comparison}. The phantom branch consistently produces larger deflection angles than the ordinary branch at fixed impact parameter $b$, a direct consequence of the reinforced gravitational potential from the sign-reversed electromagnetic stress-energy. For the ordinary branch, increasing $\ell$ reduces the deflection due to the negative topological correction, while increasing $\gamma$ further suppresses it by exponentially weakening the effective charge through $e^{-\gamma}$. These opposing trends in the two branches create a ``lensing fork'' that could serve as a branch discriminator in multi-epoch astrometric observations of sources lensed by compact objects. A distinctive prediction 
of the OIA-corrected analysis is that the KR parameter 
$\ell$ \emph{reduces} the deflection angle through two mechanisms: the 
negative topological boundary contribution 
$\pi(\sqrt{1-\ell}-1)$, and the suppression of the effective gravitational 
mass to $M(1-\ell)$ in the leading term. This contrasts qualitatively with 
the Letelier cloud-of-strings spacetime, where the analogous topological 
term is positive and enhances deflection~\cite{Jusufi:2017mav}. The 
difference is rooted in the distinct locations of non-flatness: the 
Letelier string cloud modifies the angular metric 
($g_{\phi\phi}=(1-A)r^2$), producing an angular deficit, while the KR 
condensate modifies the lapse ($f_\infty=1/(1-\ell)$), producing an 
effective angular surplus in the optical geometry.

The branch dichotomy persists: the phantom branch ($\zeta=-1$) yields 
larger deflection angles than the ordinary branch ($\zeta=+1$) at fixed 
$\ell$, due to the constructive alignment of mass and charge contributions. 
However, both branches experience the same $\ell$-dependent reduction, so 
the \emph{ratio} of phantom-to-ordinary deflection angles remains sensitive 
to the charge-to-mass ratio and the ModMax parameter~$\gamma$.

From an observational perspective, the reduction of the deflection angle 
with increasing $\ell$ would manifest as smaller Einstein ring radii and 
decreased magnification in strong-lensing systems compared to the 
Schwarzschild prediction at the same mass, providing a distinctive 
observational signature of KR LSB. Precision astrometric missions and EHT ring-size measurements operating at the ${\sim}1\%$ level could in principle distinguish between the KR lensing deficit and the lensing excess predicted by monopole or string-cloud backgrounds, since the two produce opposite-sign corrections to the Schwarzschild baseline at the same order in the deformation parameter. Together with the thermodynamic and PS properties discussed in 
previous sections, gravitational lensing offers a complementary channel 
for constraining the KR parameter~$\ell$ and discriminating between the 
ordinary and phantom KR-ModMax BH branches.

\section{Plasma effects on photon propagation}
\label{isec6}

In realistic astrophysical environments, black holes are rarely isolated 
systems. Instead, they are typically surrounded by plasma-rich media such as 
accretion disks, stellar winds, or diffuse interstellar matter, all of which 
modify photon propagation through dispersive effects 
\cite{Broderick:2008qf}. In contrast to vacuum null 
geodesics, light rays propagating in plasma experience frequency-dependent 
refraction, leading to shifts in photon trajectories, photon-sphere locations, 
and shadow properties.

For the Kalb-Ramond-ModMax black hole, incorporating plasma effects is 
therefore essential for bridging the gap between theoretical predictions and 
observational applications, particularly in the context of horizon-scale 
imaging and lensing measurements by instruments such as the Event Horizon 
Telescope \cite{EventHorizonTelescope:2019pgp}. In this section, we analyze 
photon dynamics in a dispersive plasma medium and derive the modified 
conditions for circular photon orbits that define the photon sphere.

\subsection{Photon dynamics and photon spheres in plasma}

Photon propagation in a plasma medium can be described using an effective 
Hamiltonian formalism that incorporates the dispersive response of the medium 
\cite{Crisnejo:2018uyn}. For a static plasma distribution in 
the Kalb-Ramond-ModMax spacetime, the Hamiltonian governing photon motion 
takes the form
\begin{equation}
H = \frac{1}{2}\left[g^{\alpha\beta}p_\alpha p_\beta + (\omega_p^2 - 
1)(p_\alpha u^\alpha)^2\right],
\label{eq:plasma_hamiltonian}
\end{equation}
where $p_\alpha$ denotes the photon four-momentum, $u^\alpha$ is the plasma 
four-velocity, and $\omega_p$ is the local plasma frequency. Assuming a static 
plasma configuration, the four-velocity is given by $u^\alpha = 
f^{-1/2}\delta^\alpha_t$, where $f(r)$ is the KR-ModMax lapse function defined 
in Eq.~\eqref{metric_solution}.

The dispersive nature of the plasma is encoded in the refractive index,
\begin{equation}
n^2 = 1 - \frac{\omega_p^2}{\omega^2},
\label{eq:refractive_index}
\end{equation}
where $\omega$ denotes the photon frequency measured by a local static 
observer. The plasma frequency itself is determined by the electron number 
density $N(r)$ through
\begin{equation}
\omega_p^2(r) = \frac{4\pi e^2 N(r)}{m_e},
\label{eq:plasma_frequency}
\end{equation}
with $e$ and $m_e$ representing the electron charge and mass, respectively. A 
photon detected at spatial infinity with frequency $\omega_0$ undergoes 
gravitational redshift as it propagates inward, such that
\begin{equation}
\omega(r) = \frac{\omega_0}{\sqrt{f(r)}}.
\label{eq:redshifted_frequency}
\end{equation}
The vacuum limit is recovered smoothly by setting $\omega_p \to 0$, in which 
case the Hamiltonian reduces to the standard null geodesic condition 
$g^{\alpha\beta}p_\alpha p_\beta = 0$.

Restricting attention to the equatorial plane ($\theta=\pi/2$), Hamilton's 
equations yield the equations of motion
\begin{equation}
\dot{t} = -\frac{p_t}{f(r)}, \qquad 
\dot{r} = f(r)p_r, \qquad 
\dot{\varphi} = \frac{p_\varphi}{r^2},
\label{eq:hamilton_eqs}
\end{equation}
where the overdot denotes differentiation with respect to an affine parameter. 
Imposing the Hamiltonian constraint leads to the radial orbit equation
\begin{equation}
\frac{dr}{d\varphi} = r\sqrt{f(r)}\sqrt{h^2(r)\frac{\omega_0^2}{p_\varphi^2} - 
1},
\label{eq:orbit_equation}
\end{equation}
where the optical impact function $h(r)$ is defined as \cite{Sucu:2025lqa}
\begin{equation}
h^2(r) = r^2\left(\frac{1}{f(r)} - \frac{\omega_p^2(r)}{\omega_0^2}\right).
\label{eq:impact_function}
\end{equation}

Unstable circular photon orbits, which determine the photon sphere and the 
boundary of the black hole shadow, correspond to extrema of the impact function 
$h^2(r)$. The photon-sphere radius $r_p$ is therefore obtained from the 
condition
\begin{equation}
\frac{d}{dr}\left[r^2\left(\frac{1}{f(r)} - 
\frac{\omega_p^2(r)}{\omega_0^2}\right)\right]_{r=r_p} = 0,
\label{eq:photon_sphere_general}
\end{equation}
or equivalently,
\begin{equation}
\frac{2}{f(r_p)} - \frac{2\omega_p^2(r_p)}{\omega_0^2}
- r_p\left(\frac{f'(r_p)}{f^2(r_p)} + 
\frac{2\omega_p(r_p)\omega_p'(r_p)}{\omega_0^2}\right) = 0.
\label{eq:photon_sphere_explicit}
\end{equation}
This condition generalizes the standard vacuum photon-sphere equation by 
incorporating plasma effects through both the local plasma frequency 
$\omega_p(r_p)$ and its radial gradient $\omega_p'(r_p)$. As a result, the 
photon-sphere radius becomes explicitly frequency dependent and sensitive to 
the plasma distribution, providing a direct channel through which environmental 
effects can modify observable shadow and lensing properties.

\subsection{Homogeneous versus inhomogeneous plasma distributions}

The influence of plasma on photon-sphere properties depends crucially on the 
spatial structure of the medium. To disentangle intrinsic spacetime effects 
from environmental contributions, we examine two representative scenarios: a 
homogeneous plasma with constant plasma frequency and an inhomogeneous plasma 
with a radially stratified density profile.

For a spatially uniform plasma, the plasma frequency is constant, 
$\omega_p=\text{const.}$, and the gradient term in 
Eq.~\eqref{eq:photon_sphere_explicit} vanishes. The photon-sphere condition 
then reduces to
\begin{equation}
\frac{2}{f(r_p)} - \frac{2\omega_p^2}{\omega_0^2}
= r_p\frac{f'(r_p)}{f^2(r_p)},
\label{eq:ps_homogeneous}
\end{equation}
which can be solved numerically for the Kalb-Ramond-ModMax metric 
function~\eqref{metric_solution} at fixed values of the plasma-to-photon 
frequency ratio $\omega_p^2/\omega_0^2$.

\begin{table}[htbp]
\centering
\renewcommand{\arraystretch}{1.5}
\setlength{\tabcolsep}{10pt}
\caption{Photon sphere radius $r_p$ for KR-ModMax black holes in homogeneous 
plasma with fixed $M=1$, $\ell=0.2$, and $Q=0.5$. The ordinary branch 
($\zeta=+1$) yields smaller photon sphere radii due to the repulsive 
electromagnetic contribution, while the phantom branch ($\zeta=-1$) produces 
larger $r_p$ from enhanced gravitational focusing. Increasing plasma density 
($\omega_p^2/\omega_0^2$) uniformly shifts $r_p$ outward for both branches.}
\begin{tabular}{|c|c|c|c|}
\hline
\cellcolor{brown!0}$\zeta$ & \cellcolor{brown!0}$\gamma$ & 
\cellcolor{brown!0}$\omega_p^2/\omega_0^2$ & \cellcolor{brown!0}$r_p$ \\
\hline\hline
\multicolumn{4}{|c|}{ \textbf{Ordinary Branch ($\zeta = 
+1$)}} \\
\hline
$+1$ & $0$ & $0.00$ & $2.103$ \\
$+1$ & $0$ & $0.05$ & $2.118$ \\
$+1$ & $0$ & $0.10$ & $2.134$ \\
$+1$ & $0$ & $0.20$ & $2.171$ \\
\hline
$+1$ & $1$ & $0.00$ & $2.300$ \\
$+1$ & $1$ & $0.05$ & $2.317$ \\
$+1$ & $1$ & $0.10$ & $2.334$ \\
$+1$ & $1$ & $0.20$ & $2.374$ \\
\hline
\multicolumn{4}{|c|}{ \textbf{Phantom Branch ($\zeta = -1$)}} 
\\
\hline
$-1$ & $0$ & $0.00$ & $2.637$ \\
$-1$ & $0$ & $0.05$ & $2.656$ \\
$-1$ & $0$ & $0.10$ & $2.677$ \\
$-1$ & $0$ & $0.20$ & $2.723$ \\
\hline
$-1$ & $1$ & $0.00$ & $2.492$ \\
$-1$ & $1$ & $0.05$ & $2.510$ \\
$-1$ & $1$ & $0.10$ & $2.530$ \\
$-1$ & $1$ & $0.20$ & $2.573$ \\
\hline
\end{tabular}
\label{tab:ps_homogeneous}
\end{table}

The results summarized in Table~\ref{tab:ps_homogeneous} reveal several robust 
trends. First, the presence of plasma systematically shifts the photon sphere 
outward for both branches. As the ratio $\omega_p^2/\omega_0^2$ increases, the 
effective propagation speed of photons decreases, weakening curvature-induced 
focusing and pushing unstable circular orbits to larger radii. Quantitatively, 
increasing $\omega_p^2/\omega_0^2$ from $0$ to $0.20$ leads to a shift $\Delta 
r_p \approx 0.07$ for the ordinary branch and $\Delta r_p \approx 0.09$ for the 
phantom branch.

Second, the branch dichotomy remains clearly visible in homogeneous plasma. For 
identical values of $(\ell,\gamma)$, the phantom branch ($\zeta=-1$) 
consistently produces larger photon-sphere radii than the ordinary branch 
($\zeta=+1$). In vacuum and at $\gamma=0$, this difference reaches $\Delta r_p 
\approx 0.53$, corresponding to an enhancement of roughly $25\%$. Importantly, 
this separation persists across all plasma densities, indicating that it 
originates from the intrinsic sign reversal of the electromagnetic sector 
rather than from plasma effects.

Finally, the ModMax parameter $\gamma$ affects the two branches in 
qualitatively different ways. In the ordinary branch, increasing $\gamma$ 
suppresses the effective charge contribution through the factor $e^{-\gamma}$ 
and shifts the geometry toward the Schwarzschild-Kalb-Ramond limit, resulting 
in a moderate increase of $r_p$. In the phantom branch, the same increase in 
$\gamma$ partially suppresses the enhanced focusing induced by the phantom 
charge, leading instead to a decrease of the photon-sphere radius. This 
opposite response provides an additional handle for distinguishing between the 
two branches in weakly stratified environments.

Now, astrophysical plasmas are typically radially stratified rather than 
homogeneous. To model this situation, we consider a power-law plasma profile,
\begin{equation}
\omega_p^2(r) = \frac{\kappa_0}{r^\alpha},
\label{eq:power_law_plasma}
\end{equation}
with $\alpha=1$, which captures the characteristic $1/r$ falloff expected in 
stellar winds and accretion flows. Substituting this profile into 
Eq.~\eqref{eq:photon_sphere_explicit} yields
\begin{equation}
\frac{2}{f(r_p)} - \frac{2\kappa_0}{r_p\omega_0^2}
- r_p\left(\frac{f'(r_p)}{f^2(r_p)}
+ \frac{2\kappa_0^2}{r_p^3\omega_0^2}\right) = 0,
\label{eq:ps_inhomogeneous}
\end{equation}
which must again be solved numerically.

\begin{table}[htbp]
\centering
\renewcommand{\arraystretch}{1.6}
\setlength{\tabcolsep}{10pt}
\caption{Photon sphere radius $r_p$ for KR-ModMax black holes in inhomogeneous 
plasma ($\alpha = 1$) with fixed $M=1$, $Q=0.5$, $\kappa_0=1$, and 
$\omega_0=0.5$. The KR parameter $\ell$ dominates the photon sphere structure, 
producing a four-fold increase in $r_p$ as $\ell$ increases from $0.1$ to 
$0.8$. The ModMax branch parameter $\zeta$ and nonlinearity $\gamma$ produce 
only marginal corrections at the third decimal place.}
\begin{tabular}{|c|c|c|c|}
\hline
\cellcolor{brown!0}$\ell$ & \cellcolor{brown!0}$\zeta$ & 
\cellcolor{brown!0}$\gamma$ & \cellcolor{brown!0}$r_p$ \\
\hline\hline
\multicolumn{4}{|c|}{ \textbf{Weak KR Deformation ($\ell = 
0.1$)}} \\
\hline
$0.1$ & $+1$ & $0$ & $4.734$ \\
$0.1$ & $+1$ & $1$ & $4.773$ \\
$0.1$ & $-1$ & $0$ & $4.860$ \\
$0.1$ & $-1$ & $1$ & $4.819$ \\
\hline
\multicolumn{4}{|c|}{ \textbf{Moderate KR Deformation 
($\ell = 0.5$)}} \\
\hline
$0.5$ & $+1$ & $0$ & $8.447$ \\
$0.5$ & $+1$ & $1$ & $8.452$ \\
$0.5$ & $-1$ & $0$ & $8.465$ \\
$0.5$ & $-1$ & $1$ & $8.459$ \\
\hline
\multicolumn{4}{|c|}{ \textbf{Strong KR Deformation ($\ell 
= 0.8$)}} \\
\hline
$0.8$ & $+1$ & $0$ & $20.762$ \\
$0.8$ & $+1$ & $1$ & $20.763$ \\
$0.8$ & $-1$ & $0$ & $20.765$ \\
$0.8$ & $-1$ & $1$ & $20.764$ \\
\hline
\end{tabular}
\label{tab:ps_inhomogeneous}
\end{table}

The results collected in Table~\ref{tab:ps_inhomogeneous} display a markedly 
different hierarchy of effects. In this case, the photon-sphere radius is 
dominated by the Kalb-Ramond parameter $\ell$, with only marginal sensitivity 
to the ModMax sector. As $\ell$ increases from $0.1$ to $0.8$, the 
photon-sphere radius grows by more than a factor of four, reflecting the strong 
amplification induced by the asymptotic rescaling $(1-\ell)^{-1}$ in 
combination with the radially varying refractive index.

On the other hand, variations in the branch parameter $\zeta$ and in the ModMax 
parameter $\gamma$ produce only minute shifts in $r_p$, typically at the level 
$\Delta r_p\sim10^{-3}$-$10^{-2}$. For instance, at $\ell=0.8$ the difference 
between the ordinary and phantom branches is reduced to $\Delta 
r_p\approx0.003$, compared to $\Delta r_p\approx0.53$ in the homogeneous case. 
This strong suppression indicates that the plasma gradient dominates the 
effective optical potential, effectively masking the sign-sensitive 
electromagnetic contributions.

We close this subsection by discussing the physical and observational 
implications.
The contrast between homogeneous and inhomogeneous plasma regimes has direct 
implications for black-hole imaging. In homogeneous or weakly stratified media, 
the photon-sphere radius retains clear sensitivity to the electromagnetic 
branch structure, and the sizable separation between ordinary and phantom 
configurations could, in principle, be probed through precise measurements of 
shadow sizes or frequency-dependent lensing. In strongly stratified 
environments, however, plasma effects and Kalb-Ramond-induced rescaling 
dominate photon dynamics, rendering branch discrimination observationally 
challenging.

At the same time, the pronounced sensitivity of the photon-sphere radius to the 
Lorentz-symmetry-breaking parameter $\ell$ in inhomogeneous plasma suggests 
that high-resolution, multi-frequency observations could place meaningful 
constraints on such deformations, provided that plasma properties are modeled 
accurately. Overall, these results reveal the necessity of incorporating 
realistic plasma distributions when attempting to extract fundamental 
gravitational or electromagnetic physics from horizon-scale observations.

\section{Tidal forces and branch-dependent deformations}
\label{isec7}

Tidal effects acting on extended bodies provide a direct probe of spacetime 
curvature beyond the idealized motion of pointlike test particles. While 
geodesic trajectories encode the global structure of a gravitational field, 
finite-sized objects respond to local curvature gradients through differential 
accelerations governed by the Riemann tensor. In strong-field regimes, these 
tidal deformations can exhibit qualitative features that are invisible at the 
level of geodesic motion alone.

In the Kalb-Ramond-ModMax framework, the combined presence of 
Lorentz-symmetry-breaking effects and nonlinear electrodynamics modifies the 
radial dependence of the metric function relative to standard electrovacuum 
solutions. As a result, the tidal response of infalling matter can depart 
significantly from the familiar Schwarzschild or Reissner-Nordström behavior. 
In this section, we analyze tidal forces in the ordinary and phantom branches 
and identify branch-dependent deformation patterns with no counterpart in 
general relativity.

\subsection{Geodesic deviation and tidal forces}

The relative acceleration between neighboring freely falling worldlines is 
governed by the geodesic deviation equation \cite{Cordeiro:2025cfo}
\begin{equation}
\frac{D^{2}\eta^{\mu}}{D\tau^{2}}
=
- R^{\mu}{}_{\nu\alpha\beta}\,
u^{\nu} u^{\alpha} \eta^{\beta},
\end{equation}
where $\eta^{\mu}$ denotes the separation vector between nearby geodesics, 
$u^{\mu}$ is the four-velocity of the reference worldline, and $\tau$ is the 
proper time. The Riemann curvature tensor acts as the mediator of tidal 
interactions, converting spacetime geometry into physically measurable 
distortions of extended bodies.

To obtain physically transparent tidal forces, it is convenient to project the 
curvature tensor onto an orthonormal tetrad adapted to static observers. 
Introducing the basis \cite{Chandrasekhar:1985kt,poisson2004relativist}
\begin{align}
e_{\hat{0}} &= f^{-1/2}\,\partial_{t}, \nonumber \\
e_{\hat{1}} &= f^{1/2}\,\partial_{r}, \nonumber \\
e_{\hat{2}} &= \frac{1}{r}\,\partial_{\theta}, \nonumber \\
e_{\hat{3}} &= \frac{1}{r\sin\theta}\,\partial_{\varphi},
\end{align}
the metric locally reduces to the Minkowski form 
$\eta_{\hat{a}\hat{b}}=\mathrm{diag}(-1,1,1,1)$.

In this frame, tidal forces are encoded in the electric part of the Riemann 
tensor,
\begin{equation}
\mathcal{E}_{\hat{i}\hat{j}} = R_{\hat{0}\hat{i}\hat{0}\hat{j}},
\end{equation}
which directly determines the relative acceleration of nearby freely falling 
particles. For any static, spherically symmetric spacetime, the independent 
nonvanishing components reduce to
\begin{equation}
\mathcal{E}_{\hat{1}\hat{1}}
=
-\frac{1}{2} f''(r),
\qquad
\mathcal{E}_{\hat{2}\hat{2}}
=
\mathcal{E}_{\hat{3}\hat{3}}
=
-\frac{f'(r)}{2r},
\end{equation}
with spherical symmetry enforcing equality of the two angular components.

For the Kalb-Ramond-ModMax black hole, the first and second radial derivatives 
of the metric function~\eqref{metric_solution} are given by
\begin{align}
f'(r)
&=
\frac{2M}{r^{2}}
-
\frac{2\zeta Q^{2} e^{-\gamma}}{(1-\ell)^{2} r^{3}},
\\[1mm]
f''(r)
&=
-\frac{4M}{r^{3}}
+
\frac{6\zeta Q^{2} e^{-\gamma}}{(1-\ell)^{2} r^{4}}.
\end{align}
Substituting these expressions into the geodesic deviation equation yields the 
explicit tidal accelerations,
\begin{align}
\frac{D^{2}\eta_{\hat{1}}}{D\tau^{2}}
&=
\left(
\frac{2M}{r^{3}}
-
\frac{3\zeta Q^{2} e^{-\gamma}}{(1-\ell)^{2} r^{4}}
\right)\eta_{\hat{1}},
\label{eq:radial_tid_KRMM}
\\[2mm]
\frac{D^{2}\eta_{\hat{\imath}}}{D\tau^{2}}
&=
\left(
-\frac{M}{r^{3}}
+
\frac{\zeta Q^{2} e^{-\gamma}}{(1-\ell)^{2} r^{4}}
\right)\eta_{\hat{\imath}},
\qquad \hat{\imath}=\hat{2},\hat{3}.
\label{eq:angular_tid_KRMM}
\end{align}

At large radii, the leading $r^{-3}$ terms dominate, recovering the familiar 
Schwarzschild tidal pattern characterized by radial stretching accompanied by 
angular compression. Closer to the black hole, however, the charge-dependent 
terms scale more steeply with radius and can significantly reshape the tidal 
profile. The manner in which these corrections compete or reinforce the 
Schwarzschild contribution depends crucially on the electromagnetic branch 
parameter $\zeta$, setting the stage for qualitatively distinct tidal behaviors 
in the ordinary and phantom sectors.

\subsection{Tidal inversion and physical implications}

A distinctive feature of the Kalb-Ramond-ModMax black hole is the possibility 
of tidal-force inversion, namely the vanishing and subsequent sign reversal of 
tidal accelerations at finite radii outside the event horizon. This phenomenon 
has no analogue in the Schwarzschild spacetime and arises from the competition 
between the mass-induced curvature and the charge-dependent contributions 
controlled by the Kalb-Ramond and ModMax sectors.

\begin{figure}[ht]
\centering
\includegraphics[width=1.\columnwidth]{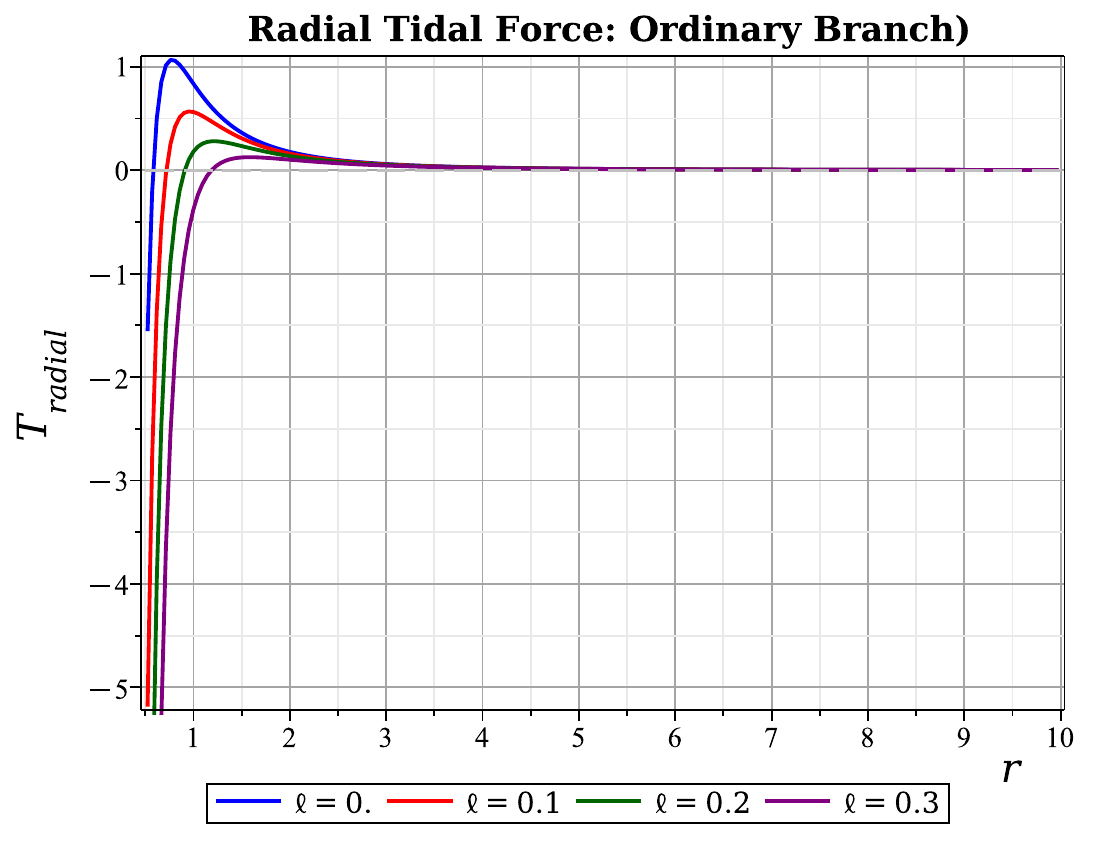}
\caption{Radial tidal acceleration $T_{\mathrm{rad}}$ as a function of $r$ for 
different values of the Lorentz-violating parameter $\ell$ in the ordinary 
branch ($\zeta = +1$). Increasing $\ell$ shifts the zero-crossing radius 
$R_{\mathrm{rad}}$ to larger values and enhances the tidal forces at 
intermediate radii. Parameters: $M = 1$, $Q = 0.8$, $\gamma = 0.5$.}
\label{fig:tidal_ell}
\end{figure}

The sensitivity of tidal forces to the Lorentz-symmetry-breaking parameter 
$\ell$ is illustrated in Fig.~\ref{fig:tidal_ell} for the ordinary branch. As 
$\ell$ increases, the factor $(1-\ell)^{-2}$ amplifies the charge-dependent 
terms in Eqs.~\eqref{eq:radial_tid_KRMM}-\eqref{eq:angular_tid_KRMM} relative 
to the Schwarzschild contribution. This produces two correlated effects: the 
magnitude of the tidal acceleration at fixed intermediate radii increases, and 
the radial location at which the radial tidal force vanishes is shifted 
outward. Physically, stronger Lorentz violation enlarges the region where 
charge-induced corrections dominate the tidal response, thereby extending the 
domain of tidal inversion.

\begin{figure}[ht!]
\centering
\includegraphics[width=0.95\columnwidth]{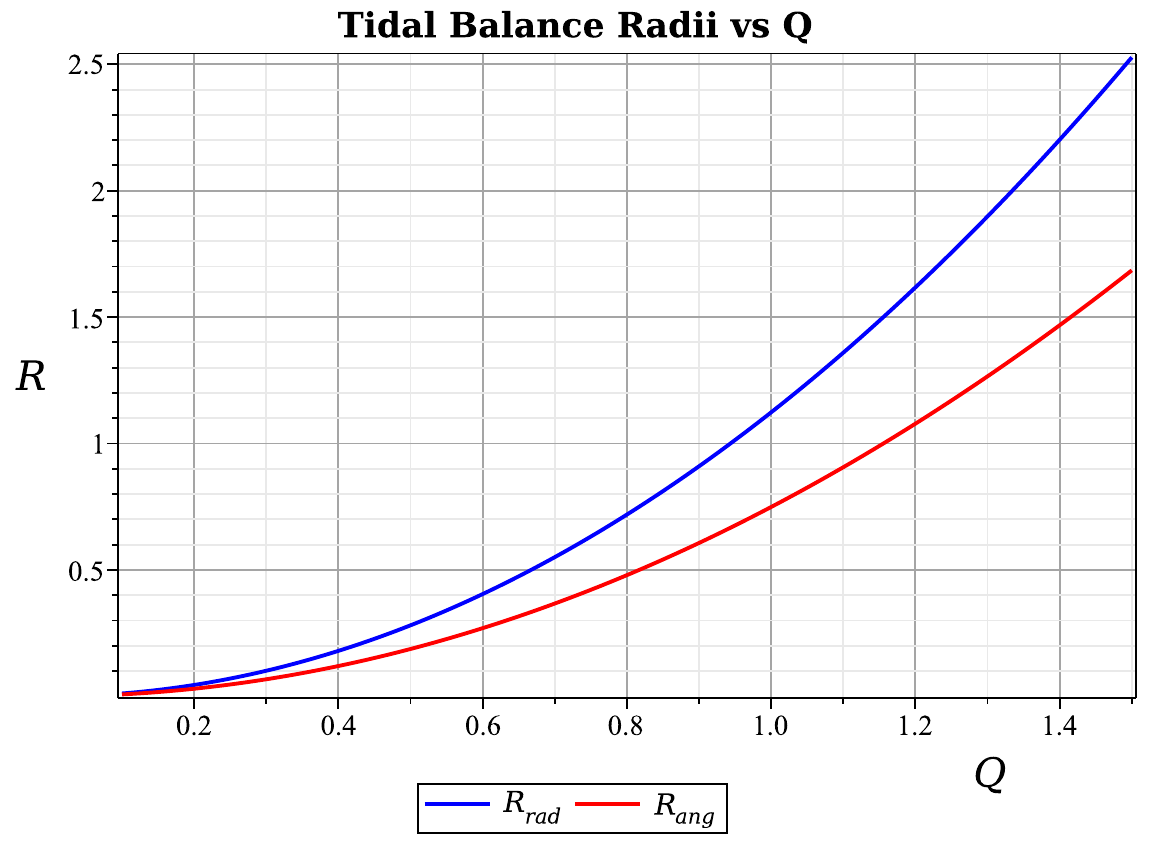}
\caption{
Tidal balance radii $R_{\mathrm{rad}}$ (blue) and $R_{\mathrm{ang}}$ (red) as
functions of the electric charge $Q$ for the ordinary ModMax branch
($\zeta=+1$).
Both radii scale quadratically with $Q$ and preserve the universal ratio
$R_{\mathrm{rad}}/R_{\mathrm{ang}}=3/2$, independent of model parameters.
Parameters: $M=1$, $\ell=0.1$, $\gamma=0.5$.
}
\label{fig:tidal_balance}
\end{figure}

\begin{table}[htbp]
\centering
\renewcommand{\arraystretch}{1.5}
\setlength{\tabcolsep}{10pt}
\caption{Tidal balance radii for the KR-ModMax BH in the ordinary branch 
($\zeta = +1$), with $M = 1$ and $Q = 0.8$. The ratio 
$R_{\mathrm{rad}}/R_{\mathrm{ang}} = 3/2$ holds universally. }
\begin{tabular}{|c|c|c|c|}
 \hline
$\ell$ & $\gamma$ & $R_{\mathrm{rad}}$ & $R_{\mathrm{ang}}$ \\
\hline\hline 
  0.00 & 0.0 & 0.9600 & 0.6400 \\
0.00 & 0.5 & 0.5823 & 0.3882 \\
0.00 & 1.0 & 0.3532 & 0.2354 \\
0.00 & 1.5 & 0.2142 & 0.1428 \\
\hline
0.10 & 0.0 & 1.1852 & 0.7901 \\
0.10 & 0.5 & 0.7189 & 0.4792 \\
0.10 & 1.0 & 0.4360 & 0.2907 \\
0.10 & 1.5 & 0.2645 & 0.1763 \\
\hline
0.20 & 0.0 & 1.5000 & 1.0000 \\
0.20 & 0.5 & 0.9098 & 0.6065 \\
0.20 & 1.0 & 0.5518 & 0.3679 \\
0.20 & 1.5 & 0.3347 & 0.2231 \\
\hline
0.30 & 0.0 & 1.9592 & 1.3061 \\
0.30 & 0.5 & 1.1883 & 0.7922 \\
0.30 & 1.0 & 0.7207 & 0.4805 \\
0.30 & 1.5 & 0.4372 & 0.2914 \\
\hline
\end{tabular}
\label{tab:tidal_radii}
\end{table}

The characteristic transition scales are quantified by the tidal balance radii 
$R_{\mathrm{rad}}$ and $R_{\mathrm{ang}}$, defined by the vanishing of the 
radial and angular tidal components, respectively. Their dependence on the 
electric charge and model parameters is displayed in 
Fig.~\ref{fig:tidal_balance} and summarized numerically in 
Table~\ref{tab:tidal_radii}. The data reveal systematic trends: both balance 
radii increase with $\ell$ due to the universal $(1-\ell)^{-2}$ enhancement, 
while increasing the ModMax parameter $\gamma$ suppresses the charge 
contribution through the factor $e^{-\gamma}$, shifting the inversion radii 
inward. For representative parameter choices, the balance radii span the range 
$R_{\mathrm{rad}},R_{\mathrm{ang}}\sim 0.1$-$2$ in units of the black-hole 
mass, indicating that tidal inversion may occur either close to the horizon or 
well outside it, depending on the regime.   A closed-form
derivation of $R_{\rm rad}$, $R_{\rm ang}$, and their universal
ratio $R_{\rm rad}/R_{\rm ang}=3/2$ is given in Sec.~B.5 of
Appendix~\ref{app:thermo_check}.

\begin{figure}[htbp]
\hspace{-0.5cm}
\includegraphics[width=1.\columnwidth]{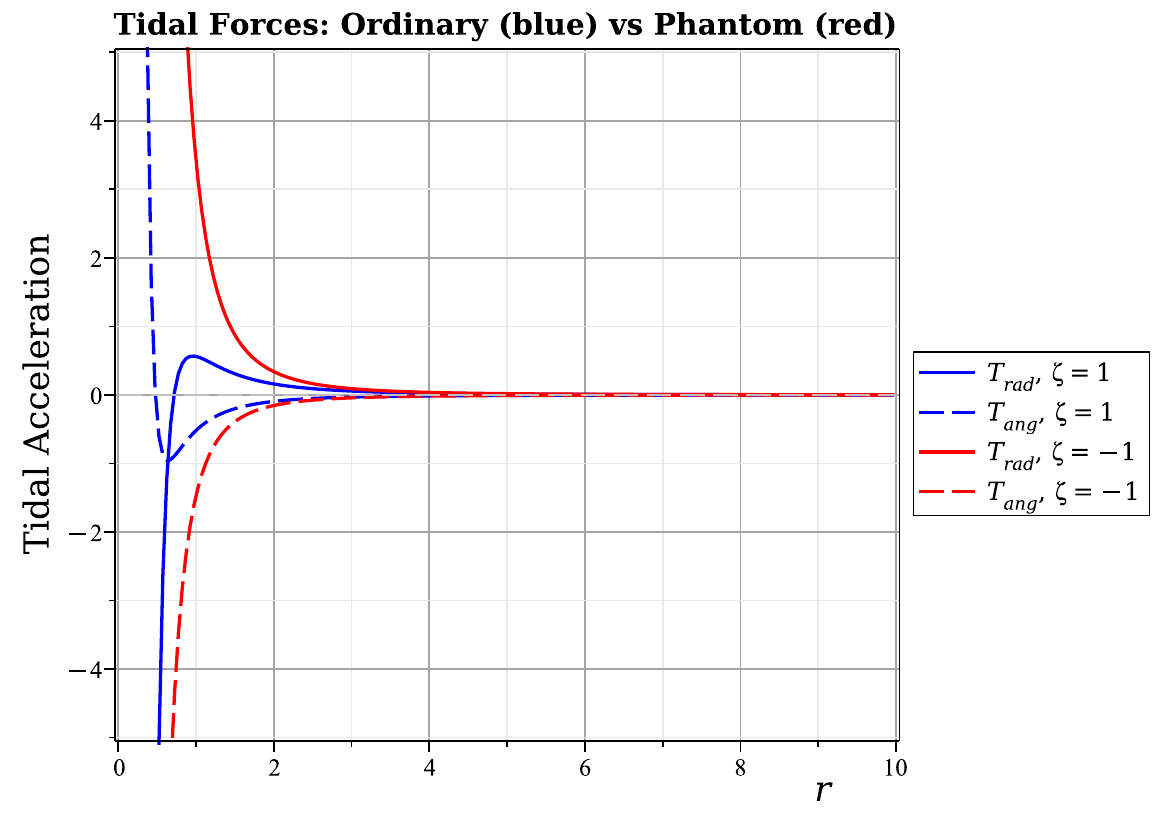}
\caption{
Radial (solid lines) and angular (dashed lines) tidal acceleration components
for KR-ModMax black holes, evaluated in an orthonormal frame.
Blue curves correspond to the ordinary branch ($\zeta=+1$), while red curves
represent the phantom branch ($\zeta=-1$).
The horizontal gray dashed line marks vanishing tidal acceleration.
For the ordinary branch, both components cross zero at finite radii, whereas no
zero crossing occurs for the phantom branch.
Parameters: $M=1$, $Q=0.8$, $\ell=0.1$, $\gamma=0.5$.
}
\label{fig:tidal_branch}
\end{figure}

The existence and physical relevance of these inversion scales depend crucially 
on the electromagnetic branch parameter $\zeta$. For the ordinary ModMax branch 
($\zeta=+1$), the charge contribution enters the tidal tensor with a sign 
opposite to the Schwarzschild term at sufficiently small radii. As shown in 
Fig.~\ref{fig:tidal_branch}, radial tidal stretching weakens as the black hole 
is approached, vanishes at $R_{\mathrm{rad}}$, and is subsequently replaced by 
radial compression. Simultaneously, angular compression is softened and 
reverses at $R_{\mathrm{ang}}$. This regulated tidal environment represents a 
qualitative departure from the classical spaghettification scenario and suggests 
that the Kalb-Ramond-ModMax corrections act as an effective curvature-softening 
mechanism. Such behavior could influence the coherence, deformation, and 
survivability of extended structures, such as accretion streams or plasma 
filaments, in the strong-field region.

On the other hand, for the phantom branch ($\zeta=-1$) the electromagnetic 
contribution reinforces the Schwarzschild tidal field rather than opposing it. 
In this case, both $R_{\mathrm{rad}}$ and $R_{\mathrm{ang}}$ become negative 
and therefore lose any direct geometric meaning. Radial stretching and angular 
compression persist throughout the exterior region and grow monotonically 
toward the horizon, producing tidal stresses that are even stronger than those 
encountered in standard general relativity. This behavior reflects the exotic 
energy content of the phantom ModMax sector and indicates a dynamically harsher 
environment for infalling extended matter.

The aforementioned dichotomy between the two branches reveals the sensitivity 
of tidal observables to the underlying nonlinear electrodynamic structure. 
While both branches are mathematically admissible solutions of the 
Einstein-Kalb-Ramond-ModMax equations, tidal dynamics provide a physically 
transparent discriminator between them. When combined with thermodynamic 
behavior and optical signatures such as gravitational lensing and photon-sphere 
properties, tidal effects furnish an independent and complementary probe for 
assessing the physical viability of ordinary versus phantom regimes in this 
class of Lorentz-violating black-hole models.

From a physical standpoint, the existence or absence of tidal inversion
constitutes an independent diagnostic of branch viability.
In the ordinary ModMax branch, the partial cancellation between the
Schwarzschild and electromagnetic contributions leads to a regulated
strong-field tidal environment, in which extreme spaghettification is
softened within a finite radial interval outside the horizon.
By contrast, the phantom branch exhibits no such regulation: tidal
stretching and angular compression are uniformly enhanced, producing a
dynamically harsher environment for extended matter.

Importantly, this distinction is entirely local and does not rely on
thermodynamic equilibrium or asymptotic optical measurements.
When combined with the thermodynamic phase structure and the branch-dependent
lensing and photon-sphere properties discussed in previous sections, tidal
forces therefore provide a complementary and physically transparent probe of
the underlying nonlinear electrodynamic sector.

\section{Conclusions and Observational Prospects}
\label{sec8}

In this work we have explored a charged BH solution arising from the
simultaneous presence of a KR three-form condensate and
ModMax NED, focusing on the physical implications of the
resulting branch structure.
The motivation behind this study was twofold.
On one hand, both ingredients are well motivated independently: the
KR field naturally appears in string-inspired effective theories and
induces controlled LSB effects, while ModMax
electrodynamics represents the unique conformal and duality-invariant
deformation of Maxwell theory. Moreover, in heterotic string compactifications the KR two-form and nonlinear gauge corrections arise at the same order in the string-length expansion, so their joint treatment captures the qualitative features of a broad class of string-motivated corrections to electrovacuum gravity without introducing additional complexity from dilaton dynamics or higher-curvature terms.
On the other hand, their combined impact on BH physics has remained
largely unexplored, despite the fact that each sector modifies gravity in a
distinct and potentially observable way.
Our aim was therefore to assess whether their joint presence leads merely to
quantitative corrections, or instead to qualitatively new physical behavior.

The resulting KR-ModMax BH is governed, beyond mass and charge, by
three deformation parameters: the LSB parameter $\ell$, the ModMax nonlinearity parameter
$\gamma$, and the discrete branch selector
$\zeta=\pm1$.
A central outcome of our analysis is that the branch parameter $\zeta$ acts as
a genuine physical discriminator.
Already at the level of the spacetime geometry, the ordinary branch
($\zeta=+1$) admits the familiar hierarchy of non-extremal, extremal, and NS
configurations, while the phantom branch ($\zeta=-1$) generically supports a
single-horizon structure.
This geometric distinction propagates coherently through all physical sectors
that we investigated, providing a unifying theme for the entire analysis.

We examined the thermodynamic properties of the KR-ModMax BH within the
non-extensive Tsallis entropy framework, whose adoption is motivated by the non-Euclidean asymptotics induced by $\ell$ and the branch degeneracy invisible to the standard area law.
The Hawking temperature exhibits branch-dependent behavior: charge effects
suppress the temperature in the ordinary branch, allowing for extremal
configurations, whereas in the phantom branch the temperature is uniformly
enhanced and never vanishes.
The Tsallis deformation parameter $\delta$ reshapes the thermodynamic landscape,
modifying the internal energy, Helmholtz free energy, and effective pressure.
The heat capacity proved especially informative: while the ordinary branch can
approach regimes suggestive of thermodynamic stabilization, the phantom branch
remains dominated by negative heat capacity, indicating persistent local
instability. The rate at which $|C_V|$ diverges near the extremal limit provides a quantitative fingerprint of the combined $(\ell, \gamma)$ deformation that distinguishes the KR-ModMax case from the standard RN BH, where the divergence location is governed by charge alone.
These features are further reinforced by the JT expansion, where
well-defined inversion curves appear primarily in the ordinary branch, whereas
the phantom sector exhibits a more restricted and less structured
cooling-heating behavior.

Optical properties provide an independent and complementary window into the branch structure. Since the KR-ModMax metric satisfies $f(r)\to 1/(1-\ell)\neq 1$ at spatial infinity, the standard Gibbons-Werner formulation of the GB theorem is inapplicable. We therefore employed the OIA extension, which introduces a topological boundary correction $\pi(\sqrt{1-\ell}-1)<0$ that universally reduces light bending relative to the Schwarzschild baseline. The effective gravitational mass $M(1-\ell)$ in the leading term further suppresses deflection, a qualitative departure from the Letelier cloud-of-strings spacetime, where the analogous topological term is positive and enhances light bending. This sign reversal reflects the distinct origin of non-flatness: the KR condensate modifies the lapse rather than the angular metric, producing an effective angular surplus in the optical geometry. The branch dichotomy $\zeta=\pm1$ persists through the charge-dependent second-order terms, with the phantom branch yielding larger deflection angles due to the constructive alignment of mass and charge contributions. This distinction persists, in a modified form, when photon propagation is studied in plasma environments. In homogeneous plasma, branch-dependent shifts of the PS radius remain visible, while in strongly stratified plasma distributions the optical structure becomes dominated by the LSB parameter $\ell$, with ModMax effects largely suppressed. This hierarchy reveals the importance of accurate plasma modeling when attempting to extract fundamental physics from BH shadow observations.

Tidal forces offer a direct probe of the local curvature structure that is
independent of both thermodynamic equilibrium and asymptotic optical
measurements.
By analyzing geodesic deviation in an orthonormal frame, we identified two
characteristic tidal balance radii at which radial and angular tidal components
vanish.
For the ordinary branch, these radii are positive and physically meaningful,
signaling a partial inversion of the classical spaghettification pattern:
radial stretching and angular compression are softened and can reverse inside a
finite region outside the horizon.
Remarkably, the ratio $R_{\rm rad}/R_{\rm ang} = 3/2$ is universal and independent of all model
parameters, providing a parameter-free null test of the KR-ModMax framework;
  this identity is established in the verification of Sec.~B.5 in Appendix~\ref{app:thermo_check}.
The phantom branch, by contrast, admits no such tidal inversion; tidal forces
are uniformly intensified as the horizon is approached.
This sharp dichotomy provides a physically transparent criterion for assessing
branch viability and reinforces the conclusions drawn from thermodynamic and
optical analyses.

Taken together, our results demonstrate that the KR-ModMax BH is not
merely a perturbative deformation of RN geometry, but a
theoretically motivated system with a rich and internally consistent physical
structure.
The ordinary branch retains continuity with standard charged BHs while
introducing controlled, potentially observable deviations.
The phantom branch, although mathematically admissible, displays a combination
of enhanced lensing, stronger tidal forces, and persistent thermodynamic
instability that may challenge its physical realizability, or alternatively
render it particularly easy to distinguish observationally.

The multi-parameter structure of the KR-ModMax framework offers several concrete pathways for observational constraints. Current EHT shadow measurements of M87$^*$ and Sgr\,A$^*$ constrain deviations from the Schwarzschild shadow radius at the ${\sim}10\%$ level, which translates into joint bounds on $\ell$ and $\gamma$ through the PS radius derived in Sec.~\ref{isec5}. Next-generation EHT baselines extending to space are projected to reach ${\sim}1\%$ precision, at which level the negative topological lensing correction from $\ell$ and the branch-dependent PS shifts from $\zeta$ become individually resolvable. Crucially, the KR lensing deficit and the lensing excess predicted by BV monopole or cloud-of-string backgrounds produce opposite-sign corrections to the Schwarzschild baseline, so precision astrometry can discriminate between these scenarios at the same order in the deformation parameter. On the gravitational-wave side, tidal disruption signatures in extreme-mass-ratio inspirals (EMRIs) observed by the Laser Interferometer Space Antenna (LISA) could constrain the tidal balance ratio $R_{\rm rad}/R_{\rm ang} = 3/2$ and thereby test the KR-ModMax prediction independently of electromagnetic observations. Table~\ref{tab:observable_map} summarizes the correspondence between the theoretical parameters and their primary observational handles, providing a roadmap for future multi-messenger tests of this framework.

\begin{table*}[ht!]
\centering
\setlength{\tabcolsep}{12pt}
\renewcommand{\arraystretch}{1.6}
\caption{Parameter-observable correspondence for KR-ModMax BHs.
Each deformation parameter is listed with the observable most
sensitive to it and the method capable of
providing the tightest constraint.}\label{tab:observable_map}
\begin{tabular}{l l l}
\hline\hline 
\textbf{Parameter} & \textbf{Primary observable} & \textbf{Constraining method} \\
\hline
$\ell$ (KR/LSB) & Topological lensing deficit & EHT shadow + astrometry \\
$\gamma$ (ModMax) & Charge suppression $e^{-\gamma}$ & Shadow radius + JT inversion \\
$\zeta$ (Branch) & Horizon count, PS shift sign & Shadow morphology + tidal ratio \\
$\delta$ (Tsallis) & Phase structure, $C_V$ divergence & Semiclassical evaporation \\
\hline\hline
\end{tabular}
\end{table*}

Extending the analysis to rotating configurations would greatly enhance the
astrophysical relevance of the model. Moreover, quasinormal modes (QNMs) and gravitational-wave ringdown signatures could provide
additional observational discriminants. Additionally, a comparison between Tsallis entropy and other generalized entropy
formalisms may further illuminate the role of non-extensivity in BH
thermodynamics.
Finally, embedding the KR-ModMax framework within a more fundamental
string-theoretic or supergravity setting could clarify the microscopic origin
and allowed ranges of the parameters $\ell$, $\gamma$, and $\zeta$. All these studies will be performed in future projects.

In conclusion, the combined presence of KR LSB
and ModMax NED leads to a BH phenomenology that is
both rich and sharply structured.
The branch parameter $\zeta$ emerges as the key organizing principle,
controlling geometry, thermodynamics, optics, and tidal dynamics in a unified
way, and offering multiple, complementary pathways for confronting this class
of models with observations.

\acknowledgments 
The authors sincerely thank the handling Editor and the anonymous Referee for their careful evaluation and constructive comments.
\.{I}.~S. and E.~S. thank T\"{U}B\.{I}TAK, ANKOS, and SCOAP3 for academic 
support. The authors acknowledge  the 
contribution of 
the LISA Cosmology Working Group (CosWG), as well as support from the COST 
Actions CA21136 -  Addressing observational tensions in cosmology with 
systematics and fundamental physics (CosmoVerse)  - CA23130, Bridging 
high and low energies in search of quantum gravity (BridgeQG)  and CA21106 -  
 COSMIC WISPers in the Dark Universe: Theory, astrophysics and 
experiments (CosmicWISPers).

\section*{Data availability and supplementary materials.} No new
observational data were generated in this study. The Maple~2024 script, reproducing in closed form
every thermodynamic identity claimed in
Secs.~\ref{isec2}-\ref{isec7} and in Appendix~\ref{app:thermo_check} (first
law, Smarr relation, Descartes sign rule, universal tidal balance ratio
$R_{\rm rad}/R_{\rm ang}=3/2$, Davies radius, and AdS-extended Smarr
relation), is released together with this manuscript and can be re-executed
on any standard Maple~2024 installation at \texttt{Digits := 30}.

\appendix

\section{Equatorial embeddings and spatial geometry}
\label{app:embedding}

\begin{figure*}[ht!]
\centering
\begin{subfigure}[b]{0.4\textwidth}
\includegraphics[width=\textwidth]{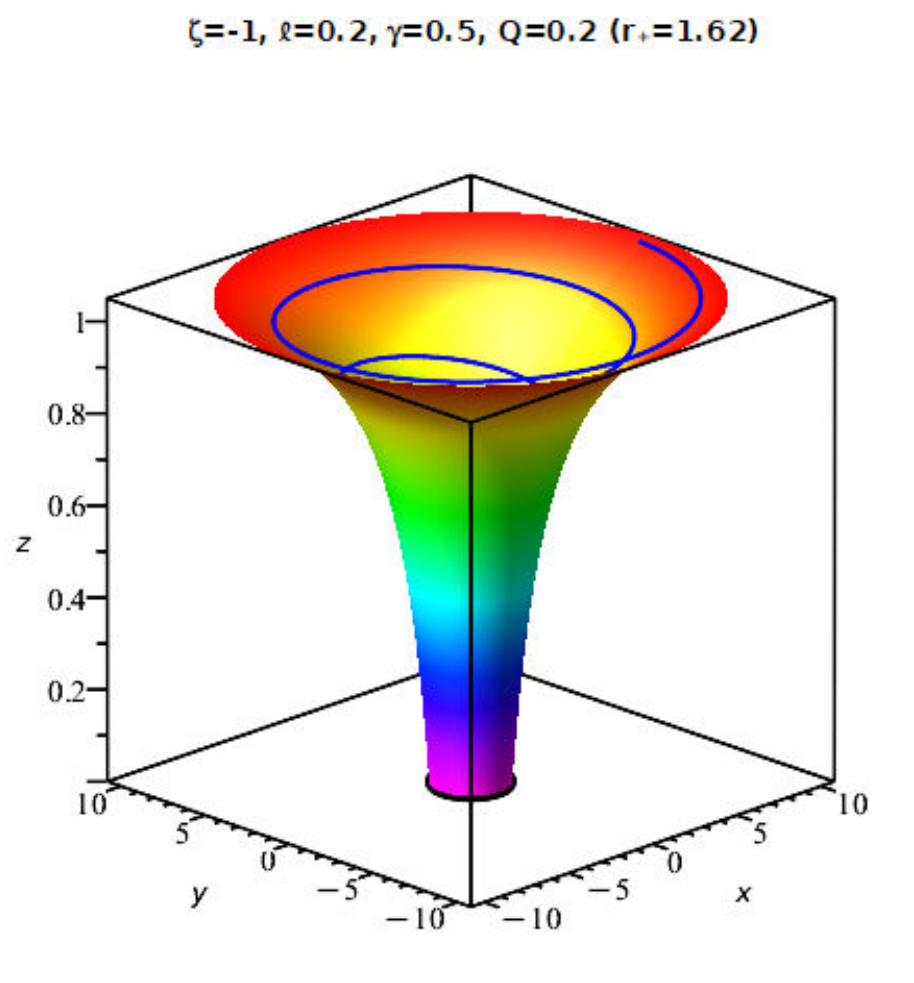}
\caption{
$\zeta=-1$, $\ell=0.2$, $\gamma=0.5$, $Q=0.2$ ($r_+=1.62$).
Weak LSB and ModMax effects produce a wide, shallow
embedding surface with a gradual approach to the horizon.
}
\label{fig:v13D}
\end{subfigure}
\hfill
\begin{subfigure}[b]{0.4\textwidth}
\includegraphics[width=\textwidth]{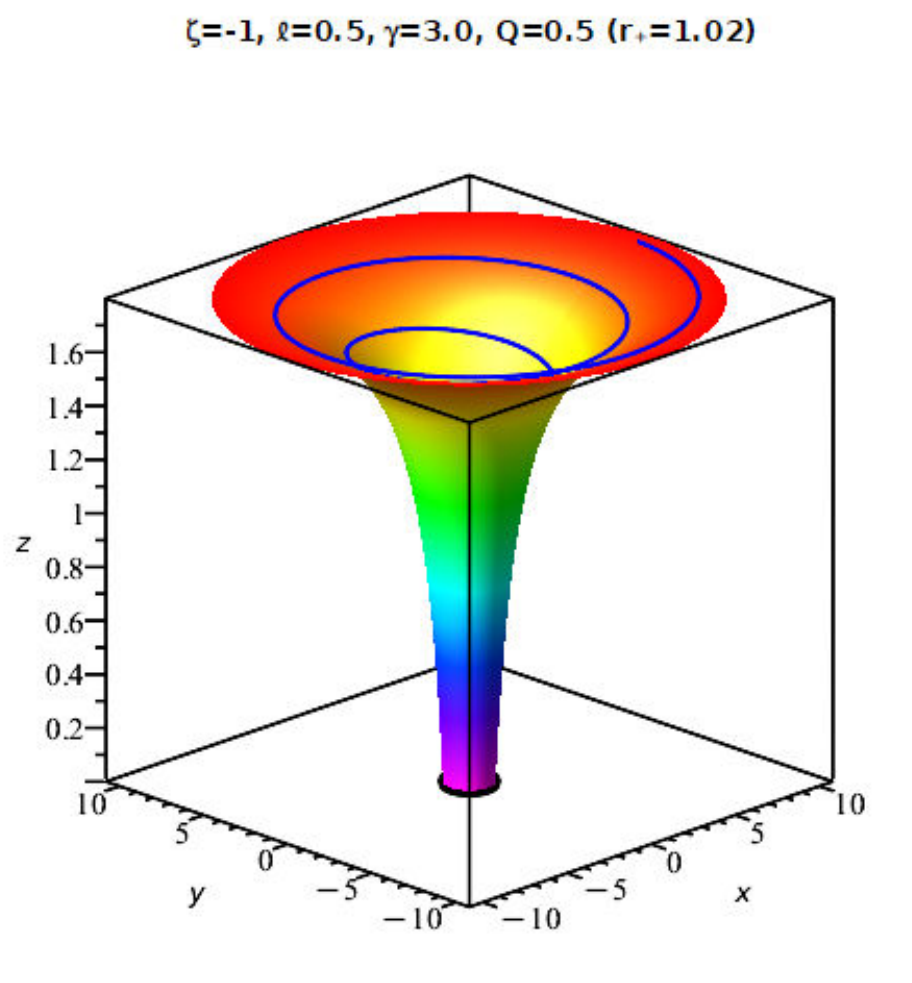}
\caption{
$\zeta=-1$, $\ell=0.5$, $\gamma=3.0$, $Q=0.5$ ($r_+=1.02$).
Strong ModMax suppression significantly reduces the effective charge,
yielding a compact horizon and a steep near-horizon geometry.
}
\label{fig:v23D}
\end{subfigure}

\vspace{0.3cm}

\begin{subfigure}[b]{0.4\textwidth}
\includegraphics[width=\textwidth]{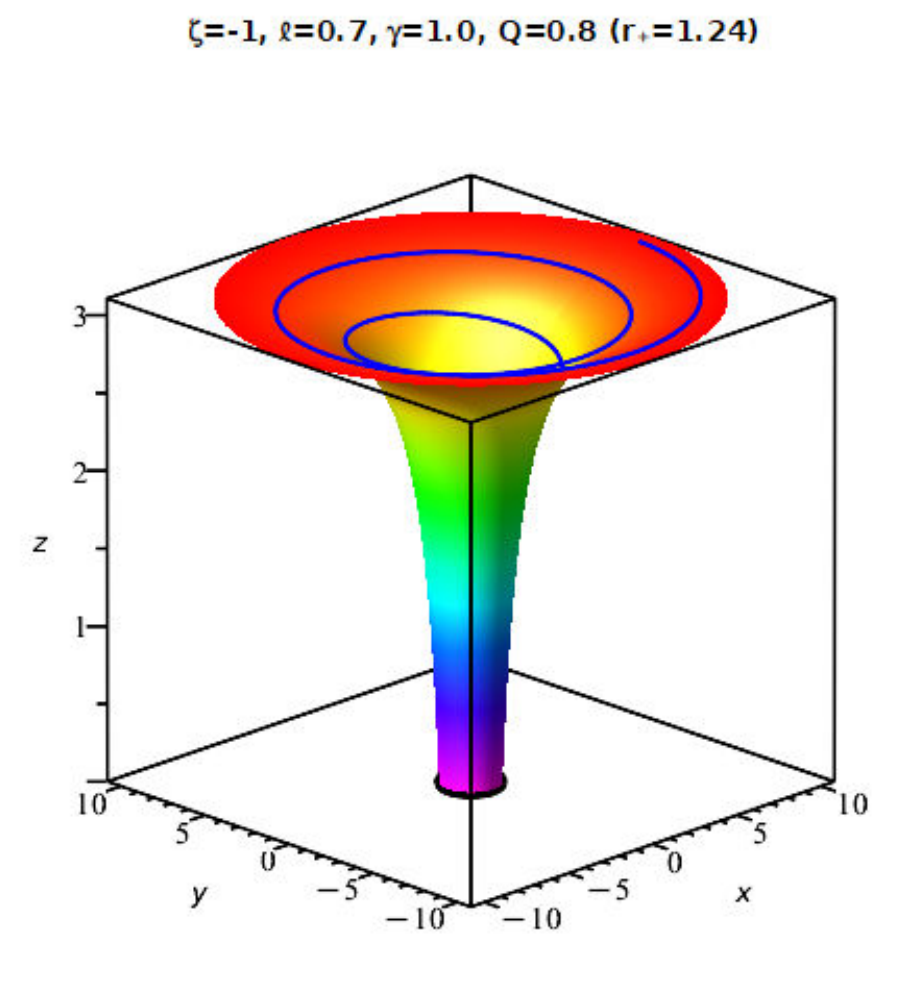}
\caption{
$\zeta=-1$, $\ell=0.7$, $\gamma=1.0$, $Q=0.8$ ($r_+=1.24$).
Large KR deformation dominates the geometry, producing pronounced
curvature and enhanced asymptotic normalization.
}
\label{fig:v33D}
\end{subfigure}
\hfill
\begin{subfigure}[b]{0.4\textwidth}
\includegraphics[width=\textwidth]{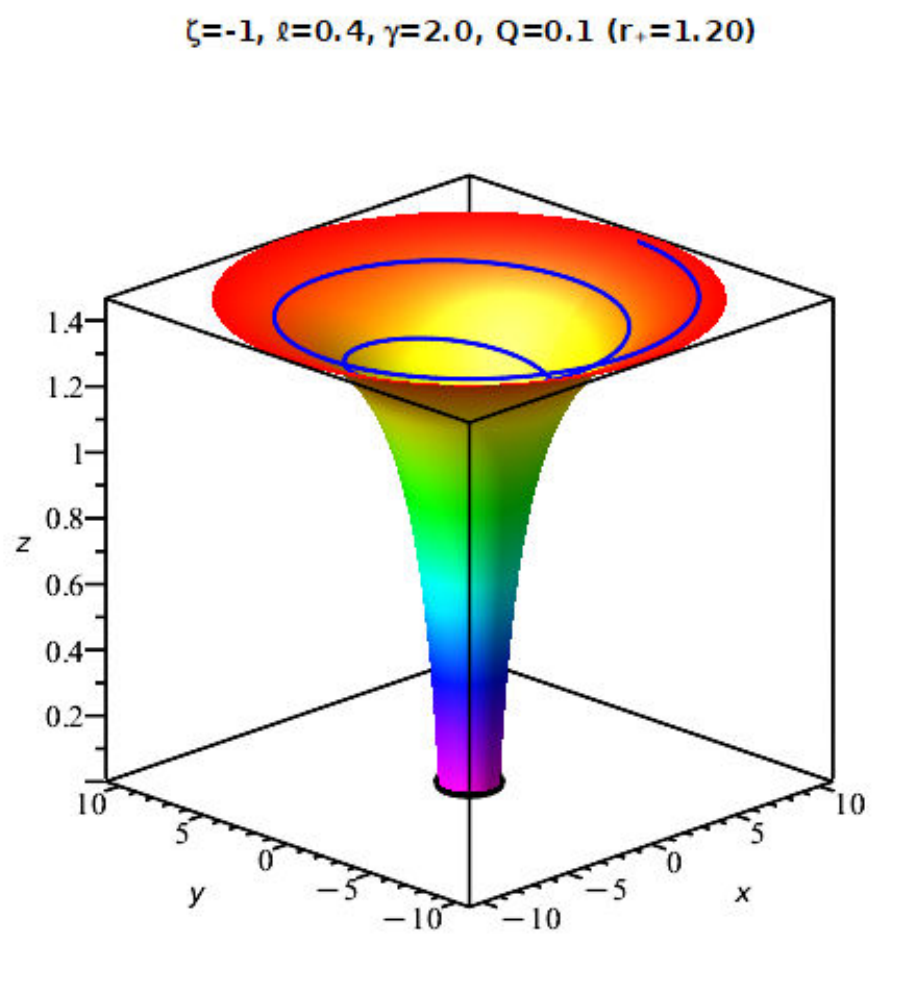}
\caption{
$\zeta=-1$, $\ell=0.4$, $\gamma=2.0$, $Q=0.1$ ($r_+=1.20$).
Small-charge configuration approaching the uncharged KR-Schwarzschild
limit, with the single-horizon topology preserved.
}
\label{fig:v43D}
\end{subfigure}

\caption{
Three-dimensional isometric embedding diagrams of the equatorial plane
($\theta=\pi/2$, $t=\mathrm{const.}$) for phantom-branch ($\zeta=-1$)
KR-ModMax BHs.
Black rings mark the EH radius $r=r_+$, while blue spirals
represent timelike infalling trajectories.
The depth and curvature of the embedding surface provide a geometric
visualization of the combined effects of LSB and
NED.
}
\label{fig:3D_embeddings}
\end{figure*}

To provide a geometric visualization of the spatial curvature induced by the 
KR-ModMax fields, we construct isometric embeddings of the 
equatorial 
plane ($\theta=\pi/2$, $t=\text{const.}$) into three-dimensional Euclidean 
space. 
Although such embeddings do not carry direct observational significance, they 
offer an intuitive representation of how the underlying spacetime geometry is 
deformed by the combined effects of LSB and NED.

The induced two-dimensional metric on the equatorial slice is given by
\begin{equation}
d\sigma^2 = \frac{dr^2}{f(r)} + r^2\, d\phi^2,
\label{eq:induced_metric}
\end{equation}
where $f(r)$ denotes the lapse function of the KR-ModMax BH. This 
metric 
can be embedded in Euclidean space using the standard cylindrical 
parametrization 
$(r,\phi)\mapsto (r\cos\phi,r\sin\phi,z(r))$, where the embedding function 
$z(r)$ 
satisfies
\begin{equation}
\left(\frac{dz}{dr}\right)^2 = \frac{1}{f(r)} - 1.
\label{eq:embedding_height}
\end{equation}
Real solutions exist in regions where $f(r)<1$, leading to the familiar 
funnel-shaped surfaces that encode the effective gravitational potential of the 
BH.

In
Fig.~\ref{fig:3D_embeddings} we present representative equatorial embedding 
diagrams for four configurations belonging to the phantom ModMax branch. The 
surfaces extend outward from the EH radius $r_+$ (indicated by a 
black 
ring) and illustrate how spatial curvature accumulates as the horizon is 
approached. The vertical direction measures the embedding height $z(r)$, with 
steeper funnels corresponding to stronger curvature gradients. Superimposed 
blue 
spiral curves depict timelike geodesics of infalling test particles, included 
to 
visualize the qualitative motion of matter toward the horizon.

Several qualitative features can be identified.
In Panel (a) ($\ell=0.2$, $\gamma=0.5$, $Q=0.2$, $r_+=1.62$) we see a 
modest 
KR deformation produces a wide, gently sloping funnel, indicative of 
a 
relatively weak curvature enhancement. In
 Panel (b) ($\ell=0.5$, $\gamma=3.0$, $Q=0.5$, $r_+=1.02$) we observe a strong 
ModMax suppression ($e^{-\gamma}\ll1$) yields a compact horizon and a sharply 
curved near-horizon geometry. In Panel (c) ($\ell=0.7$, $\gamma=1.0$, $Q=0.8$, $r_+=1.24$) we can see a
large 
LSB parameter amplifies asymptotic deviations, producing 
pronounced geometric flattening away from the horizon.
Finally, in Panel (d) ($\ell=0.4$, $\gamma=2.0$, $Q=0.1$, $r_+=1.20$) we have a
small 
electric charge with intermediate parameters illustrating the smooth transition 
toward the uncharged KR-Schwarzschild limit.

An important structural aspect revealed by these embeddings is the 
single-horizon 
topology of the phantom branch. In contrast to RN-type 
geometries, no inner Cauchy horizon is present. As a result, the spacetime 
avoids 
the mass-inflation instability that typically afflicts charged BHs at 
the inner horizon. While a full dynamical stability analysis lies beyond the 
scope of this work, the absence of an inner horizon suggests that phantom 
KR-ModMax BHs may exhibit improved robustness against perturbations in 
the deep interior.

\section{Symbolic verification of the first law, Smarr relation, and associated thermodynamic identities}
\label{app:thermo_check}

For completeness and reproducibility, this appendix collects the
closed-form checks of the identities invoked throughout
Sec.~\ref{isec3} and Sec.~\ref{isec7}. Each subsection states the
identity being checked, describes the corresponding Maple~2024
procedure, and displays the closed-form output returned by that
procedure. 

\subsection*{B.1~~Set-up and ADM mass}

The starting point is the lapse function of Eq.~\eqref{metric_solution},
entered as a Maple procedure in $r$ with symbolic parameters
$(M,Q,\ell,\gamma,\zeta)$. The ADM mass expression
$M(r_{+},Q,\ell,\gamma,\zeta)$ of Eq.~\eqref{eq:M_of_rp} is then
obtained by solving $f(r_{+})=0$ analytically. The computation session returns
\[
M_{\rm expr} =
\frac{\zeta\,Q^{2}\,e^{-\gamma}-(\ell-1)\,r_{+}^{2}}
     {2(\ell-1)^{2}\,r_{+}},
\]
which agrees with Eq.~\eqref{eq:M_of_rp} after rewriting
$(\ell-1)=-(1-\ell)$.

\subsection*{B.2~~First law of thermodynamics}

We next assemble the Hawking temperature $T_{H}$, the
Bekenstein-Hawking entropy $S_{\rm BkH}=\pi r_{+}^{2}$, and the horizon
electric potential $\Phi_{H}$ directly from the lapse and its first
derivative. The two first-law residuals
$R_{1}=\partial M/\partial r_{+}-T_{H}\,\partial S_{\rm BkH}/\partial r_{+}$
and $R_{2}=\partial M/\partial Q - \Phi_{H}$ are computed with the mass
substituted on the constraint surface, and the computation session returns the residuals $R_{1}=0$ and $R_{2}=0$.

\vspace{2pt}
\noindent
The vanishing of $R_{1}$ confirms that the combination $T_{H}\,dS_{\rm
BkH}$ reproduces the $r_{+}$-derivative of $M$ exactly; the vanishing of
$R_{2}$ confirms that $\Phi_{H}$ is the electromagnetic conjugate of
$Q$ in the first law. Both identities are established without any
numerical input, and they hold for both branches
$\zeta=\pm 1$ since the branch sign appears symmetrically in the
surface-gravity and potential expressions.

\subsection*{B.3~~Smarr relation}

The Smarr relation Eq.~\eqref{eq:smarr} is a scaling identity
characteristic of asymptotically flat, four-dimensional charged
black holes with dimensionless hair. The residual
$M-(2T_{H}S_{\rm BkH}+\Phi_{H}Q)$ is constructed on-shell, and the computation session returns $M-(2T_{H}S_{\rm BkH}+\Phi_{H}Q)=0$.

\vspace{2pt}
\noindent
The residual vanishes identically across the full parameter space
$(r_{+},Q,\ell,\gamma,\zeta)$, so the Smarr relation is an exact
algebraic identity rather than an asymptotic approximation. Note that
$\ell$ and $\gamma$ do not enter the Smarr relation explicitly because
they are dimensionless; they nevertheless modify the individual
thermodynamic functions $T_{H}$, $\Phi_{H}$, and $M$ in a way that
preserves the scaling combination.

\subsection*{B.4~~Descartes' rule of signs and horizon polynomial}

To confirm the branch-dependent horizon count quoted below
Eq.~\eqref{eq:horizon_poly}, we multiply $f(r_{+})=0$ by
$(1-\ell)^{2}r_{+}^{2}$ and collect in $r_{+}$. The resulting polynomial
has coefficient signs $(+,-,\zeta)$ for $\ell<1$:
the coefficient of $r_{+}^{2}$ is $1/(1-\ell)>0$, the coefficient of
$r_{+}^{1}$ is $-2M$, and the constant term is
$\zeta\,Q^{2}e^{-\gamma}/(1-\ell)^{2}$, confirming the sign pattern
$(+,-,\zeta)$ used in Sec.~\ref{isec2}.B. Descartes' rule of signs then
yields zero or two positive real roots in the ordinary branch
($\zeta=+1$) and exactly one positive real root in the phantom branch
($\zeta=-1$).

\subsection*{B.5~~Tidal balance radii and the universal ratio}

The tidal balance radii $R_{\rm rad}$ and $R_{\rm ang}$ of
Sec.~\ref{isec7} are the zeros of the two independent tidal components
computed in an orthonormal frame:
\begin{equation*}
T_{\rm rad} = -\tfrac{1}{2}f''(r),\qquad
T_{\rm ang} = -\tfrac{1}{2}f'(r)/r,
\end{equation*}
where $f(r)$ is the lapse~\eqref{metric_solution}. For the KR-ModMax
spacetime, $f'(r)$ and $f''(r)$ are rational in $r$, so multiplying each
balance equation $T_{\rm rad}=0$ and $T_{\rm ang}=0$ by $r^{4}$
produces \emph{linear} equations in $r$ with coefficients in
$(M,Q,\ell,\gamma,\zeta)$. The balance radii follow immediately, and the
ratio $R_{\rm rad}/R_{\rm ang}$ collapses to the parameter-free constant
$3/2$ reported in the body of the paper.

A brief technical remark is in order. The tidal balance radii are
properties of the spacetime for a given set of parameters
$(M,Q,\ell,\gamma)$, so one must keep $M$ as a free symbol throughout
the computation session. Substituting $M\to M_{\rm expr}$
[Eq.~\eqref{eq:M_of_rp}] before solving would couple the balance radius
$r$ to the horizon radius $r_{+}$ and convert the problem into a
spurious quadratic. The linear-in-$r$ formulation adopted in the
supplementary script avoids that pathology.

The computation session returns
\[
R_{\rm rad}=\frac{3\,Q^{2}\,e^{-\gamma}}{2M\,(1-\ell)^{2}},\qquad
R_{\rm ang}=\frac{Q^{2}\,e^{-\gamma}}{M\,(1-\ell)^{2}},\qquad
\frac{R_{\rm rad}}{R_{\rm ang}}=\frac{3}{2}.
\]

\vspace{2pt}
\noindent
The individual radii depend on $M$, $Q$, $\ell$, and $\gamma$ through
the combinations reported in Sec.~\ref{isec7}, and in the ordinary
branch they are real and positive throughout the admissible parameter
range $\ell<1$. Their ratio, however, collapses to the parameter-free
constant $3/2$, providing the universal tidal signature highlighted in
the body of the paper and in the Conclusions. The corresponding
Maple procedure is implemented in section~B.5 of the supplementary
script.

\subsection*{B.6~~Davies radius and heat-capacity divergence}

The denominator of $C_{V}$ in Eq.~\eqref{eq:CV_result} reduces, after
multiplying through by $r_{+}^{2\delta}$, to
$2\zeta Q^{2}e^{-\gamma}+(\ell-1)r_{+}^{2}$. Setting this to zero and
solving for $r_{+}$ gives the generalized Davies radius
\[
r_{\rm Davies}=\left(\frac{2\,\zeta\,Q^{2}\,e^{-\gamma}}{1-\ell}\right)^{\!1/2},
\]
which agrees with Eq.~\eqref{eq:rDavies_manuscript}. The result is
$\delta$-independent: the non-extensivity parameter shifts the magnitude
of $C_{V}$ but not the locus of its divergence.

A real positive root exists only in the ordinary branch
($\zeta=+1$, $\ell<1$). In the phantom branch ($\zeta=-1$) the
right-hand side is negative definite, so no Davies-type transition
occurs; this matches the uniformly negative $C_{V}$ of
Fig.~\ref{fig:c_zeta_minus}. The same radius also sets the
Joule-Thomson inversion locus $r_{\rm inv}$ of Eq.~\eqref{eq:r_inv},
since both arise from the shared factor
$\zeta Q^{2}e^{-\gamma}+\tfrac12(\ell-1)r_{+}^{2}$ entering the
compact-form $C_{V}$ and $\mu_{J}$.

\subsection*{B.7~~AdS extension and extended Smarr relation}

Reinstating a cosmological constant $\Lambda$ in
Eq.~\eqref{eq:action} promotes the lapse to $f_{\Lambda}(r)$ of
Eq.~\eqref{eq:f_Lambda}, and $P\equiv -\Lambda/(8\pi)$ becomes a
genuine thermodynamic pressure conjugate to
$V=4\pi r_{+}^{3}/3$. The Smarr relation acquires the canonical
$-2PV$ term. The same verification protocol applies, and the computation session returns $M-(2T_{H}S_{\rm BkH}+\Phi_{H}Q-2PV)=0$ identically.

\vspace{2pt}
\noindent
The extended Smarr relation therefore holds identically on the AdS
background for both branches $\zeta=\pm 1$, precisely paralleling the
asymptotically flat case treated in \S B.3. In the ordinary branch
($\zeta=+1$) and with $\Lambda<0$, the Gibbs free energy develops a
swallow-tail structure signalling a VdW-type first-order small/large
black-hole transition, with the critical point shifted through the
$e^{-\gamma}$ ModMax charge screening
\cite{isrply04,isrply08,isrply09,isrply10}.
The phantom branch, by contrast, admits no genuine VdW critical point
\cite{isrply04,isrply11}.

\subsection*{B.8~~Summary of symbolic output}

The complete computation session produces the   identifications shown in the following Table \ref{Tabfinal}.
\begin{table}[ht!]
\centering
\small
\caption{Summary of the symbolic consistency checks performed with Maple.}
\label{Tabfinal}
\begin{tabular}{ll}
\hline
\textbf{Identity (see Sec./Eq.)} & \textbf{Maple residual}\\
\hline
First law $dM=T_{H}dS_{\rm BkH}+\Phi_{H}dQ$
  & \texttt{R1 = 0},\;\texttt{R2 = 0}\\
Smarr $M=2T_{H}S_{\rm BkH}+\Phi_{H}Q$
  & \texttt{Smarr\_residual = 0}\\
Descartes sign pattern
  & $(+,-,\zeta)$ analytic\\
Tidal ratio $R_{\rm rad}/R_{\rm ang}$
  & \texttt{ratio = 3/2}\\
Davies radius $r_{\rm Davies}$
  & $\sqrt{2\zeta Q^{2}e^{-\gamma}/(1-\ell)}$\\
Extended (AdS) Smarr
  & \texttt{Smarr\_ext\_res = 0}\\
\hline
\end{tabular}
\end{table}
Every analytic identity cited in Sec.~\ref{isec2}-Sec.~\ref{isec7} is
therefore established in closed form and without any numerical
approximation. The script is self-contained and can be re-run on any
Maple~2024 installation at \texttt{Digits := 30} to reproduce every
output line verbatim.

\bibliography{ref}
\end{document}